\documentclass[aps,prd,10pt,twocolumn,showpacs,groupedaddress,amsmath,amssymb,superscriptaddress,nofootinbib,floatfix,noeprint]{revtex4-2} 

\usepackage[dvipsnames]{xcolor}
\usepackage{array}
\usepackage{graphicx}
\usepackage{dcolumn}
\usepackage{multirow} 
\usepackage{makecell}
\usepackage{bm}
\usepackage[colorlinks,linkcolor=blue,urlcolor=blue,citecolor=blue]{hyperref}
\usepackage{amsmath}
\usepackage{longtable}
\usepackage{color}
\usepackage[utf8]{inputenc}
\usepackage[bottom]{footmisc}
\usepackage[T1]{fontenc}
\usepackage{mathptmx}
\usepackage{url}
\usepackage{relsize}
\usepackage{orcidlink}
\usepackage{tabularx}

\newcommand\apjl{Astrophys. J. Lett.}
\newcommand\apjs{Astrophys. J. Suppl. Ser.}

\newcommand\mnras{Mon. Not. R. Astron. Soc.}
\newcommand\aap{Astron. Astrophys.}

\newcommand\nphysa{Nucl. Phys. A}
\newcommand\epja{Eur. Phys. J. A}
\newcommand\plb{Phys. Lett. B}
\newcommand\jpg{J. Phys. G}
\newcommand\ppnp{Prog. Part. Nucl. Phys}

\newcommand\skipme[1]{}

\begin{document}

\title{Correlated and uncorrelated Monte Carlo neutron capture rate variations\\ in weak \textit{r}-process simulations}

\author{Atul~Kedia\,\orcidlink{0000-0002-3023-0371}}
\email{akedia@nccu.edu}
\affiliation{Department of Physics and Astronomy, North Carolina State University, Raleigh, North Carolina 27695, USA}
\affiliation{Department of Mathematics and Physics, North Carolina Central University, Durham, North Carolina 27707, USA}
\author{Jeffrey~M.~Berryman\,\orcidlink{0000-0002-9663-693X}}
\affiliation{Lawrence Livermore National Laboratory, Livermore, California 94550, USA}
\author{Jonathan~Cabrera~Garcia\,\orcidlink{0009-0006-7257-913X}}
\affiliation{Department of Physics and Astronomy, University of Notre Dame, Notre Dame, Indiana 46556, USA}
\author{Jutta~E.~Escher\,\orcidlink{0000-0002-0829-9153}}
\affiliation{Lawrence Livermore National Laboratory, Livermore, California 94550, USA}
\author{Oliver~C.~Gorton\,\orcidlink{0000-0003-3643-9640}}
\affiliation{Lawrence Livermore National Laboratory, Livermore, California 94550, USA}
\author{Erika~M.~Holmbeck\,\orcidlink{0000-0002-5463-6800}}
\affiliation{Lawrence Livermore National Laboratory, Livermore, California 94550, USA}
\author{Gail~C.~McLaughlin\,\orcidlink{0000-0001-6811-6657}}
\affiliation{Department of Physics and Astronomy, North Carolina State University, Raleigh, North Carolina 27695, USA}
\author{Cole~D.~Pruitt\,\orcidlink{0000-0003-0607-9461}}
\affiliation{Lawrence Livermore National Laboratory, Livermore, California 94550, USA}
\author{Andre~Sieverding\,\orcidlink{0000-0001-8235-5910}}
\affiliation{Lawrence Livermore National Laboratory, Livermore, California 94550, USA}
\author{Rebecca~Surman\,\orcidlink{0000-0002-4729-8823}}
\affiliation{Department of Physics and Astronomy, University of Notre Dame, Notre Dame, Indiana 46556, USA}

\date{\today}

\begin{abstract}

Reliable predictions of weak rapid neutron capture (\emph{r}-process) abundances require a systematic treatment of nuclear physics uncertainties, especially neutron capture rates far from stability. We employ new neutron capture rates from cross sections calculated with Yet Another Hauser-Feshbach Code (\texttt{YAHFC}) using an uncertainty-quantified Koning-Delaroche potential modified for use on neutron-rich systems. Using these rates as a baseline, we perform Monte-Carlo studies with independent rate variations (uncorrelated Monte Carlo) and find correlations between specific neutron capture rates and the resulting elemental abundances for the three weak r-process scenarios: two separate simulations of neutron star merger remnant accretion disks and a simulation of a magnetorotational supernova. We discuss the underlying nuclear dynamics that give rise to these correlations and the role of astrophysical conditions in them. We demonstrate how reducing the uncertainty in these rates would improve the prospects for conducting precision \emph{r}-process studies in the future. We additionally present a correlated Monte Carlo study, which incorporates the full covariance matrix that describes the relationships between individual neutron capture rates that arise from an uncertainty-quantified optical potential. We find that the magnitude of the uncertainty in the abundance pattern is similar to that produced by an uncorrelated Monte Carlo that employs only the on-diagonal components of the covariance matrix. We show how correlations restructure how the abundances co-vary, but do not necessarily decrease the overall uncertainty envelope.

\end{abstract}

\maketitle

\section{Introduction}
\label{sec:introduction}
Understanding the astrophysical origins of the heavy elements has been a longstanding challenge. While it has been known since the 1950s that about half of the Solar system abundances of the elements heavier than iron were formed via rapid neutron capture (\emph{r}-process) nucleosynthesis \citep{BBFH1957}, important questions remain in both the nuclear physics of the \emph{r} process and the astrophysical site or sites of production. The past decade has seen significant advances in both areas. The first active \emph{r}-process site was identified in the multimessenger neutron star merger event GW170817 \citep{LIGO-GW170817} accompanied by the \emph{r}-process-powered kilonova AT2017gfo \citep{LIGO-GW170817-MMA}, and recently the number of well-characterized metal-poor, \emph{r}-process enhanced stars has been doubled by the \emph{R}-Process Alliance \citep{Hansen+2018,Sakari+2018,Holmbeck+2020,Bandyopadhyay+2024}. 
Potential astrophysical sites such as binary neutron star mergers \citep{Meyer1989,Just+2015,Bovard+2017,Radice:2018ghv,Fujibayashi+2023,Foucart:2024kci,PhysRevResearch.5.013168,PhysRevResearch.6.033078,Fryer_2024,PhysRevResearch.5.043106}, black-hole neutron star mergers \citep{Surman:2008qf,Darbha+2021,Curtis+2023,Wanajo+2024}, collapsars \citep{Surman+2006,Siegel+2019,Miller+2020}, magnetorotational supernovae \citep{Nishimura+2015,Mosta+2018,Reichert:2020mjo,Zha+2024,Liu:2025auu}, magnetar winds \citep{Patel+2025}, and more exotic scenarios \citep{Fuller+2017,Fischer+2020} have been explored. Efforts to connect the nucleosynthetic yields of these events to observables such as stellar abundances and light curves are impeded by uncertainties in the astrophysical conditions of the potential sites as well as in the nuclear physics of the thousands of nuclear species far from stability that participate in the $r$ process \cite{Zhu:2020eyk,Barnes:2020nfi,Lund:2022bsr,Holmbeck:2022mog}.

On the nuclear physics side, extensive experimental and theoretical efforts have gone into understanding the nuclear properties of the extremely neutron-rich species populated in an \emph{r}-process event. These efforts have been particularly successful in the lighter-mass regions of the nuclear chart, where measured values are now available for many of the masses \citep{Reiter+2020,Li+2022} and $\beta$-decay half lives \citep{Yokoyama+2019,Xu+2023} required for \emph{r}-process calculations. This low-mass range of the \emph{r}-process abundance pattern is dubbed the ``weak \emph{r} process.'' It can occur in a broader range of astrophysical conditions compared to the ``main'' \emph{r} process, which is responsible for elements up to and including uranium and thorium. Given the impressive pace of experimental progress, we look forward to the near future when the ground state nuclear data (masses and halflives) for a weak \emph{r} process are known, and the major remaining nuclear physics uncertainties lie in the neutron-induced reaction rates.

Determining neutron capture rates is challenging both experimentally and theoretically.  Experimentally, direct measurements of neutron-induced reactions on radioactive isotopes remain a formidable technical challenge, due to the short life times of the target nuclei~\citep{Reifarth+2017}.  Reliable calculations are also extremely challenging, since the majority of the nuclei involved in the astrophysical \emph{r} process are far enough away from stability that one has to carefully consider the validity of the models originally developed to describe capture on stable and near-stable species.

Theoretical predictions of neutron capture rates play a critical role in nuclear astrophysics simulations, as many rates have not been measured.  Hauser-Feshbach statistical reaction calculations are widely used 
in this context 
and significant effort has been devoted to improving the models that enter the computation of reaction rates~\cite{Capote:09, Hilaire:21}.  Structural properties, such as level densities (LDs) and $\gamma$-ray strength functions (GSFs) can now be predicted using microscopic nuclear structure theories~\cite{Alhassid:08, Alhassid:2015a, Shimizu:16, Karampagia:20, Ormand:20, Tsoneva:2016aa, Goriely:18a, Goriely:19a, Goriely:2019aa}.  Optical-model potentials (OMPs) used are often still phenomenological, with efforts underway focusing on improved descriptions away from the valley of stability~\cite{HebbornNPDH2022, Pruitt:25cnr, Dickhoff:19}, integration of microscopic structure components~\cite{Bauge:00a, Blanchon:15, Bostrom:25}, and quantification of uncertainties~\cite{CatacoraRios:21, 2023PhRvC.107a4602P,Beyer-PRC-2025, Dimitrakopoulos:2025}. 

Much progress has been made to obtain experimental constraints for statistical reaction calculations of neutron capture.  For unstable isotopes, indirect methods, such as the surrogate reaction method~\cite{Escher:12rmp, Escher:16a, Escher:2025puf} and the Oslo and Shape methods~\cite{Larsen:19, Ingeberg:2020aa, Wiedeking:21} have become an important tools.  The surrogate reactions method has been used in regular and inverse kinematics to obtain neutron capture cross sections from charged-particle inelastic scattering and transfer reactions~\cite{Escher:18prl, Ratkiewicz:19prl, PerezSanchez:20, Sguazzin:25prl, 2026PhLB..87640433T}.  A variant of the Oslo method, the $\beta$-Oslo method, has been applied to determine LDs and GSFs for several compound nuclei of interest, including $^{75}$Ge \citep{Spyrou:2014}, $^{69}$Ni \citep{Liddick:2016}, $^{68}$Ni \citep{Spyrou:2017}, and $^{87}$Kr \citep{Mucher:2023}.  Combining the LDs and GSFs and a neutron-nucleus OMP enables the calculation of neutron capture rates.
We look forward to the potent combination of these techniques with the expanded reach of, e.g., the Facility for Rare Isotope Beams~\citep[FRIB;][]{Horowitz+2019,FRIB-motivation-2024}. In light of these new capabilities, here we revisit the role of neutron capture reactions in a weak \emph{r} process.

Many studies have been devoted to understanding the role of neutron capture rates in the main $r$-process
\citep{Surman+2001,Surman+2009,Arcones+2011,Mumpower+2012,Mumpower+2015}, adding to the rich literature on $r$-process sensitivities to mass models and $\beta$-decay rates.Recently, uncertainties in \emph{r}-process abundance patterns stemming from modern mass models have been investigated \cite{2025EPJA...61...48M, 2025A&A...694A.180M}. Reference~\cite{2025A&A...694A.180M} conducted uncorrelated Hartree-Fock-Bogoliubov (HFB) mass model parameter variations and showed that the local variations of parameters in a single HFB mass model lead to smaller uncertainties than model uncertainties, i.e., switching from one HFB model to another.

However, investigation of neutron capture in the weak \emph{r} process has been sparser.
Previously, \citet{Surman:2014} identified the individual isotopes whose neutron capture rates produce the strongest impact on the overall weak \emph{r}-process isotopic abundance pattern through sensitivity studies in which individual reaction rates were varied. In addition, \citet{Vescovi+2022} conducted individual neutron capture rate sensitivity studies using individual astrophysical trajectories with parameters consistent with neutron star merger environments, locating neutron capture rates that have the largest impact on the overall abundance pattern and heating rate. Another approach is to perform Monte Carlo rate variations and look for abundance-rate correlations. This method has been successfully applied to investigate reaction sensitivities of the $\nu$p-process~\citep{Nishimura:2019} and the ($\alpha$, $n$) reactions that shape weak \emph{r}-process conditions within $\alpha$-rich freeze-out \citep{2020PhRvC.101e5807B, 2022ApJ...935...27P}. 

In this paper, we take a first step towards a complete account of nuclear physics uncertainties as they impact models of the weak \emph{r} process. We address two questions: (1) How might improvements to neutron capture reaction rates for a few select nuclei reduce the uncertainty of predicted isotopic abundances, even if we cannot yet model the underlying correlations among rates? (2) What more can we learn about uncertainty in abundance patterns by explicitly propagating physics-based correlations among capture rates all the way through element synthesis calculation? To answer these questions, we present a series of Monte Carlo studies that use different assumptions about physics-based correlations among neutron capture rates. We focus on neutron capture rate variations only while keeping all other nuclear inputs fixed (such as $\beta$-decay rates, $\beta$-delayed neutron emission probabilities and nuclear masses). A comprehensive study would necessarily involve varying all relevant nuclear physics inputs to neutron capture rate calculations, but such an approach is not currently feasible with well-quantified uncertainties.  We take a first step towards answering the above questions by focusing on one model input---the OMP---for which systematic uncertainties have already been established.

We choose sets of astrophysical conditions extracted from hydrodynamical simulations of neutron star merger accretion disk ejecta and ejecta from magnetohydrodynamical (MHD) supernovae. We focus on the resulting $r$-process elemental patterns, and we use uncorrelated Monte Carlo studies to identify neutron capture rates that correlate strongly with elemental yields. Our neutron capture rate calculations use an uncertainty-quantified OMP, which we subsequently use to extract correlations between individual neutron capture rates. We then perform an additional set of Monte Carlo studies, incorporating these correlations when generating the rate variations, and compare the outcomes with those from the uncorrelated studies.

This paper is organized as follows: In Section~\ref{sec:nuclear-astrophysics-model} we detail the three sets of astrophysical conditions adopted for our studies, and we describe the methodology of our neutron capture rate calculations, their variations, and our analysis. In Section~\ref{sec:uncorrelated} we present the results of our Monte Carlo studies, which employ uncorrelated rate variations. We focus first on the correlations between individual neutron capture rates and elemental abundances as determined by the first set of Monte Carlo studies.
We then turn to uncorrelated rate variations, where a reduced variation distribution width is adopted for the 35 most consequential neutron capture rates. In Section \ref{sec:correlated}, we compare the results of uncorrelated rate variations with correlated rate variations. 
We provide concluding remarks in Section~\ref{sec:conclusion}.

\section{Method}
\label{sec:nuclear-astrophysics-model}

In this study, we consider covariance matrices that describe uncertainties in neutron capture rates on individual nuclei as well as the correlations between these rates.  We assess the impact of \emph{correlated}, i.e., using a full covariance matrix,
and of \emph{uncorrelated}, i.e., using a purely on-diagonal covariance matrix, neutron-capture reaction rates on the weak \emph{r} process, beginning with the nuclear physics underlying statistical calculations of neutron-capture cross sections. 
By comparing these two cases, we examine to what extent constraints near stability may inform astrophysical observables that are influenced by nuclei farther from stability. We analyze the effect of correlated and uncorrelated rates in the context of the weak \emph{r} process by running
nucleosynthesis simulations for three sets of astrophysical conditions, as described in Sec.\ \ref{subsec:astro-model}, using the nuclear data choices described in Sec.\ \ref{subsec:nuclear-model} as a starting point. Our Monte Carlo simulations involve sampling neutron capture rates using a variety of covariance matrices as described in Sec.\ \ref{subsec:MC-sampling}. 

\subsection{Astrophysics: Weak \emph{r}-process conditions}
\label{subsec:astro-model}

We explore three sets of astrophysical simulations spanning thermodynamic conditions and event histories representative of the weak \emph{r} process as presently understood, for example in Refs. \cite{Sprouse+2024,Bernuzzi+2025}, providing our analysis with an appropriate spread of bases for judging the impact of uncertain reaction rates. We choose a subset of simulation tracer particles (the time evolution of temperature and density for individual mass packets in the simulation) from each of the three simulations rather than the entire tracer particle set, which may be in the thousands, in order to save on computational time, which we justify below.
Nucleosynthesis is calculated from these ``thermodynamic trajectories'' in a standard, post-processed way using the nuclear reaction network code \texttt{PRISM} \citep{PRISM}. Our three simulation choices are as follows:

\begin{enumerate}
    \item the postmerger disk wind simulations from~\citet{2024ApJ...964..111L}, specifically
    from the ``b10'' simulation that includes a strong magnetic field, $\beta =  P_{gas}/P_{magnetic} = 10$. We choose 10 trajectories whose summed abundance pattern 
    approximates the full yields of the weak \emph{r}-process component of this simulation. We use the designation ``L+24'' to refer to this event. 
    \item the disk wind simulation from \citet{2014MNRAS.441.3444M} as in \citet{2017MNRAS.472..904L}. We conduct our study for a sample of 10 trajectories selected by \citet{Holmbeck:2022mog} that fit the \emph{r}-process abundance pattern of the \emph{r}-process enhanced metal-poor star HD~222925 \citep{2022ApJS..260...27R}. We use the designation ``MF14'' for these trajectories.
    \item the magnetorotational supernova simulation from \citet{Reichert:2020mjo}, specifically ``35OC-Rs.'' We include the 10 representative trajectories each from the weak \emph{r} process and the Fe-weak \emph{r}-process groups for a total of 20. These were placed into groups by Ref.~\cite{Reichert:2020mjo} at the time the simulation was performed.  As weak \emph{r}-process elements are synthesized both groups, we need to consider them together. We refer to this combined set of simulations with the prefix ``R+21.''
\end{enumerate}

Table \ref{table:astrophysics} lists the trajectories used in this study along with some of their pertinent evolution parameters. One of these is the electron fraction, $Y_e = \frac{n_p}{n_p+n_n}$. The events have weighted average electron fractions at 10 GK of $\langle Y_e \rangle_{\mathrm{L+24}} = 0.27$, $\langle Y_e \rangle_{\mathrm{MF14}} = 0.26$, and $\langle Y_e \rangle_{\mathrm{R+21}} = 0.37$. Weights are relative contributions by each trajectory within the respective scenario.  Note that although some authors create representative trajectories with equal weights and others use unequal weights, this choice should not affect our conclusions.

\begin{table}[!htbp]
    \renewcommand{\arraystretch}{1.2}
    \caption{Astrophysical scenarios and individual trajectories. The labels ``L+24,'' ``MF14,'' and ``R+21'' refer to the simulations of \citet{2024ApJ...964..111L}, \citet{2014MNRAS.441.3444M}, and \citet{Reichert:2020mjo}, respectively.     The ``Identifier'' column indicates how the trajectory is labeled in the original source, $Y_e$ is the electron fraction at 10 GK, $s$ is the entropy per baryon in units of Boltzmann constant $k_B$ at 10 GK calculated based on Eq.\ (25) of \citet{2004ApJ...603..611S}, and $t_{1\,\mathrm{GK}}-t_{2.5\,\mathrm{GK}}$ is the time interval between temperatures of 2.5 GK and 1.0 GK. The column labeled ``Weight'' gives the relative mass weight of each trajectory. Within each scenario, all weights sum to unity.}
    \begin{tabular}{@{\extracolsep{5pt}}c c c c c c @{}}
        \hline \hline
        \multirow{2}{*}{Label} & \multirow{2}{*}{Identifier} & \multirow{2}{*}{$Y_e$} & $s$ & $t_{1\,\mathrm{GK}}-t_{2.5\,\mathrm{GK}}$& \multirow{2}{*}{Weight} \\
         & & & [$k_B$] & [s] \\ \hline
        L+24-01 & 00349547 & 0.30 & 19 & 0.072 & 0.124\\
        L+24-02 & 00349580 & 0.22 & 16 & 0.084 & 0.017\\
        L+24-03 & 00570724 & 0.25 & 16 & 0.072 & 0.237\\
        L+24-04 & 00681327 & 0.33 & 19 & 0.085 & 0.092\\
        L+24-05 & 01013135 & 0.23 & 15 & 0.044 & 0.081\\
        L+24-06 & 01444920 & 0.28 & 18 & 0.069 & 0.172\\
        L+24-07 & 01580853 & 0.27 & 17 & 0.071 & 0.076\\
        L+24-08 & 01691441 & 0.31 & 22 & 0.073 & 0.054\\
        L+24-09 & 01788947 & 0.27 & 17 & 0.055 & 0.085\\
        L+24-10 & 02347330 & 0.31 & 19 & 0.085 & 0.061\\ \hline
        MF14-01 & 00618 & 0.30 & 20 & 3.3 & 0.100\\
        MF14-02 & 01400 & 0.34 & 28 & 0.81 & 0.100\\
        MF14-03 & 01559 & 0.29 & 18 & 1.1 & 0.100\\
        MF14-04 & 02731 & 0.29 & 20 & 0.52 & 0.100\\
        MF14-05 & 06181 & 0.29 & 18 & 0.70 & 0.100\\
        MF14-06 & 07455 & 0.29 & 17 & 0.70 & 0.100\\
        MF14-07 & 08530 & 0.12 & 12 & 0.63 & 0.100\\
        MF14-08 & 09283 & 0.14 & 12 & 0.51 & 0.100\\
        MF14-09 & 09641 & 0.40 & 24 & 0.10 & 0.100\\
        MF14-10 & 09774 & 0.15 & 13 & 0.061 & 0.100\\ \hline
        R+21-01 & weak-r-1 & 0.30 & 18 & 12 & 0.027\\
        R+21-02 & weak-r-2 & 0.24 & 13 & 2.8 & 0.027\\
        R+21-03 & weak-r-3 & 0.23  & 13 & 1.6 & 0.027\\
        R+21-04 & weak-r-4 & 0.33 & 19 & 2.0 & 0.027\\
        R+21-05 & weak-r-5 & 0.30 & 20 & 4.4 & 0.027\\
        R+21-06 & weak-r-6 & 0.31 & 18 & 0.96 & 0.027\\
        R+21-07 & weak-r-7 & 0.30 & 19 & 7.1 & 0.027\\
        R+21-08 & weak-r-8 & 0.37 & 23 & 1.5 & 0.027\\
        R+21-09 & weak-r-9 & 0.36 & 22 & 0.85 & 0.027\\
        R+21-10 & weak-r-10 & 0.36 & 25 & 5.0 & 0.027\\
        R+21-11 & fe-weak-r-1 & 0.38 & 21 & 2.4 & 0.073\\
        R+21-12 & fe-weak-r-2 & 0.41 & 24 & 1.6 & 0.073\\
        R+21-13 & fe-weak-r-3 & 0.35 & 19 & 3.0 & 0.073\\
        R+21-14 & fe-weak-r-4 & 0.34 & 21 & 1.2 & 0.073\\
        R+21-15 & fe-weak-r-5 & 0.37 & 24 & 2.5 & 0.073\\
        R+21-16 & fe-weak-r-6 & 0.44 & 25 & 2.3 & 0.073\\
        R+21-17 & fe-weak-r-7 & 0.39 & 23 & 1.6 & 0.073\\
        R+21-18 & fe-weak-r-8 & 0.40 & 25 & 3.1 & 0.073\\
        R+21-19 & fe-weak-r-9 & 0.44 & 21 & 2.7 & 0.073\\
        R+21-20 & fe-weak-r-10 & 0.43 & 22 & 2.8 & 0.073\\
        \hline \hline
    \end{tabular}
    \label{table:astrophysics}
\end{table}

\subsection{Nuclear physics inputs}
\label{subsec:nuclear-model}

For our nucleosynthesis simulations, the nuclear datasets we employ include the JINA Reaclib database \citep{cyburt2010jina} for charged particle reactions, the NUBASE 2020 database \citep{2021ChPhC..45c0001K} for $\beta$-decay properties, and AME2020 \citep{2021ChPhC..45c0003W} for nuclear masses (both measured and recommended values). Where experimental $\beta$-decay rates and the AME2020
masses are not available, we use FRDM2012 separation energies \citep{FRDM} and $\beta$-decay rates and $\beta$-delayed neutron emission probabilities from \citet{2019ADNDT.125....1M}. 
Masses are also among the required inputs for neutron-capture cross-section calculations. For our case,
we ensure a smooth transition to the region beyond the AME2020 masses as well as consistency with the separation energies we use in Portable Routines for Integrated nucleoSynthesis Modeling (\texttt{PRISM}) by combining the heaviest AME2020 mass along each isotopic chain with the one-neutron separation energies ($S_n$) implied by FRDM2012 for species without AME2020 masses.

As we are studying the weak \emph{r} process, we focus on isotopes with $Z = 20$ to 60.
We evaluate neutron capture cross sections in the Hauser-Feshbach (HF) statistical formalism with the code Yet Another Hauser-Feshbach Code (\texttt{YAHFC}) \citep{YAHFC}, which uses Monte Carlo sampling to simulate ensembles of compound-nuclear decays.
The HF cross section for neutron-induced reactions can be
written schematically as
\begin{equation}\label{eq: hf}
	\sigma_{n,x}(E_n) = 
    \sum_{J^\pi}
	\sigma_{n}^{CN}(E,J^\pi)
    G_{x}^{CN}(E,J^\pi),
\end{equation}
where $\sigma_{n}^{CN}(E,J^\pi)$ is the compound nucleus (CN) formation cross
section; $G_{x}^{CN}(E,J^\pi)$ is the CN decay probability for a given channel
$x = \gamma, n, 2n, ...$; $E$ and $J^\pi$ are the energy and spin-parity of the
CN, while $E_n$ is the energy of the incident neutron. The formation cross
section is given by
\begin{equation}\label{eq:formation}
    \sigma_{n}^{CN}(E,J^\pi)
    = \pi \lambdabar^2 g_J T_n^{CN}(E),
\end{equation}
where $\lambdabar$ is the de Broglie wavelength of the neutron and $g_J$ is a statistical factor accounting for angular momentum of the neutron, target, and CN.
The \emph{transmission coefficient} $T_n^{CN}(E)$, which describes the interaction between the neutron and the CN, is obtained from a suitable OMP.
We omit details pertaining to sums over spins and
orbital angular momentum channels, as well as other standard correction terms (e.g., width fluctuations); for a full treatment of the formalism see the
review \cite{Escher:12rmp}.
The CN decay
probability, which is assumed to be independent of the formation, is given
schematically by
\begin{equation}\label{eq:decay}
    G_{x}^{CN}(E,J^\pi) = 
    \frac{\mathcal{T}_{x}(E, J^\pi)}
    {\sum_{x'} \mathcal{T}_{x'}(E, J^\pi)},
\end{equation}
where for each exit channel $x'$, the generalized transmission coefficient
$\mathcal{T}_{x'}$ is given by the integral of a particle transmission
coefficient $T_{x'}$ and residual-channel LD $\rho_{x'}(E_{x'})$:
\begin{equation}\label{eq:gent}
    \mathcal{T}_{x'}(E, J^\pi) = \int T_{x'}(E_{x'})\rho_{x'}(E_{x'})dE_{x'},
\end{equation}
where $E_{x'}$ is the nuclear excitation energy in the channel $x'$. 
The transmission coefficients and LD inputs are required for each channel entering into the
denominator of Eq.~\eqref{eq:decay}. For neutrons, the transmission
coefficients are obtained from the same OMP used to compute the CN
formation cross section. Photon transmission coefficients are computed using GSFs.

Several physics model choices underlie the HF method, including the LD prescription, GSFs \cite{Alhassid:2015aa, Hilaire:23, Goriely:25, Sun:25, 2026PhRvC.113d4327G}, and the OMP \cite{Bauge:01, Whitehead:21}. For the present study, we use the default LD and GSF forms available in the HF code \texttt{YAHFC}. The level density scheme utilized is the standard Gilbert-Cameron scheme \cite{gilbert1965composite}, i.e., a combination of a constant-temperature formula (at low excitation energies) with a Fermi-gas description (at higher energies).  The main contributions to the $\gamma$-ray strength function are the electric dipole (\textit{E}1) and magnetic (\textit{M}1) dipole components.  For the former, we select a modified Lorentzian form \cite{Goriely:19a} and the latter follows a simple Lorentzian (without an added low-energy enhancement).  Additional functional forms, as well as microscopically calculated LDs and GSFs, will be explored in future work. 
By default, \texttt{YAHFC} implements the LD and GSF models with the parametric forms and values recommended by RIPL-3 \cite{Capote:09}, as well as incorporating RIPL-3 discrete-level data.
We use the OMP to calculate not only the transmission coefficient for the incoming particle, but also the transmission coefficients, $T_{x}$, in Eqs.~\eqref{eq:formation} and \eqref{eq:gent}. 
 
We employ a variation of the Koning-Delaroche (KD) Uncertainty Quantified (KDUQ) OMP \citep{2023PhRvC.107a4602P} ``Federal'' ensemble, which
quantifies parametric uncertainties of the canonical KD OMP \cite{2003NuPhA.713..231K}. 
Because the Fermi-energy prescription in the KD OMP form is independent of neutron-proton asymmetry, it is unsuitable for neutron-rich systems at astrophysical energies, warranting a change to the KD model form. Instead of using the KD Fermi energy prescription, we directly compute Fermi energies from assessed nuclear masses as described below. We refer to this model-form variant of KD as Koning-Delaroche with Experimental Fermi energies (KDEF) \cite{osti_3022089}. We note that near $\beta$-stability, there is little deviation from the form used in KD, but near the neutron dripline, where the Fermi energy tends toward zero, the KDEF form is more physically meaningful. We refer to the combination of the KDEF model form and the KDUQ parameter vectors as KDUQEF, which is the OMP used to generate neutron-capture cross sections
in this work.

Neutron-capture cross sections calculated by \texttt{YAHFC} are converted into neutron-capture \emph{rates} for \texttt{PRISM} input by
\begin{equation}
    \lambda(T) = N_A \sigma_{\text{MACS}} v_T, \label{eqn:rates}
\end{equation}
where $\lambda(T)$ is the (reduced) reaction rate, $N_A$ is Avogadro's number, $\sigma_{\text{MACS}}$ is the Maxwellian-averaged cross section, and $ v_T$ is the thermal velocity, $v_T = \sqrt{2 k_B T/\mu}$, with $k_B$ being Boltzmann's constant and $\mu$ being the reduced mass of the compound nucleus.
The Maxwellian-averaged cross section is related to the \texttt{YAHFC}-calculated cross section through
\begin{equation}
    \sigma_{\text{MACS}} \propto \int_0^\infty E_n e^{-m_nE_n/\mu k_B T}\sigma_{n,x}(E_n)~dE_n, \label{eqn:macs}
\end{equation}
which effectively converts the energy-dependent cross section to a temperature-dependent one by assuming the particles' velocities occupy a thermal distribution.
In other words, in post-processed nucleosynthesis simulations, particles are assumed to thermalize instantly, which alleviates the need to track individual particle energies.

\begin{figure}
    \centering
    \includegraphics[width=\linewidth]{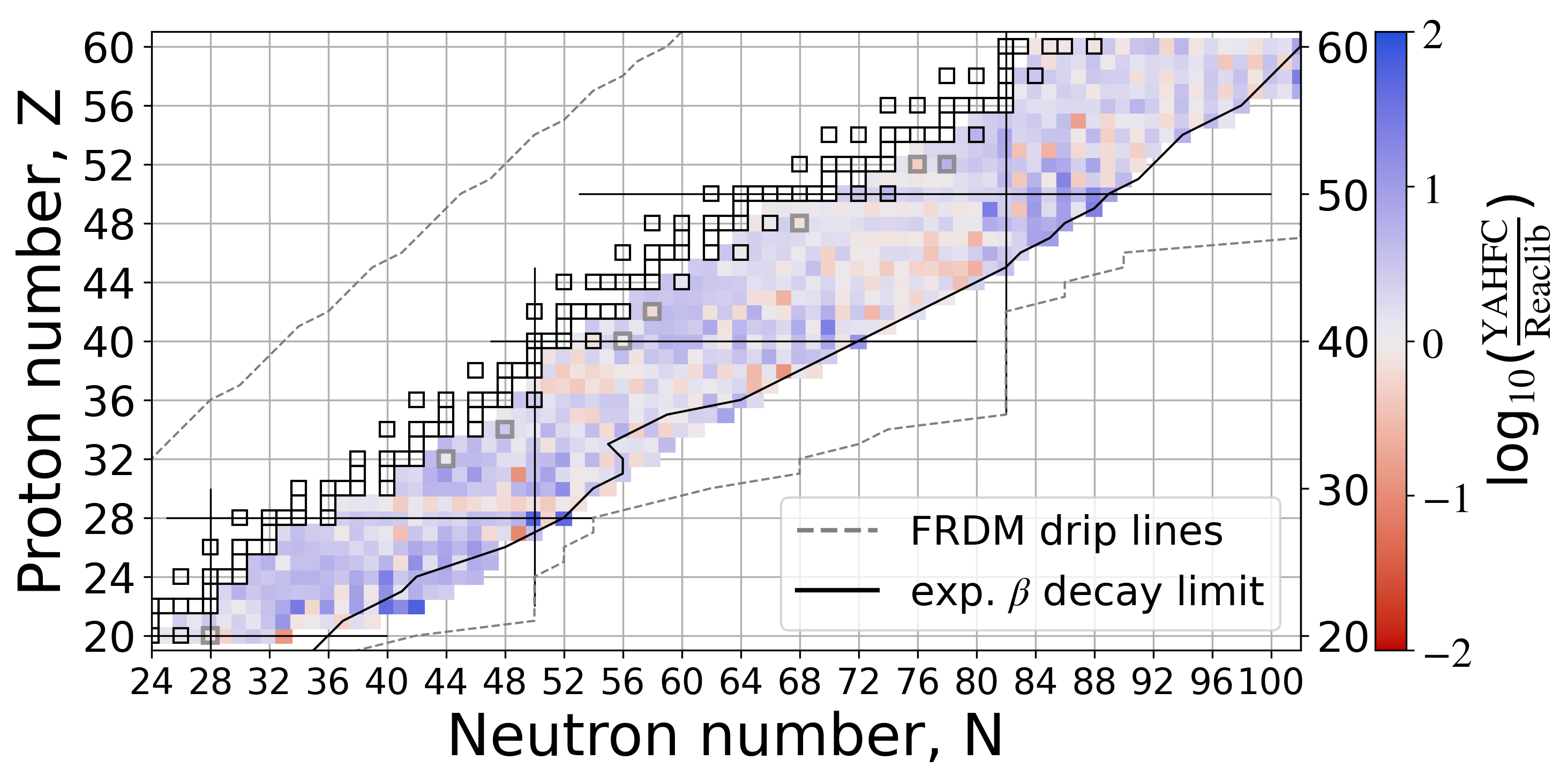}
    \includegraphics[width=\linewidth]{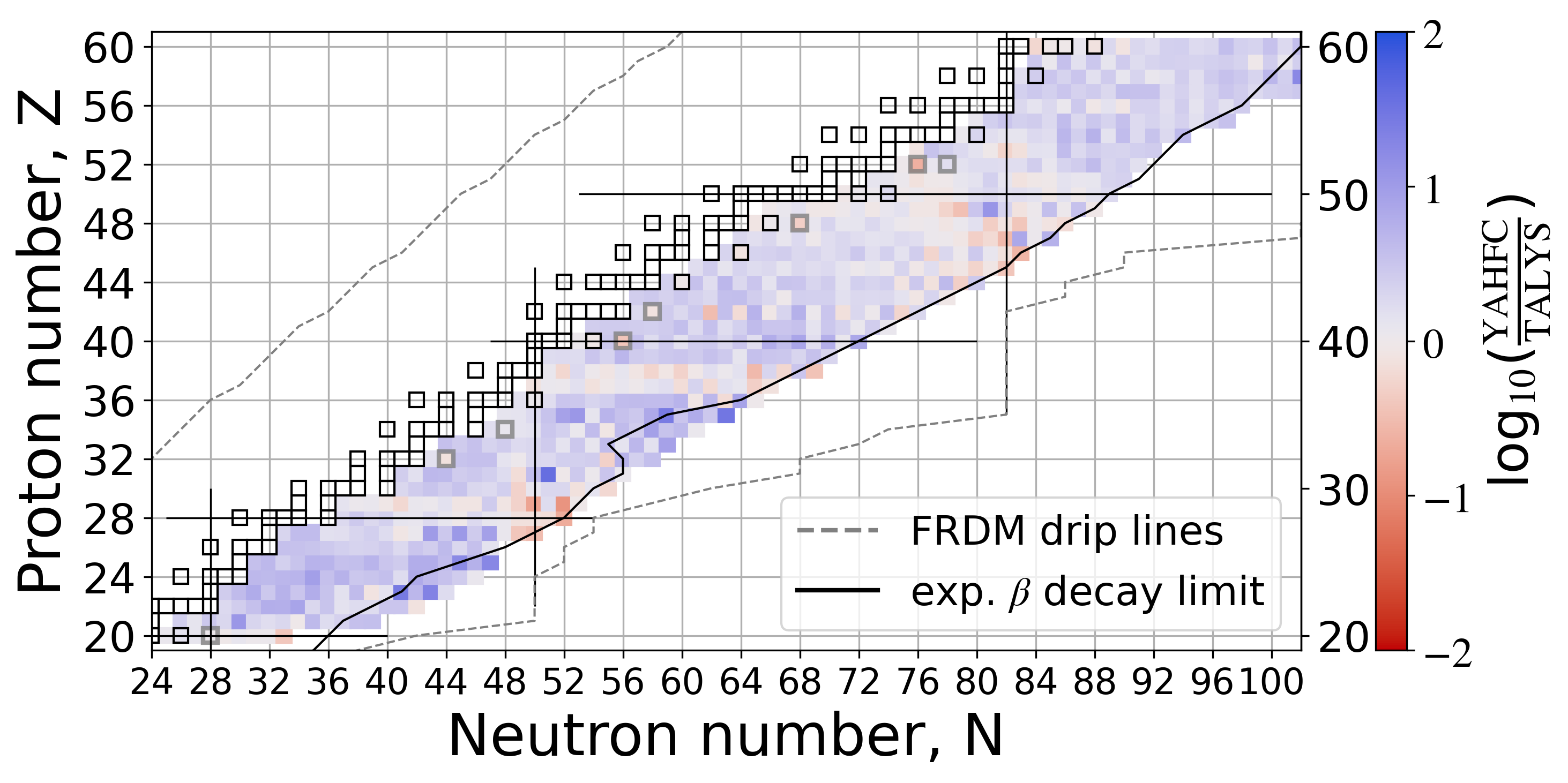}
    \caption{The logarithm of the ratio of our baseline \texttt{YAHFC} rates for $\approx$1000 nuclei to those of Reaclib (NON-SMOKER) (upper panel) and TALYS (lower panel) at 1 GK.  Our baseline rates use the KDEF optical model potential. 
    }
    \label{fig:chart-of-nuclides-included}
\end{figure}

We take 416 samples from the uncertainty-quantified OMP dataset, therefore yielding 
416 sets of transmission coefficients for each nuclear species.
With these 416 sets of transmission coefficients, we use \texttt{YAHFC} to calculate a new set of 416 neutron-capture cross section and, from it, 416 new astrophysical reaction rates for each of the $\approx$1000 nuclei in this study using Eqs.~\eqref{eqn:rates} and \eqref{eqn:macs}. We determine neutron-capture rates relevant for the weak \emph{r} process, i.e., for isotopes with $Z = 20$--60, in the temperature range of $\approx$1.16$\times10^{-3}$ GK to 11.6 GK (100 eV to 1 MeV).  These rates are therefore correlated through the underlying OMP by way of the transmission coefficients.  We note that at 1.16 GK the ratio of the KDUQEF standard deviation to KDEF rate, averaged across all isotopes, is 0.047. We show the set of nuclides for which we calculate and sample their neutron-capture rates in Fig.\ \ref{fig:chart-of-nuclides-included} and compare the KDEF reaction rate at 1 GK to the corresponding reaction rates in Reaclib (from NON-SMOKER \cite{non-smoker}) (top panel) and TALYS \citep{Koning+2023} (bottom panel). The TALYS rates are calculated from the default parameters: Fermi gas level densities, Simplified Modified Lorentzian $\gamma$-strength functions, and KD global OMPs.

The magnitude of a neutron capture rate also influences the corresponding photodissociation rate through detailed balance.
Specifically, $\lambda_{(\gamma,n)}^{Z,A+1} \propto \lambda_{(n,\gamma)}^{Z,A} e^{-S_n/k_B T}$, where $\lambda_{(\gamma,n)}^{Z,A+1}$ is the photodissociation rate of the $Z,A+1$ nucleus, $\lambda_{(n,\gamma)}^{Z,A}$ is the neutron-capture rate onto $Z$, $A$, and $S_n$ here refers to the one-neutron separation energy of the $Z$, $A+1$ nucleus. Through this expression, we can see how
increasing (decreasing) a neutron capture rate, $\lambda_{(n,\gamma)}^{Z,A}$, also increases (decreases) the reverse photodissociation rate,  $\lambda_{(\gamma,n)}^{Z,A+1}$. In the \texttt{PRISM} nucleosynthesis reaction network code, photodissociation rates are calculated dynamically to be consistent with the neutron capture rates.

\subsection{Monte Carlo studies}
\label{subsec:MC-sampling}

To understand the impact of uncertainties in neutron-capture rates on astrophysical weak \emph{r}-process yields, we conduct Monte-Carlo simulations of neutron-capture reaction rates relevant for the weak \emph{r}-process regime of nucleosynthesis, i.e., on the $\approx$1,000 isotopes in Fig.\ \ref{fig:chart-of-nuclides-included}.
The fiducial neutron-capture reaction rates are calculated using the \texttt{YAHFC} Hauser-Feshbach statistical model code \citep{YAHFC} using the KDEF OMP as described in Sec.\ \ref{subsec:nuclear-model}. As we explain later in this section, to improve the statistics 
of our uncertainty bands we use the 416 KDUQEF samples described in Sec.\ \ref{subsec:nuclear-model} to construct a covariance matrix from which we draw 5000 samples.

Reaction rates for each of the isotopes are sampled from a log-normal distribution centered at the fiducial reaction rate for any given species at 1.16 GK (0.1 MeV). We choose this temperature to roughly correspond to weak \emph{r}-process freeze-out; weighted average temperatures around freezeout, as defined in Section \ref{subsec:basecorrelations}, for each event are $\langle T_{\tau_{(n,\gamma)}\approx 2\tau_{\beta^-}, \mathrm{L+24}} \rangle = 0.58~\mathrm{GK}$, $\langle T_{\tau_{(n,\gamma)}\approx 2\tau_{\beta^-}, \mathrm{MF14}} \rangle = 1.22~\mathrm{GK}$, and $\langle T_{\tau_{(n,\gamma)}\approx 2\tau_{\beta^-}, \mathrm{R+21}} \rangle = 1.60~\mathrm{GK}$, respectively. The values of the rates at much higher or lower temperatures will have less effect on the final abundance pattern.
We scale the rates at all other temperatures by the same factor as the newly chosen rate at 1.16 GK,   
i.e., we implicitly make the simplification that the underlying energy-energy covariance matrix for the cross section is perfectly correlated. Because of the narrow window of temperature during which the neutron capture rates affect the final abundance pattern, we expect that this approximation will be sufficient for our purposes.  We leave to future work a detailed investigation of the role of the energy-energy correlation matrix.
With $\bm{y}=\log_{10}\bm{\lambda}$, we draw 5000 samples on $\bm{y}$ for a new set of rates using the probability distribution function of a multivariate Gaussian:
\begin{align}
    \mathrm{PDF}(\bm y) = & (2\pi)^{-N/2}|\bm\Sigma|^{-1/2} \times \notag \\ 
                          & \exp\left[-\frac{1}{2}( \bm y-\bm \mu_c)^\mathrm{T}\bm\Sigma^{-1}(\bm y-\bm\mu_c)\right],
    \label{eqn:probability}
\end{align}
where $\bm{\mu_c}$ are the centroids of the log-normal distribution of
the rates calculated with the baseline KDEF OMP, i.e.,
\begin{gather}
    \bm\mu_c = 
    \begin{bmatrix}
    \log_{10} \lambda_{\text{KDEF},1}\\ \log_{10}\lambda_{\text{KDEF},2} \\ \vdots\\ \log_{10}\lambda_{\text{KDEF},N}
    \end{bmatrix} ~,
\end{gather}
where the index $i= 1,2,...,N$ refers to nuclear species. For Eq.\ \ref{eqn:probability} we also construct a covariance matrix for the logarithms of the rates:
\begin{gather}
    \bm\Sigma = \begin{bmatrix} \label{eqn:matrix}
    \Sigma_{11} & \Sigma_{12} & \cdots & \Sigma_{1N}\\
    \Sigma_{21} & \Sigma_{22} &  & \vdots\\
    \vdots &  & \ddots & \\
    \Sigma_{N1} & \cdots &  & \Sigma_{NN}
    \end{bmatrix}.
\end{gather} 
For some studies in this work, we perform an uncorrelated Monte Carlo by choosing a diagonal covariance matrix such that $\Sigma_{i\neq j}=0$ in Eq.\ \ref{eqn:matrix}, and Eq.\ \ref{eqn:probability} therefore reduces to independent Gaussians, one for each nuclear species. 
We use this choice to make contact with previous studies, e.g., Refs. \cite{Mumpower+2012, Surman:2014, 2020arXiv201001698N}, and to compare to other studies in this work in which the full complexity of Eq.\ \ref{eqn:probability} is retained, i.e., by using the KDUQEF samples to construct a full covariance matrix.

In total, we perform seven Monte Carlo studies in this work, and our covariance matrices are summarized in Table \ref{tab:montecarlo-summary}.
All sets use KDEF for the centroids of the Gaussians, i.e., $\bm{\mu_c}$.
The sets differ in their choice of covariance matrix ($\bm{\Sigma}$).
For sets 1, 2A, and 2B, the KDUQEF covariance matrix is not used.
For sets 3A and 4A, only the diagonals of the KDUQEF covariance matrix are used.
Finally, sets 3B and 4B use the full KDUQEF covariance matrix.
We briefly describe each of the seven sets below.

%
%

\begin{table}
    \setlength{\tabcolsep}{16pt}
    \renewcommand{\arraystretch}{1.3}
    \caption{Summary of the sets of Monte Carlo samples discussed in Sec.~\ref{subsec:MC-sampling}.
    For each study, we summarize our choices for the diagonal ($\Sigma_{ii}$) and off-diagonal ($\Sigma_{i\neq j}$) elements of the covariance matrix used when drawing 5000 samples. Studies with $\Sigma_{i\neq j}=0$ are included when referring to ``uncorrelated Monte Carlo,'' whereas the studies with non-zero off-diagonals are ``correlated Monte Carlo'' studies.}
    \label{tab:montecarlo-summary}
    \begin{tabular}{l l l}
        \hline\hline
        \textbf{Study} & $\bm{\Sigma_{ii}}$ & $\bm{\Sigma_{i\neq j}}$\\
        \hline
        Set 1 & $(0.5)^2$ & 0 \\
        \hline
        Set 2A & $(0.1)^2$\footnote{For $Z,A$ in Table \ref{table:rate-abundance-correlation}} and $(0.5)^2$\footnote{For $Z,A$ not in Table \ref{table:rate-abundance-correlation}} & 0 \\
        Set 2B & $(0.1)^2$ & 0 \\
        \hline
        Set 3A & $\Sigma_{\rm KDUQEF}$ & 0 \\
        Set 3B & $\Sigma_{\rm KDUQEF}$ & $\Sigma_{\rm KDUQEF}$ \\
        \hline
        Set 4A & $100 \Sigma_{\rm KDUQEF}$ & 0 \\
        Set 4B & $100 \Sigma_{\rm KDUQEF}$ & $100 \Sigma_{\rm KDUQEF}$\\
        \hline\hline
    \end{tabular}
\end{table}

{\bf Set 1} {\it uncorrelated Monte-Carlo sampling, one order of magnitude uncertainty spread.} Our first covariance matrix contains only on-diagonal entries, each of which has a value of $\Sigma_{ii}=0.5^2$. This corresponds to a standard deviation of 0.5 on the log scale, which represents a half order of magnitude in the rate. As has been previously studied, neutron capture rates evaluated using different nuclear model inputs can disagree by a factor of two at stability \cite{1997PhRvC..56.1613R} but by an order of magnitude or more away from stability; see Fig.\ 2 in Ref. \citep{2020arXiv201001698N}  as well as \cite{Liddick:2016, Spyrou:2017}. Hence, our estimate of the log-normal standard deviation in this first set is exaggerated near stability but is more representative of the spread in model systematics away from stability in the region where the rates will shape \emph{r}-process yields.

\textbf{Set 2A:} {\it Uncorrelated Monte-Carlo sampling with reduced uncertainties.} In this set, we sample most of the rates as we do in the set 1 Monte Carlo study, but we take a subset of the rates that show the strongest correlations with the resulting elemental abundances
(which we determine in Sec.\ \ref{subsec:basecorrelations}) and reduce the standard deviation of the log-normal rates for each by a factor of 5, i.e., a standard deviation of 0.1, which results in $\Sigma_{ii}=0.1^2$, representing how decreased uncertainties from experiment for key nuclei may improve the predictions of the weak \emph{r} process.

\textbf{Set 2B:} {\it Uncorrelated Monte-Carlo sampling with reduced uncertainties.}
Similarly to set 2A, we reduce the standard deviation of the log-normal distribution for \textit{all} rates by a factor of 5, i.e., a standard deviation of 0.1, to compare to the relative improvement that can be gained by concerted efforts on a few nuclei (i.e., set 2A).

\begin{figure}
    \centering
    \includegraphics[width=\linewidth]{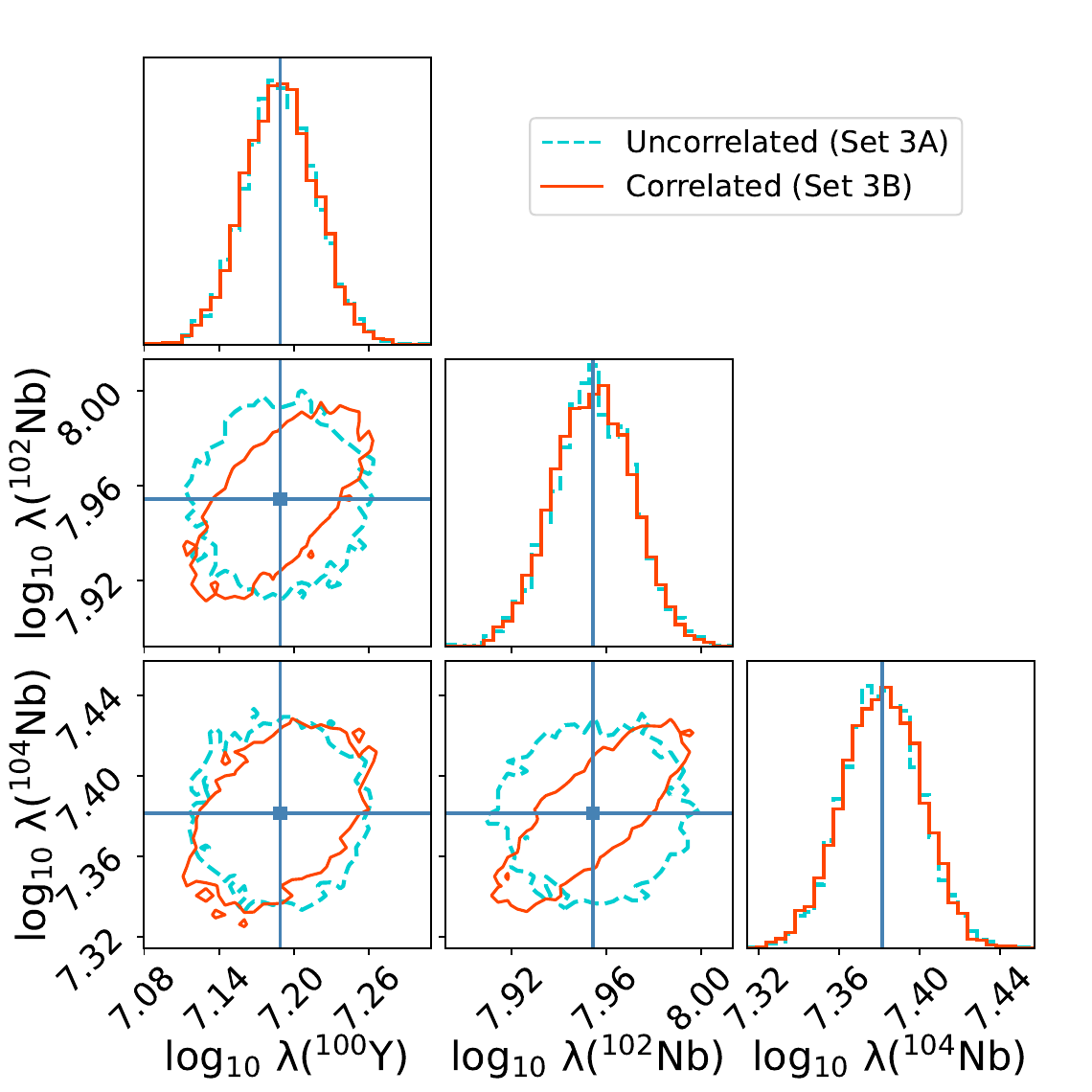}
    \caption{1D and 2D marginal distributions ("corner plot") of the Monte-Carlo samples for $\lambda_{n,\gamma}$ for $^{100}$Y, $^{102}$Nb, and $^{104}$Nb at a temperature of $T = 1.16$~GK.  The central values are the \texttt{YAHFC}/KDEF baseline rates.  The cyan (dashed) contours denote the 2-$\sigma$ range from the set 3A uncorrelated Monte Carlo simulations.  The red (solid) contours represent the 2-$\sigma$ range of the samples from the set 3B correlated Monte-Carlo simulations.} 
    \label{fig:corr-MC-rates-samples-corner-only-set-3}
\end{figure}

\textbf{Set 3A:} {\it Uncorrelated Monte-Carlo with individual uncertainties for each nuclear species based on KDUQEF.}  We take the on-diagonals of the covariance matrix to be 
\begin{align}
\Sigma_{ii} & =  \bm\sigma_i^2  \nonumber \\
            & =  \sum_{n=1}^{416}(\log_{10}\lambda_{ni,\text{KDUQEF}} - \langle \log_{10} \lambda_{i,\text{KDUQEF}}\rangle)^2/416,
\end{align}
where $\langle \log_{10}\lambda_{i,\text{KDUQEF}}\rangle$ is the average of the logarithm of the 416 KDUQEF sampled rates, $n$ is the index for the 416 KDUQEF samples, and $i$ indicates the nuclear species. We take the off-diagonal components of the covariance matrix to be zero. In other words, reaction rates on all nuclear species are varied independently at the scale of the uncertainties of KDUQEF rather than at a globally specified uniform scale as in sets 1 and 2.

\textbf{Set 3B:} {\it Correlated Monte-Carlo based on KDUQEF.}  We create a full covariance matrix based on the 416 KDUQEF samples. Unlike in set 3A, we retain the off-diagonal of the covariance matrix. 
We reiterate that we resample according to Eq.\ \ref{eqn:probability} to obtain 5000 sets of rates for increased statistics. 
In Fig.~\ref{fig:corr-MC-rates-samples-corner-only-set-3} we demonstrate the difference between using a covariance matrix with and without off-diagonals by showing 2-$\sigma$ contours of the 5000 sampled rates at $T=1.16$ GK (0.1 MeV) from sets 3A and 3B for a few isotopes.

\textbf{Set 4A:} {\it Uncorrelated Monte-Carlo with individual KDUQEF uncertainties scaled up by a factor of 10.}  We take the on-diagonals of the covariance matrix to be 
\begin{align}
\Sigma_{ii} & =  (10 \bm\sigma_i)^2 \nonumber \\
& =  100 \sum_{n=1}^{416}(\log_{10}\lambda_{in,\text{KDUQEF}} - \langle \log_{10} \lambda_{\text{KDUQEF}}\rangle)^2/416,
\end{align}
i.e., the same as set 3A but with the uncertainties from KDUQEF increased by a factor of 10.

\textbf{Set 4B:} {\it Correlated Monte-Carlo with KDUQEF uncertainties scaled up by a factor of 10.}  We again create a full covariance matrix based on the 416 KDUQEF samples (i.e., that in set 3B) but multiply the entire matrix by a factor of 100.  We again resample according to Eq.\ \ref{eqn:probability} to obtain 5000 sets of rates.

To verify that the results we find from sets 3A and 3B are not contaminated by computational noise, we use Monte-Carlo sets 4A and 4B for the purposes of confirming that the conclusions we obtain from a comparison of sets 3A and 3B are similar to what would be obtained if the covariance matrix had a KDUQEF-derived structure but substantially larger magnitude.

\begin{figure}
    \centering
    \includegraphics[width=\linewidth]{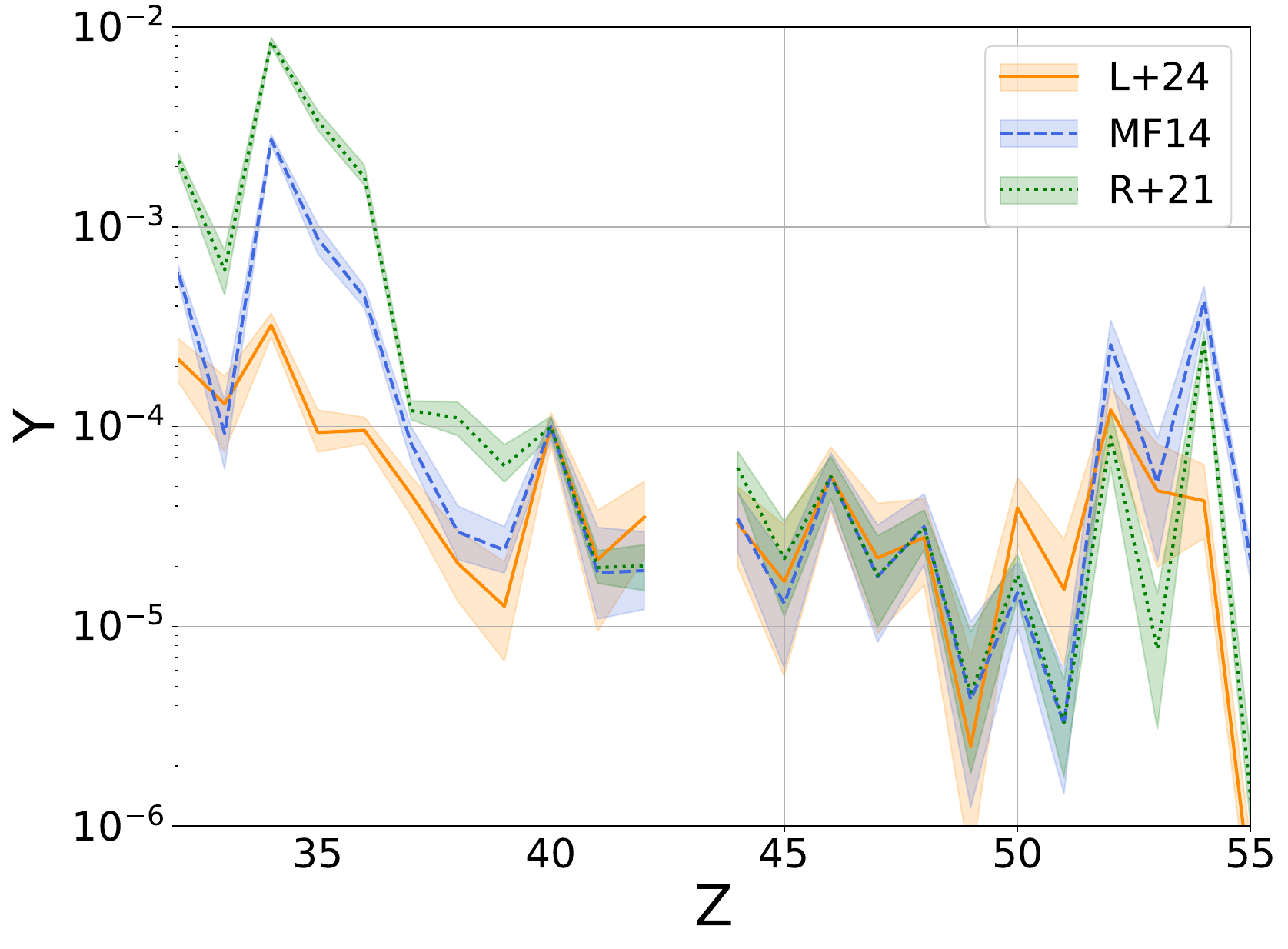}
    \caption{Final elemental yield distributions in the weak $r$-process region for the three astrophysical scenarios, L+24: orange (solid), MF14: blue (dashed) and R+21: green (dotted).  The dark line shows the log-mean abundance, and the shaded regions show the 2-$\sigma$ regions for set 1 of Monte-Carlo simulations (uncorrelated, uniform scaling).  All abundances are scaled to have the same mean Zr ($Z = 40$) abundance of $10^{-4}$. }
    \label{fig:abundance-events}
\end{figure}

\subsection{Analysis method}

%
%

\begin{figure*}
    \centering
    \includegraphics[width=0.333\linewidth]{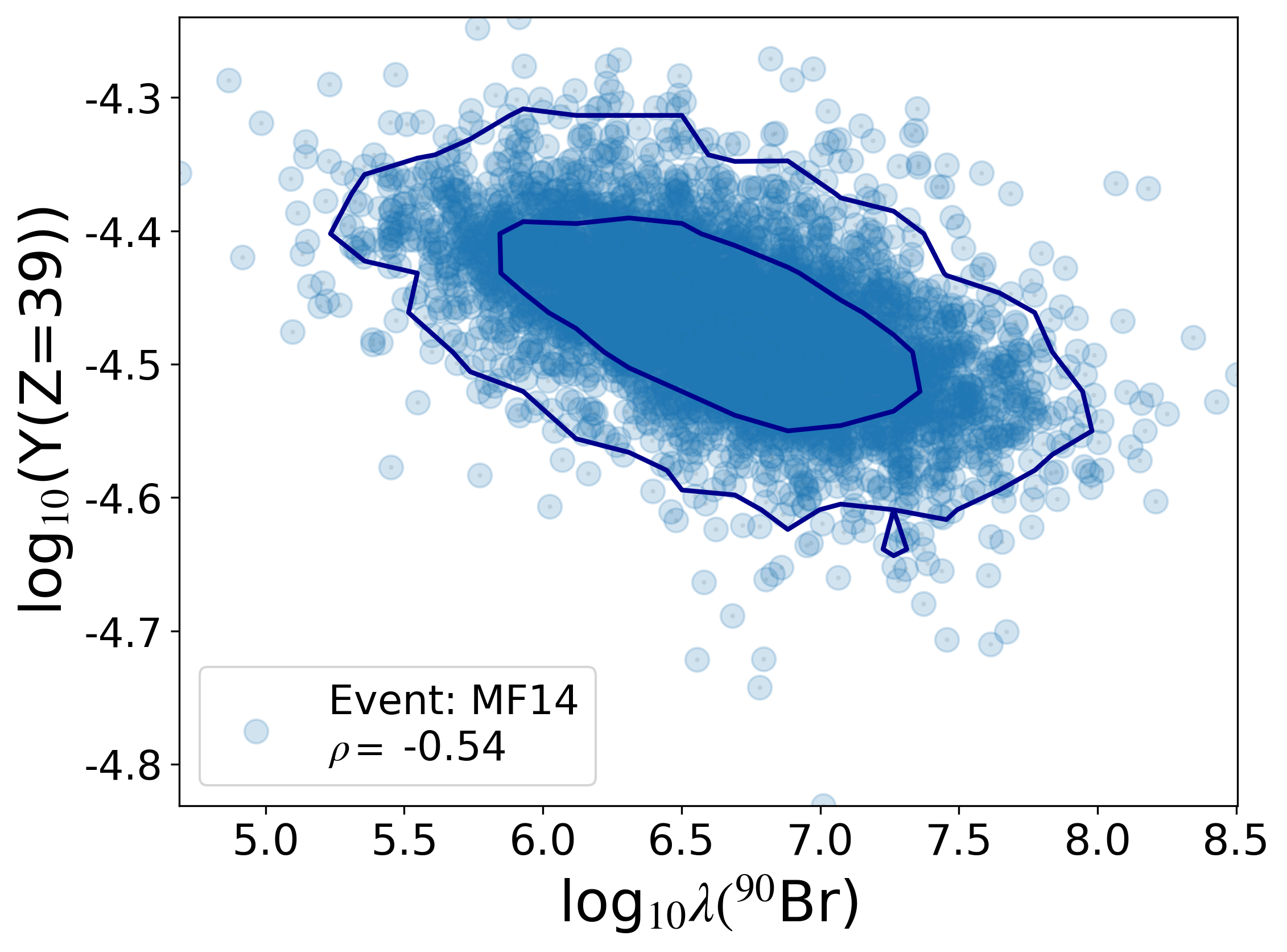}
    \includegraphics[width=0.325\linewidth]{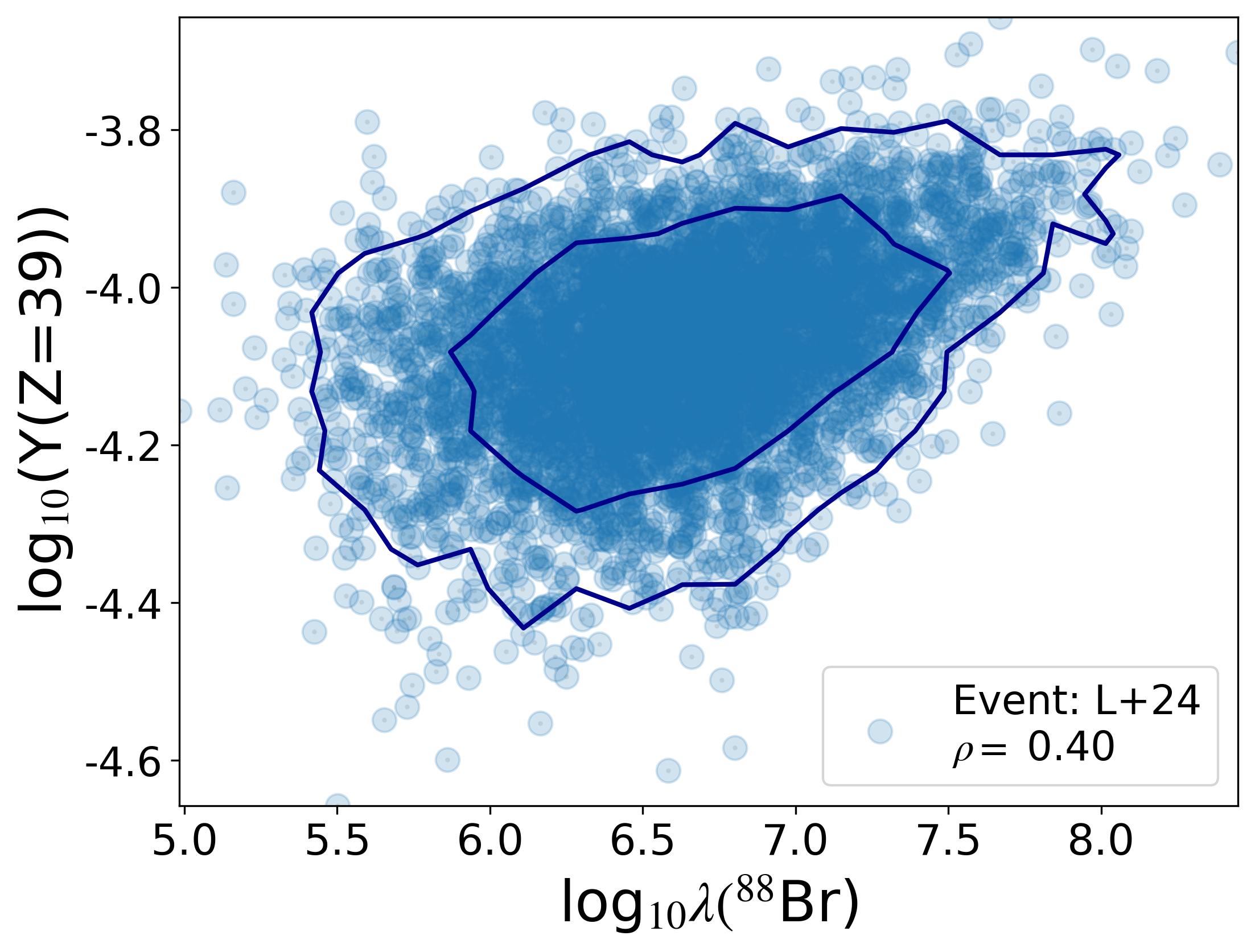}
    \includegraphics[width=0.333\linewidth]{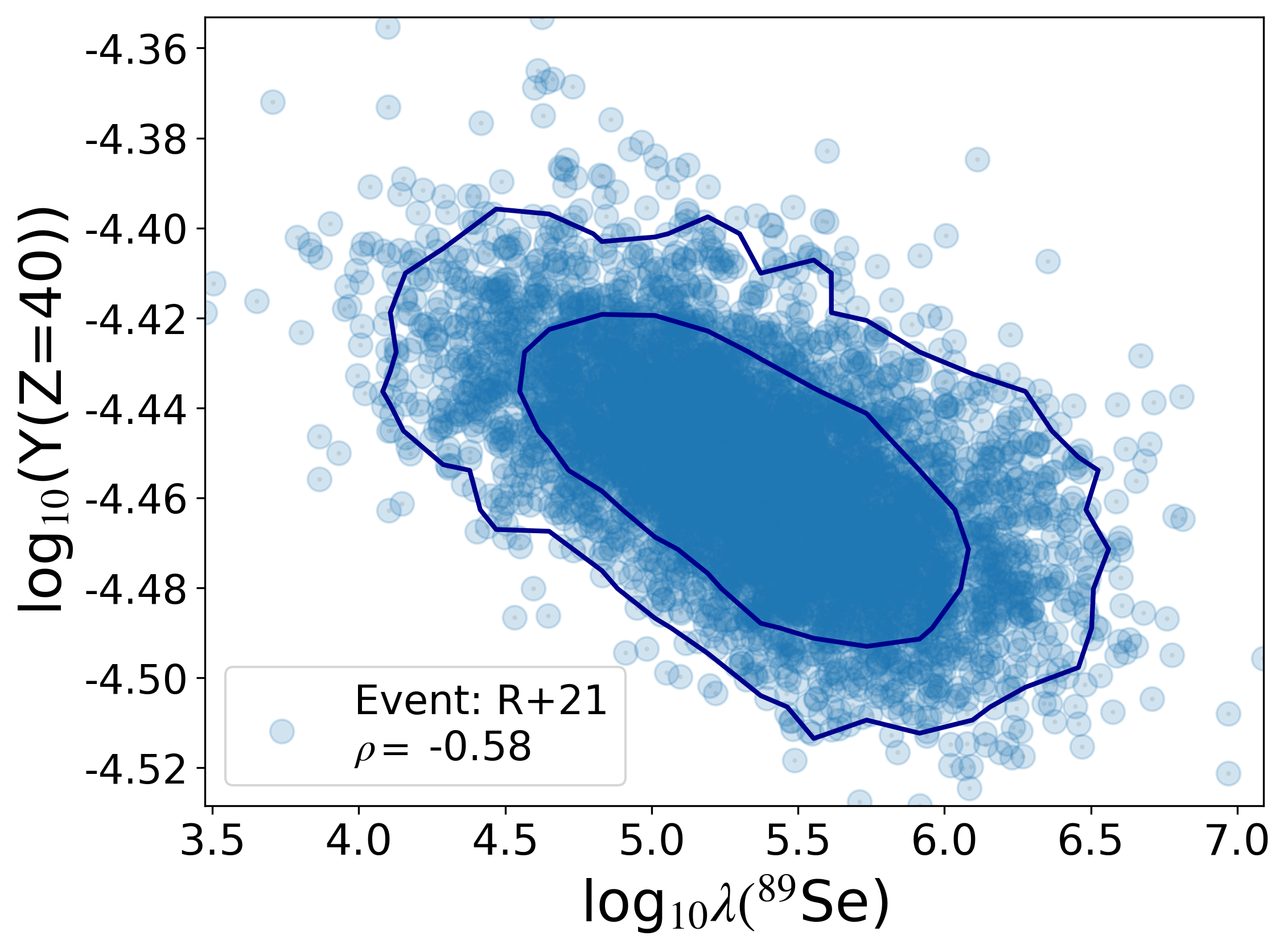}
    \caption{Selected results from our set 1 Monte-Carlo simulations. Final abundance of yttrium versus the rate of neutron capture, $\lambda_{n,\gamma}$ at a temperature of 1.16 GK on $^{90}$Br (left), and $^{88}$Br (center) obtained in the MF14 and L+24 scenarios, respectively is shown in the left and center panels.  In the right panel, the final abundance of zirconium versus the rate of neutron capture on $^{89}$Se (right) (at 1.16 GK) in the R+21 scenario is shown. The contours represent 1-$\sigma$ and 2-$\sigma$ ranges.  Each blue dot represents one of the 5000 samples. }
    \label{fig:correlation-scatter}
\end{figure*}

We run \texttt{PRISM} using each new set of rates generated according to Table \ref{tab:montecarlo-summary} for the 40 trajectories described in Sec.\ \ref{subsec:astro-model} and listed in Table \ref{table:astrophysics}.
All calculations are started at an initial temperature of 10 GK and run to a time of 1 Gyr. A nuclear statistical equilibrium (NSE) solver is employed at the start of the calculation, i.e., 10 GK, to set initial abundance distribution, and not used subsequently. 
For a given rate combination, the abundances of individual trajectories are combined using a weighted sum, where the weights are defined as in Table \ref{table:astrophysics}, resulting in a net abundance from each of the three weak \emph{r} processes, as shown in for example, Fig.\ \ref{fig:abundance-events}.  This figure is made using the rates from the set 1 Monte-Carlo.

We use Pearson correlation coefficients to interpret the results of our simulations in terms of the correlation between neutron capture rates and elemental weak \emph{r}-process abundances. 
One could also employ Spearman correlation coefficients, e.g., Ref.~\cite{2022ApJ...935...27P}; however, given the primarily linear relationship of the log-abundances and log-reaction rates, we expect that we capture the main physical relationships with Pearson correlation coefficients. The Pearson correlation coefficient is given by 
\begin{equation}
\rho_{kl} = \frac{\mathrm{cov}(a_k, b_l)}{\sigma(a_k) \sigma(b_l)}~,
\label{eq:pearson}
\end{equation}
where $\mathrm{cov}(a_k, b_l)$ is the covariance between quantity $a_k$ and quantity $b_l$,  while  $\sigma(a_k)$ is the standard deviation of $a_k$. In the following, we will compute correlation coefficients between neutron capture rates (at 1.16 GK) and final abundances.
A summary of the correlations we investigate is given in Table \ref{tab:correlation_summary}.

%
%

\begin{table}
    \renewcommand{\arraystretch}{1.6}
    \setlength{\tabcolsep}{5pt}  
    \caption{Summary of the correlations analyzed in this work. Here $Y(Z)$ and $Y(Z')$ denote the final abundances of nuclei with atomic numbers $Z$ and $Z'$.}
    \begin{tabular}{c c c c c}
        \hline\hline
        \textbf{Quantities compared} & \textbf{$a_k$} & \textbf{$k$} & \textbf{$b_l$} & \textbf{$l$} \\
        \hline
        Rate vs abundance & $\log_{10}\left[\lambda^{Z,A}_{(n,\gamma)}\right]$ & $Z, A$ & $Y(Z')$ & $Z'$ \\ 
        Abundance vs abundance & $Y(Z)$ & $Z$ & $Y(Z')$ & $Z'$ \\
        \hline\hline
    \end{tabular}
    \vspace{2pt}
    \footnotesize\emph{Note:}\; In the rate–abundance rows, the index $k$ runs over all possible combinations of atomic number $Z$ and mass number $A$, i.e., over all nuclear species.
    \label{tab:correlation_summary}
\end{table}

We illustrate examples of two negative and one positive Pearson correlations in Fig.\ \ref{fig:correlation-scatter}, where we show examples of scatter plots of abundance versus neutron capture rate for our 5000 samples of the set 1 {\it uncorrelated} Monte-Carlo.  
These were obtained through Monte-Carlo simulations with the astrophysical conditions of \citet{2024ApJ...964..111L} (L+24, center panel), \citet{2014MNRAS.441.3444M} (MF14, left panel), and \citet{Reichert:2020mjo} (R+21, right panel). We also show 1-$\sigma$ and 2-$\sigma$ contours of the 5000 samples.

\section{Uncorrelated Monte Carlo}
\label{sec:uncorrelated}
%
%

\subsection{Mechanisms by which neutron capture rates directly influence abundance yields}
\label{subsec:basecorrelations}
We begin by analyzing our set 1, uniform standard deviation of one-half order of magnitude across nuclear species, an uncorrelated Monte Carlo simulation whose final elemental abundance 2-$\sigma$ spreads are shown in Fig.\ \ref{fig:abundance-events}.  We note that the median, shown as a solid colored line in the center of each band, depends on our choice of underlying nuclear physics models and may shift in future calculations if different level densities and $\gamma$-strength functions are used.

We are interested in quantifying the effect of each nucleus's neutron-capture rate on the final abundance of each element within the context of the set 1 Monte Carlo samples.  We use the Pearson correlation coefficients, $\rho$ [Eq.\ \ref{eq:pearson}], between final elemental abundances and neutron capture rates as our quantifying measure. In Table \ref{table:abundance-rate-correlation}, we tabulate the neutron capture rates with the three largest $\rho$s, as long as  $|\rho| > 0.3$, where 0.3 corresponds to a weak, positive correlation of borderline significance.  
In the first column, we list all stable elements between selenium ($Z=34$) and xenon ($Z=54$), leaving out the unstable element technetium ($Z=43$). For each element and each of our three astrophysical scenarios, we list the 2$\sigma$ deviation of the abundance from the mean as a percentage, i.e., $\Delta=\frac{Y_{\mu+2\sigma}-Y_\mu}{Y_\mu}\times100$, the relevant neutron-capture rates and associated correlation coefficients. We include similar and more complete information in Table \ref{table:rate-abundance-correlation} of Appendix \ref{app:tables}, organized instead by neutron-capture rate. 

\begin{table*}
    \renewcommand{\arraystretch}{1.3}
    \caption{Elements that have a Pearson correlation coefficient $|\rho|>$0.30 in the set 1, uncorrelated sampling of neutron capture rates with uniform standard deviation of one half an order of magnitude, Monte Carlo simulations are listed in the first column. For each element, first given are the variations, $\Delta$, of its abundance ($\Delta = \frac{Y_{\mu+2\sigma} - Y_\mu}{Y_{\mu}}\times100$, where $Y_{\mu}$ is the mean abundance) in each scenario (L+24, MF14, and R+21) and the subsequent columns give the neutron capture rates with the three highest correlations above 0.3. }
    \begin{tabular}{@{\extracolsep{0pt}}c | c c c c | c c c c | c c c c@{}}
        \hline \hline
        \multirow{2}{*}{Element} & \multicolumn{4}{c}{Lund+24 (L+24)} & \multicolumn{4}{c}{Metzger \& Fern\'{a}ndez14 (MF14)} & \multicolumn{4}{c}{Reichert+21 (R+21)} \\
         & $\Delta$ & $^{A}$I~~~~~~~~~$\rho$ & $^{A}$I~~~~~~~~~$\rho$ & $^{A}$I~~~~~~~~~$\rho$ & $\Delta$ & $^{A}$I~~~~~~~~~$\rho$ & $^{A}$I~~~~~~~~~$\rho$ & $^{A}$I~~~~~~~~~$\rho$ & $\Delta$ & $^{A}$I~~~~~~~~~$\rho$ & $^{A}$I~~~~~~~~~$\rho$ & $^{A}$I~~~~~~~~~$\rho$ \\
        \hline
        Se  & 15 & $^{76}$Ni \hfill 0.47 & $^{76}$Cu \hfill 0.43 & -  & 6 & $^{78}$Ga \hfill -0.65 & $^{80}$Ga \hfill -0.32 & -  & 5 & $^{78}$Ga \hfill -0.59 & $^{82}$As \hfill -0.42 & $^{80}$Ga \hfill -0.35 \\
        Br  & 29 & $^{80}$Ga \hfill 0.51 & $^{78}$Ga \hfill 0.39 & $^{80}$Zn \hfill 0.31  & 18 & $^{78}$Ga \hfill 0.72 & $^{80}$Ga \hfill 0.34 & $^{81}$Ge \hfill -0.32  & 12 & $^{78}$Ga \hfill 0.68 & $^{81}$Ge \hfill -0.44 & $^{80}$Ga \hfill 0.34 \\
        Kr  & 16 & $^{83}$Ga \hfill 0.57 & $^{85}$Zn \hfill -0.40 & -  & 13 & $^{82}$As \hfill 0.86 & $^{82}$Ge \hfill 0.35 & -  & 15 & $^{82}$As \hfill 0.93 & - & - \\
        Rb  & 23 & $^{87}$As \hfill -0.49 & $^{88}$As \hfill -0.32 & -  & 21 & $^{87}$Se \hfill -0.51 & $^{85}$Ge \hfill -0.45 & -  & 12 & $^{87}$Br \hfill -0.38 & $^{85}$Se \hfill -0.37 & $^{86}$Se \hfill 0.35 \\
        Sr  & 46 & $^{89}$As \hfill -0.47 & $^{87}$As \hfill 0.40 & $^{88}$Se \hfill -0.39  & 35 & $^{88}$Br \hfill -0.58 & $^{87}$Br \hfill 0.35 & -  & 20 & $^{87}$Se \hfill -0.56 & $^{89}$Se \hfill 0.50 & $^{88}$Br \hfill -0.49 \\
        Y  & 67 & $^{88}$Br \hfill 0.40 & $^{90}$Br \hfill -0.40 & $^{88}$As \hfill 0.38  & 31 & $^{90}$Br \hfill -0.54 & $^{89}$Br \hfill -0.30 & -  & 28 & $^{88}$Br \hfill 0.61 & $^{89}$Se \hfill -0.38 & - \\
        Zr  & 18 & $^{94}$Rb \hfill -0.41 & $^{92}$Br \hfill -0.34 & $^{93}$Kr \hfill 0.34  & 12 & $^{92}$Kr \hfill -0.47 & $^{94}$Rb \hfill -0.40 & -  & 12 & $^{89}$Se \hfill -0.58 & $^{90}$Br \hfill -0.33 & - \\
        Nb  & 76 & $^{93}$Kr \hfill -0.59 & $^{94}$Br \hfill -0.51 & $^{92}$Br \hfill 0.36  & 69 & $^{93}$Kr \hfill -0.54 & $^{92}$Kr \hfill 0.50 & $^{92}$Rb \hfill 0.38  & 21 & $^{92}$Rb \hfill 0.59 & $^{93}$Kr \hfill -0.31 & $^{94}$Rb \hfill -0.30 \\
        Mo  & 51 & $^{94}$Rb \hfill 0.50 & $^{94}$Br \hfill 0.32 & $^{96}$Rb \hfill 0.31  & 56 & $^{94}$Rb \hfill 0.61 & $^{100}$Y \hfill -0.45 & -  & 27 & $^{100}$Y \hfill -0.56 & $^{97}$Sr \hfill -0.44 & - \\
        Ru  & 52 & $^{104}$Y \hfill -0.40 & $^{102}$Y \hfill -0.38 & -  & 35 & $^{104}$Nb \hfill -0.57 & - & -  & 21 & $^{104}$Nb \hfill -0.58 & $^{102}$Nb \hfill -0.46 & $^{103}$Mo \hfill 0.39 \\
        Rh  & 92 & $^{103}$Y \hfill -0.64 & $^{103}$Zr \hfill -0.36 & -  & 70 & $^{103}$Nb \hfill -0.45 & $^{103}$Y \hfill -0.42 & $^{103}$Zr \hfill -0.36  & 54 & $^{103}$Mo \hfill -0.50 & $^{102}$Nb \hfill 0.47 & $^{103}$Nb \hfill -0.42 \\
        Pd  & 40 & $^{103}$Y \hfill 0.33 & - & -  & 32 & $^{110}$Tc \hfill -0.47 & $^{104}$Nb \hfill 0.34 & -  & 26 & $^{104}$Nb \hfill 0.60 & $^{108}$Tc \hfill -0.46 & - \\
        Ag  & 88 & $^{109}$Mo \hfill -0.37 & $^{106}$Nb \hfill 0.32 & -  & 81 & $^{108}$Tc \hfill 0.37 & $^{109}$Mo \hfill -0.32 & $^{109}$Tc \hfill -0.32  & 58 & $^{108}$Tc \hfill 0.58 & $^{107}$Mo \hfill -0.41 & $^{106}$Nb \hfill 0.38 \\
        Cd  & 57 & $^{110}$Tc \hfill 0.41 & $^{109}$Nb \hfill 0.38 & $^{110}$Nb \hfill 0.34  & 46 & $^{110}$Tc \hfill 0.57 & - & -  & 23 & $^{110}$Tc \hfill 0.66 & $^{114}$Rh \hfill -0.46 & - \\
        In  & 183 & $^{114}$Rh \hfill 0.43 & $^{115}$Rh \hfill -0.38 & $^{115}$Ru \hfill -0.36  & 145 & $^{114}$Rh \hfill 0.53 & $^{115}$Rh \hfill -0.50 & -  & 101 & $^{114}$Rh \hfill 0.65 & $^{115}$Pd \hfill -0.42 & - \\
        Sn  & 42 & $^{124}$Ag \hfill -0.56 & $^{124}$In \hfill -0.33 & -  & 43 & $^{116}$Rh \hfill 0.46 & $^{124}$Ag \hfill -0.34 & -  & 27 & $^{116}$Rh \hfill 0.70 & $^{124}$Ag \hfill -0.36 & - \\
        Sb  & 78 & $^{123}$Ag \hfill -0.60 & $^{123}$Cd \hfill -0.36 & -  & 85 & $^{123}$Ag \hfill -0.51 & $^{123}$Cd \hfill -0.44 & $^{120}$Ag \hfill 0.33  & 65 & $^{123}$Cd \hfill -0.48 & $^{120}$Ag \hfill 0.42 & $^{124}$Ag \hfill -0.34 \\
        Te  & 29 & $^{127}$Cd \hfill 0.45 & $^{126}$In \hfill -0.31 & -  & 33 & $^{130}$In \hfill -0.62 & $^{129}$Cd \hfill 0.33 & -  & 34 & $^{130}$In \hfill -0.85 & $^{128}$In \hfill -0.34 & - \\
        I  & 71 & $^{127}$Cd \hfill -0.61 & $^{126}$In \hfill 0.32 & -  & 66 & $^{127}$Cd \hfill -0.71 & - & -  & 89 & $^{127}$Cd \hfill -0.74 & $^{128}$Ag \hfill -0.42 & - \\
        Xe  & 52 & $^{128}$In \hfill 0.44 & $^{128}$Ag \hfill 0.42 & $^{129}$Cd \hfill -0.38  & 18 & $^{130}$In \hfill 0.62 & $^{129}$Cd \hfill -0.38 & $^{129}$Sn \hfill -0.32  & 9 & $^{130}$In \hfill 0.86 & $^{128}$In \hfill 0.33 & - \\
        \hline \hline
    \end{tabular}
    \label{table:abundance-rate-correlation}
\end{table*}

\begin{figure*}
    \centering
    \includegraphics[width=0.48\linewidth]{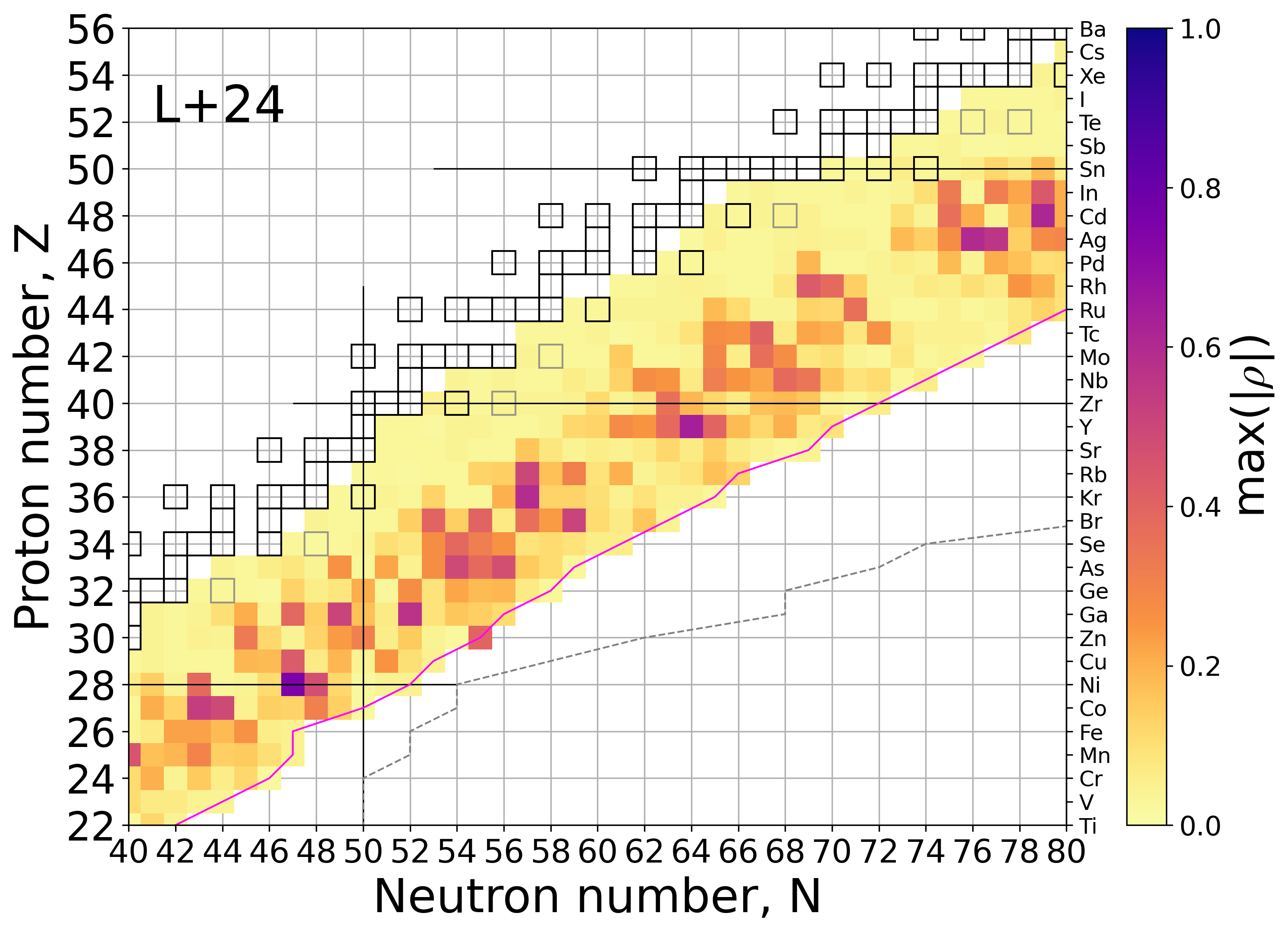}
    \includegraphics[width=0.48\linewidth]{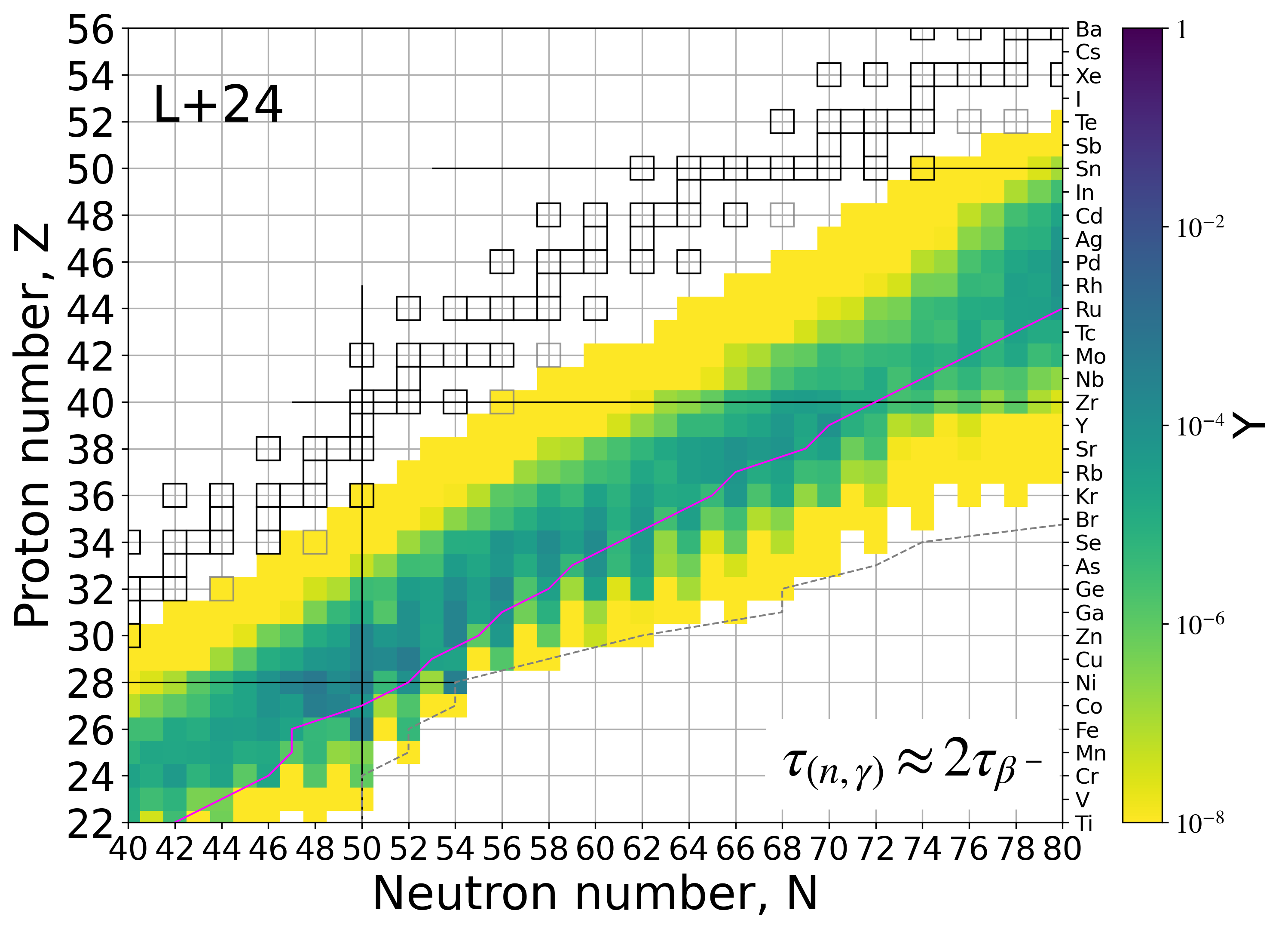}\\
    \includegraphics[width=0.48\linewidth]{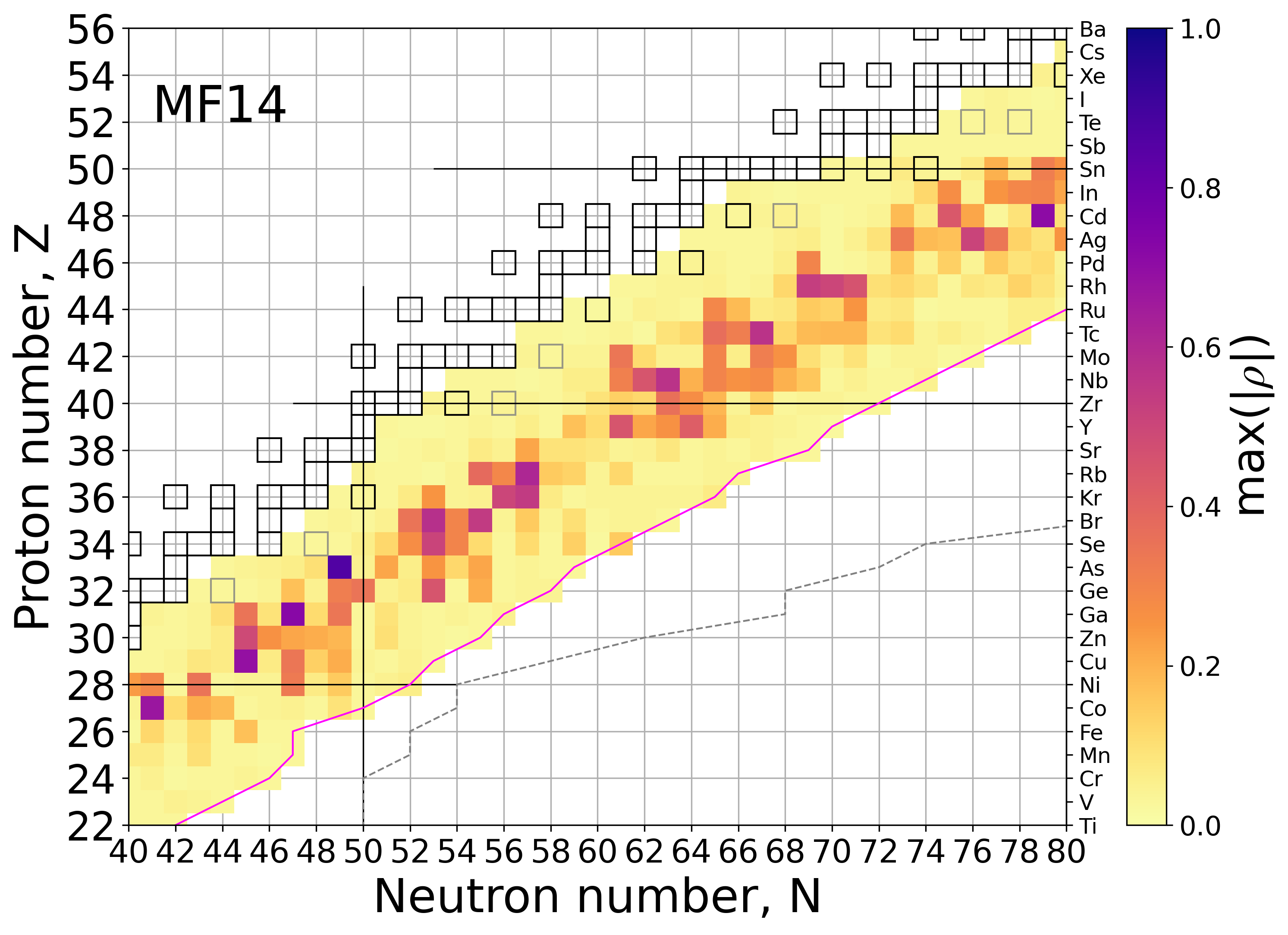}
    \includegraphics[width=0.48\linewidth]{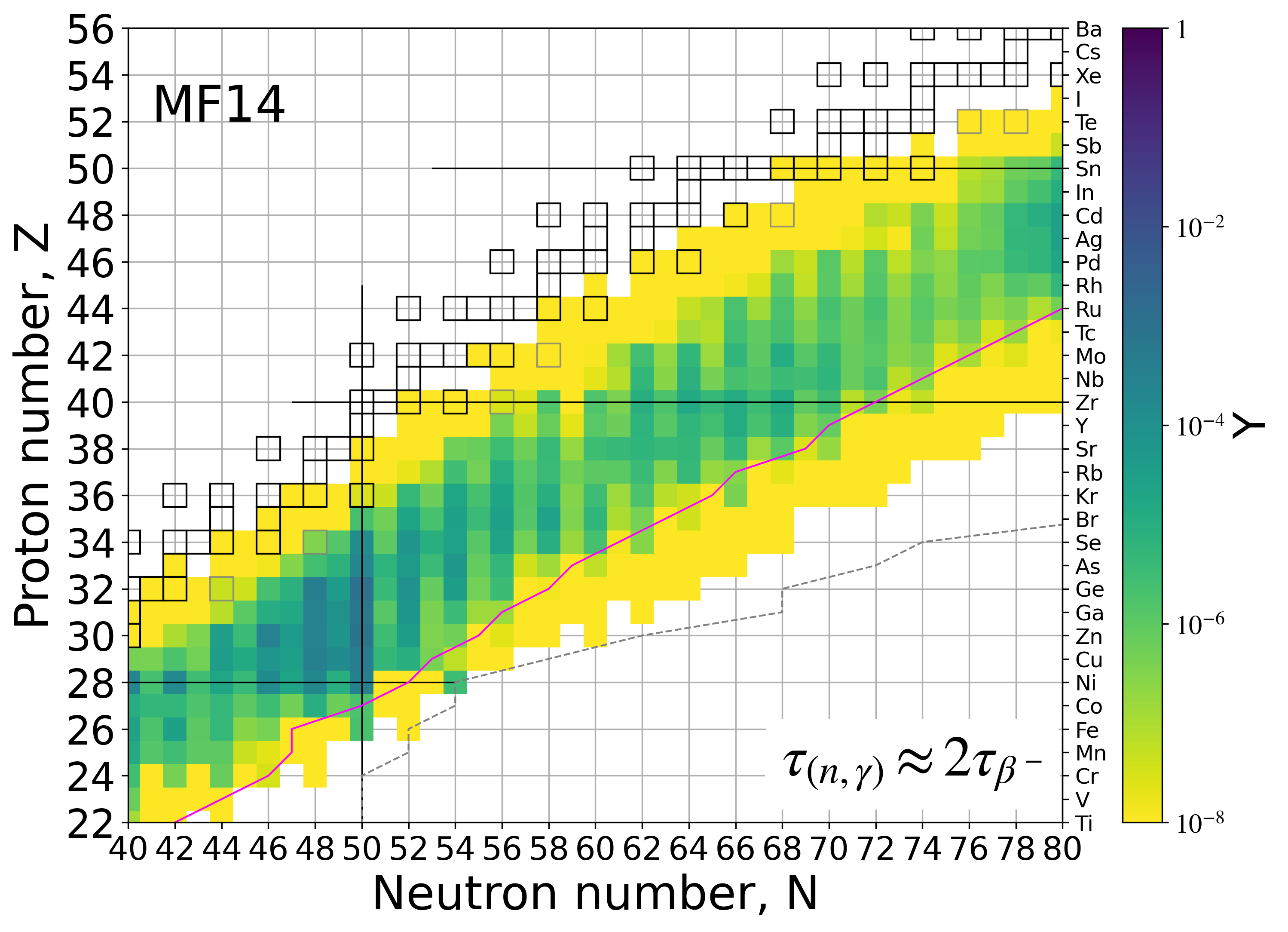}\\
    \includegraphics[width=0.48\linewidth]{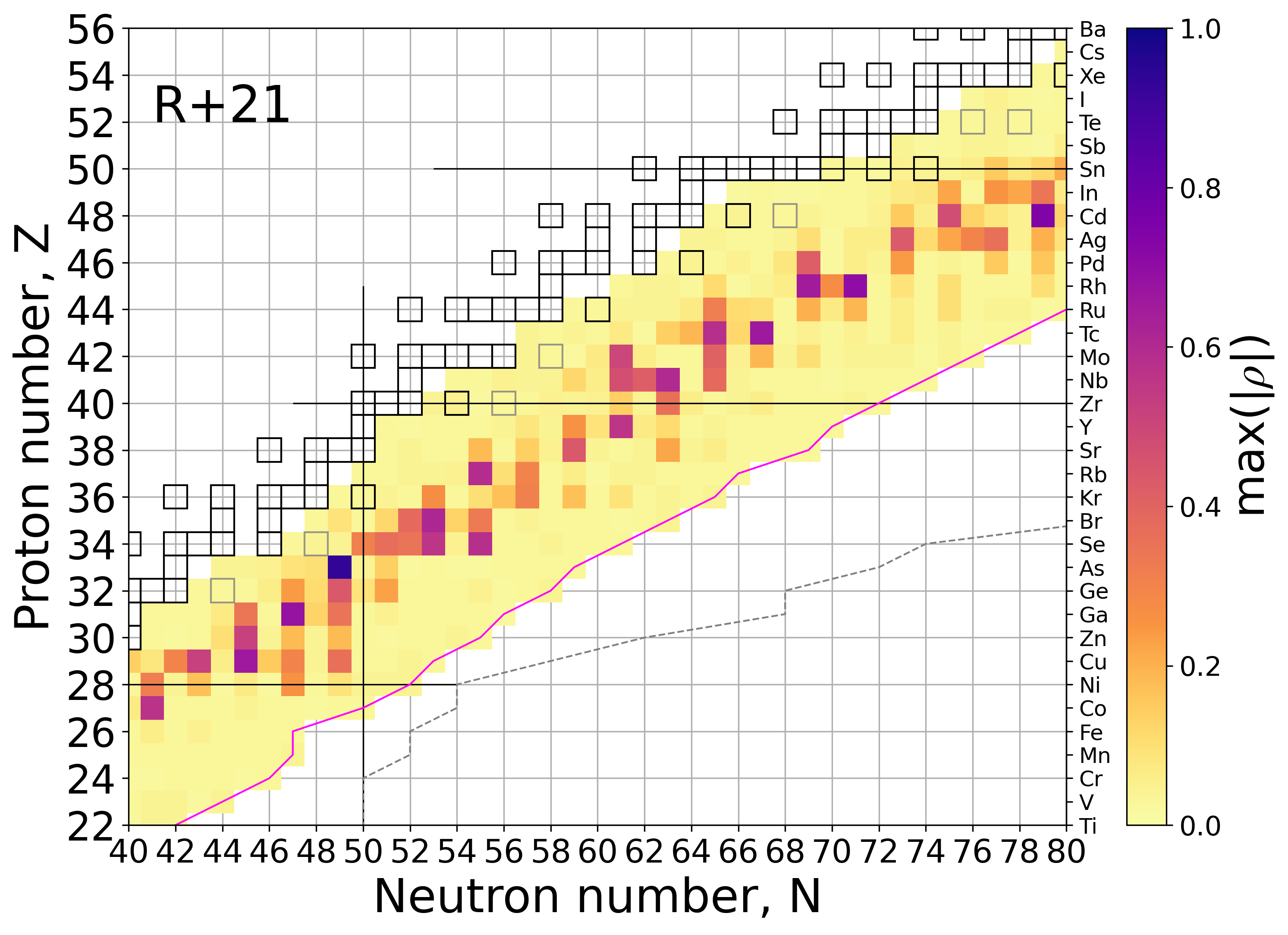}
    \includegraphics[width=0.48\linewidth]{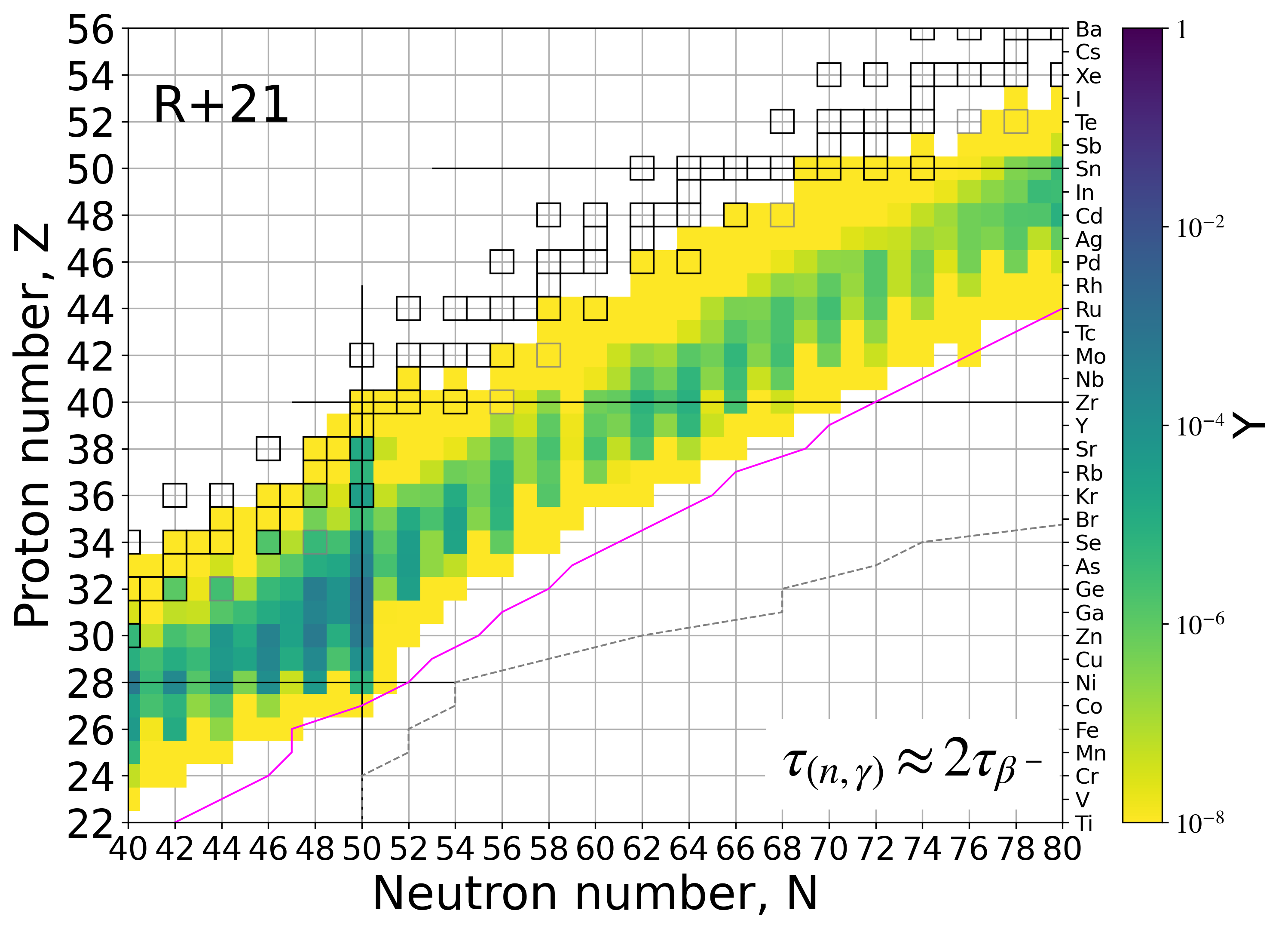}
    \caption{ Left column: Highest magnitude Pearson correlation coefficient for each neutron capture rate with {\it any} elemental abundance in the set 1 {\it uncorrelated} Monte Carlo simulations. Top, middle and bottom panels represent simulations performed with the three astrophysical scenarios: L+24, MF14, and R+21. The square boxes indicate stable isotopes (black) and long-lived isotopes (gray). The dashed gray lines indicate the FRDM neutron drip line and the solid magenta lines shows the limit of the isotopes sampled in this study. Right column:   Abundance of each nuclear species, $Y$, at the time when the abundance weighted $(n,\gamma)$ timescale becomes twice the abundance weighted $\beta$-decay timescale,  $\tau_{(n,\gamma)}\approx 2\tau_{\beta^-}$. }
    \label{fig:highest_and_freezeout}
\end{figure*}

In the left panels of Fig.\ \ref{fig:highest_and_freezeout}, we show the highest Pearson correlation coefficient that each neutron capture rate has with {\it any} of the elements for each of our astrophysical scenarios. We see in these left panels that the highest Pearson correlation coefficients constitute a band on the $N$--$Z$ plane. However, this band appears closer or farther away from stability depending on the astrophysical model. The band is the farthest from stability in the black hole accretion disk scenario of \citet{2024ApJ...964..111L} (L+24, top panel), closest to stability for the magnetohydrodynamic supernova model of \citet{Reichert:2020mjo} (R+21, bottom panel), and midway between these two cases for outflows from the black hole accretion disk of \citet{2014MNRAS.441.3444M} (MF14, middle panel).

The pattern observed in Fig.~\ref{fig:highest_and_freezeout} left panels can be understood in terms of the nuclei which are populated after freeze-out from $(n,\gamma)$--$(\gamma,n)$ equilibrium and during the time period in which the \emph{r}-process path moves back toward stability. Freeze-out typically begins when the photodissociation rates can no longer compete with neutron capture rates, often around a temperature of $T \approx 1$~GK. Before this time, the path is in equilibrium, meaning that the neutron separation energies set the abundance distribution along each isotopic chain, and the timescales for neutron capture and photodissociation are significantly shorter than that of $\beta$-decay. After $(n,\gamma)$ and $(\gamma,n)$ reactions are no longer in equilibrium, the abundance pattern is finalized by the interplay between neutron capture, $\beta$-decay, and (occasionally) photodissociation. Individual neutron capture rates have the most significant role during the portion of the freeze-out phase when the timescales for neutron capture and $\beta$-decay are similar. Therefore, the neutron capture rates of nuclei along the pathway that is populated during this time period, shown in the right panels of Fig.~\ref{fig:highest_and_freezeout}, tend to have the highest correlations with elemental abundances. Specifically, in the right panel of this figure, we plot the (combined) abundances at the time when the abundance weighted timescale for a nucleus to capture a neutron is about twice the abundance weighted timescale for a nucleus to undergo $\beta$-decay,  $\tau_{(n,\gamma)}\approx 2\tau_{\beta^-}$. This condition is achieved at different times and temperatures for different trajectories.

In the majority of the trajectories in the L+24 scenario (top right panel of Fig.\ \ref{fig:highest_and_freezeout}), the temperature and density evolve very quickly, producing ``cold’' $r$-process conditions (as described in, e.g., Ref. \cite{Mumpower+2012}). Here the system tends to drop out of ($n$,$\gamma$)-($\gamma$,$n$) equilibrium when the neutron abundance is still quite high and the $r$-process path is far from stability. In contrast, the trajectories of the R+21 scenario (bottom panel) tend to produce a classic hot $r$ process, where freeze-out is prompted by the depletion of neutrons and the result of longer lasting photodissociation reactions is that the path moves relatively
closer to stability before ($n$,$\gamma$)-($\gamma$,$n$) equilibrium fails. The set of trajectories in MF14 (middle panel) includes a larger variety of conditions, with both cold and hot $r$ processes represented, and thus, the abundance range at $\tau_{(n,\gamma)}\approx 2\tau_{\beta^{-}}$ is broadest in this scenario. In general the importance of the freeze-out phase in setting the final abundance pattern depends on how long neutron capture and $\beta$ decay compete after ($n$,$\gamma$)-($\gamma$,$n$) equilibrium fails, as well as the quantity of neutrons available during this time.

%
%

While the neutron capture rates with the highest correlation coefficients with respect to the final abundances are found along the pathway of most abundant nuclei during freezeout, it is not necessarily true that the most important capture rates are for nuclei with the highest abundances at that time. For example, for the R+21 scenario, the highest abundances are found along the $N=50$ shell closure, while few of these $N=50$ species show up with high correlation coefficients. Instead, the most impactful neutron capture rates are along the $N=49$ isotone---the species to which the abundant $N=50$ nuclei decay, and the neutron capture rate then determines whether the $N=49$ species will continue to $\beta$ decay or capture back into the closed shell. Additional examples of the mechanisms by which neutron capture rates influence the final abundance pattern are described in detail below.

Finally, we note that the broad region in the nuclear chart where individual neutron capture rates have the largest direct effect on the final abundance pattern is independent of the structure of the covariance matrix. While we will not delve into this further in the manuscript, we note that applying a full covariance matrix could indirectly link additional neutron-capture rates outside this region to the final abundance pattern via a nuclear species that does participate.

\begin{figure}
    \centering
    \includegraphics[width=\linewidth]{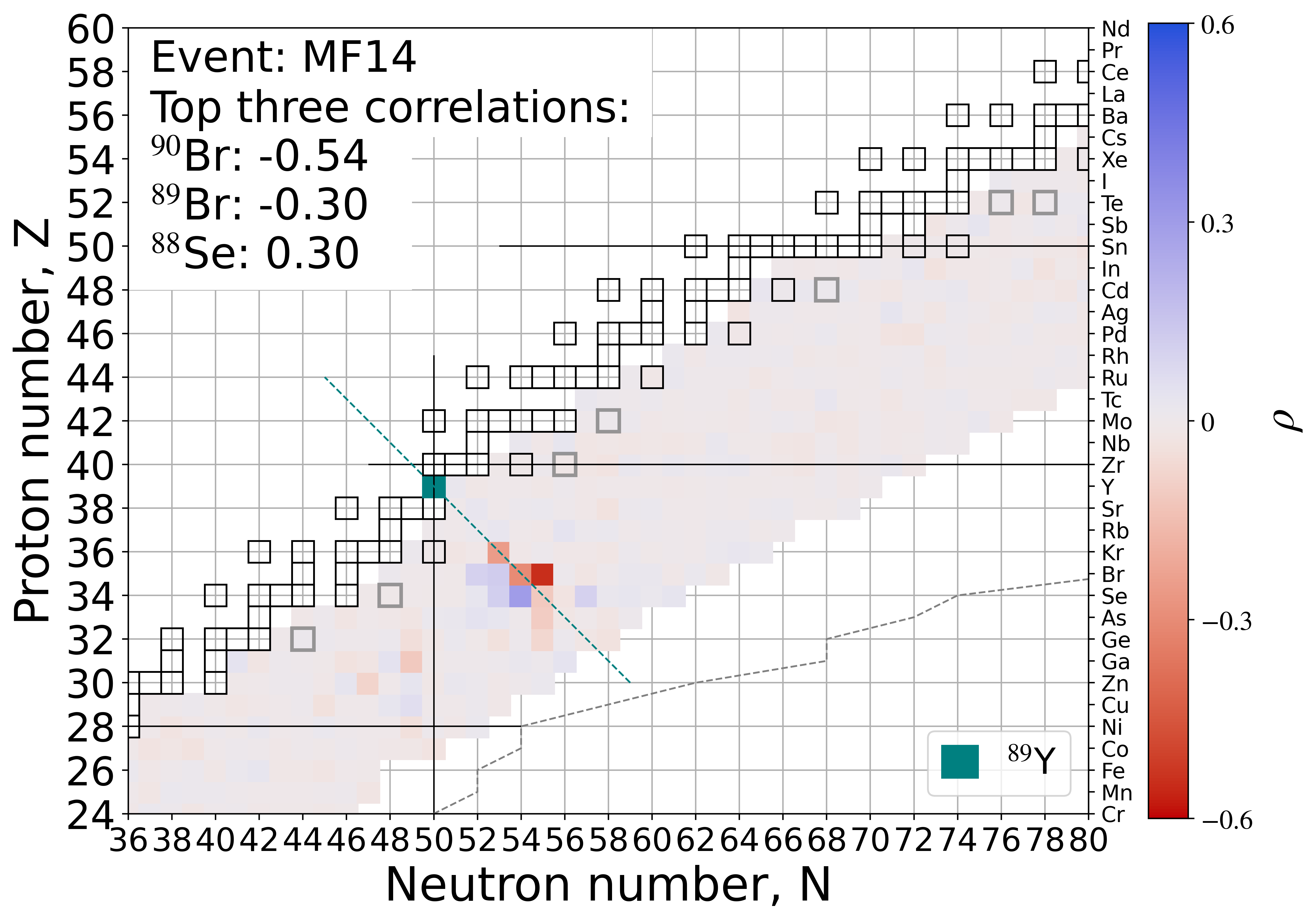}
    \includegraphics[width=\linewidth]{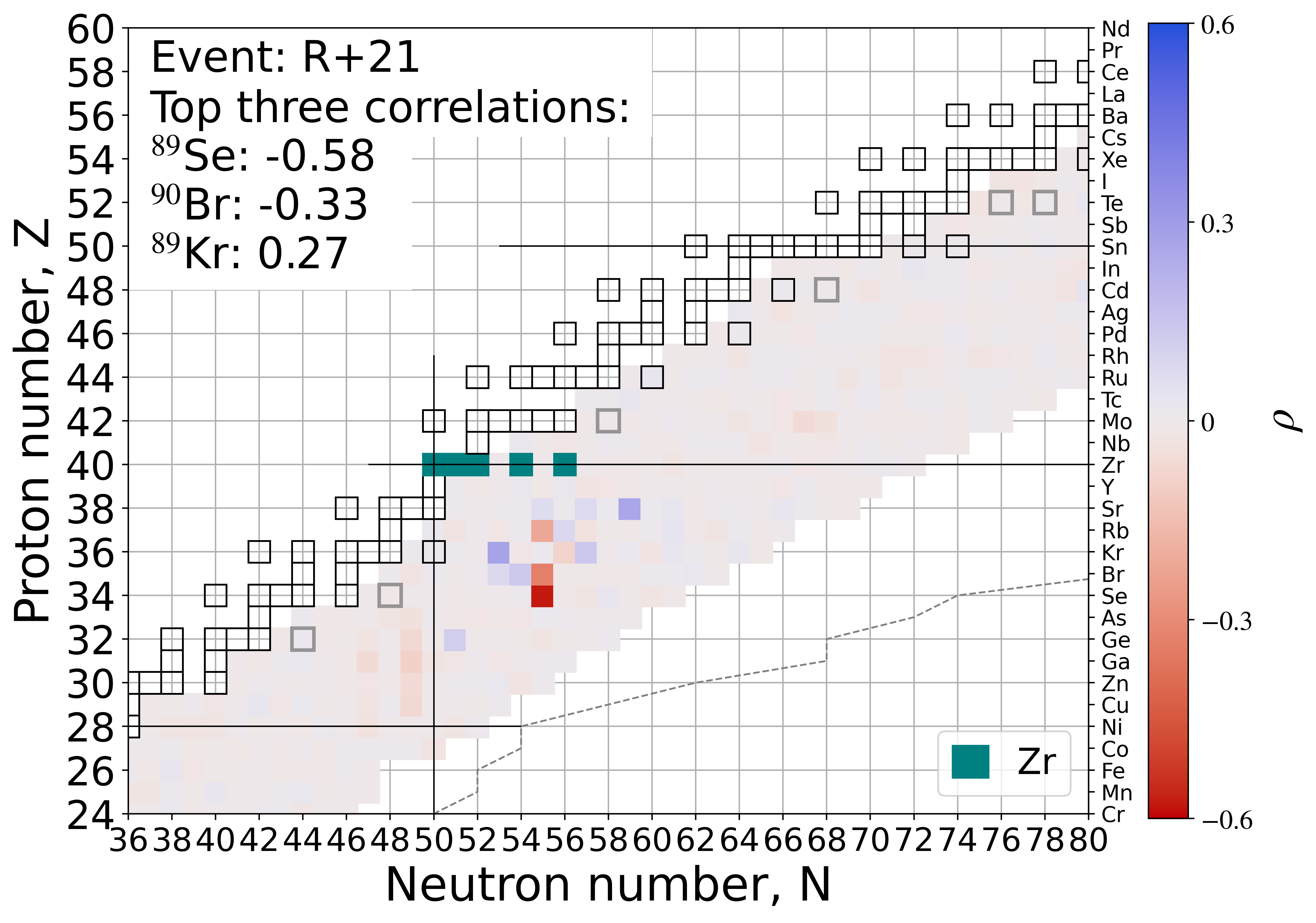}
    \caption{Pearson correlation coefficients, $\rho$, between the abundance of yttrium in the MF14 scenario (top panel), and zirconium in the R+21 scenario (bottom panel), and neutron capture isotopes included in Monte Carlo sampling.  The data used to determine the coefficients comes from the set 1 Monte Carlo simulations. The isotope(s) highlighted in teal are the stable isotope(s) of respective elements, yttrium ($^{89}$Y) and zirconium ($^{90}$Zr, $^{91}$Zr, $^{92}$Zr, $^{94}$Zr, and $^{96}$Zr). The teal dashed line in the top panel indicates the $A=89$ isobaric chain.}
    \label{fig:correlation-all-species}
\end{figure}

We now turn to discussing how the dominant neutron capture rates influence the elemental abundance pattern in the context of the same set 1 (uncorrelated, uniform standard deviation of one half order of magnitude across nuclear species) Monte Carlo simulations.
In the top panel of Fig.~\ref{fig:correlation-all-species}, we use the black hole accretion disk outflows from Ref. \citep{2014MNRAS.441.3444M} (MF14) and plot the Pearson correlation coefficient between each neutron capture rate and the element yttrium, which has only one stable isotope. The dashed green line in this panel illustrates the $A=89$ isobaric chain which includes yttrium.  We see that the neutron capture rates with the highest Pearson correlation coefficients are clustered around this line. Furthermore, the $(n, \gamma)$ rates which are on the line and to the right of the line (higher $A$) are negatively correlated with the abundance of yttrium, while those on the left (lower $A$) are positively correlated. 

%
%

%
%

\begin{figure}
    \centering
    \includegraphics[width=\linewidth]{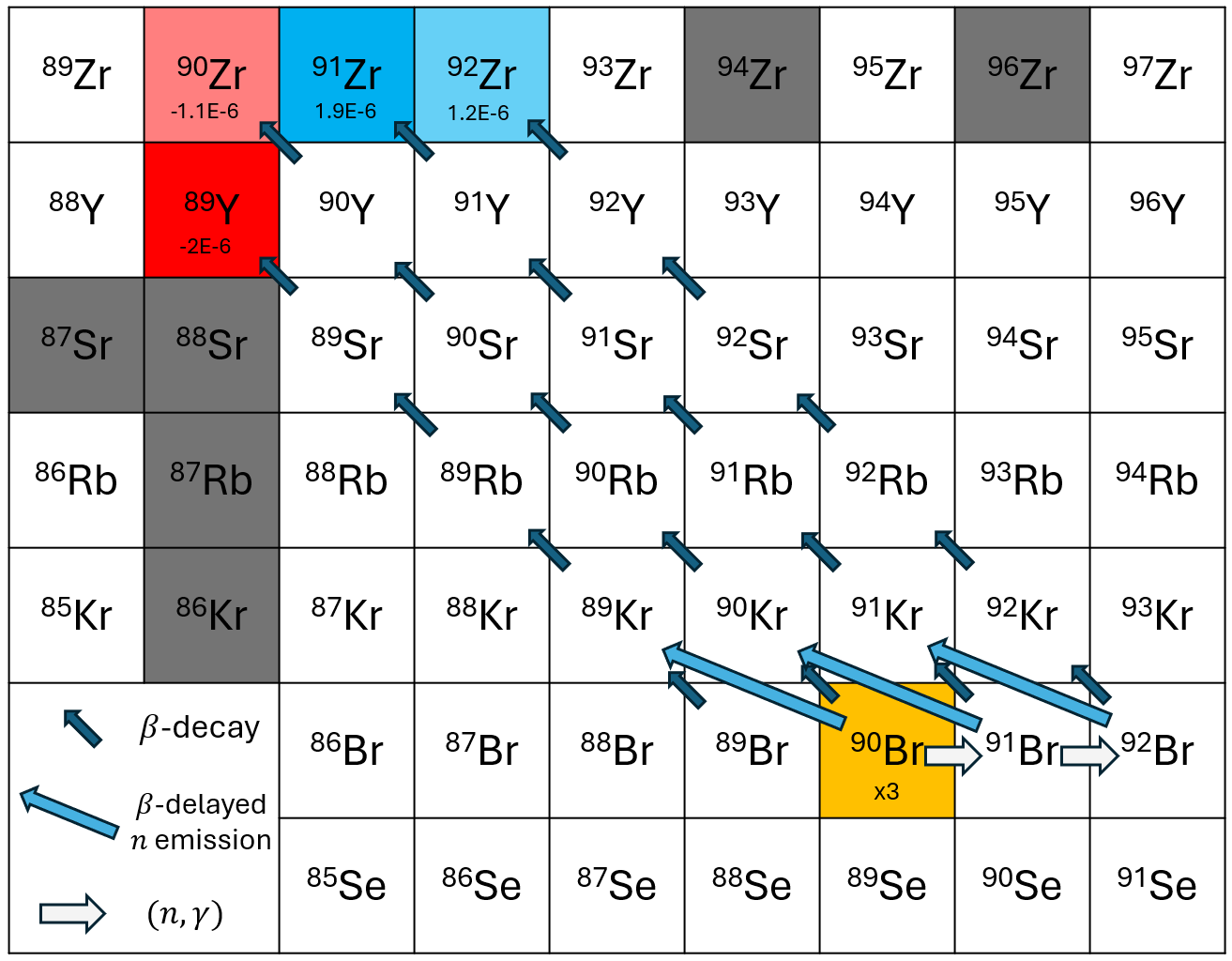}
    \caption{Schematic depiction of nuclear flows in trajectory MF14-03 that are relevant when the $^{90}$Br$(n,\gamma)$ rate is increased. The arrows represent $\beta$-decay (dark blue), $\beta$-delayed neutron emission (lighter blue), and neutron capture (white).  Blue squares represent stable isotopes that gain final abundance when the $^{90}$Br$(n,\gamma)$ rate is increased. In contrast, red squares represent stable isotopes that lose final abundance due to the rate increase.  A lighter red color indicates a smaller magnitude of correlation of the $^{90}$Br$(n,\gamma)$ rate with the final abundance of that nuclear species. 
    For this example we adjusted the $^{90}$Br(n,$\gamma$) rate upward by a factor of 3.  The resulting changes in the final abundances of yttrium and zirconium are shown in their tiles.
    Gray squares are other stable or long-lived isotopes.}
    \label{fig:Y-90Br}
\end{figure}

%
%

\begin{figure}
    \centering
    \includegraphics[width=\linewidth]{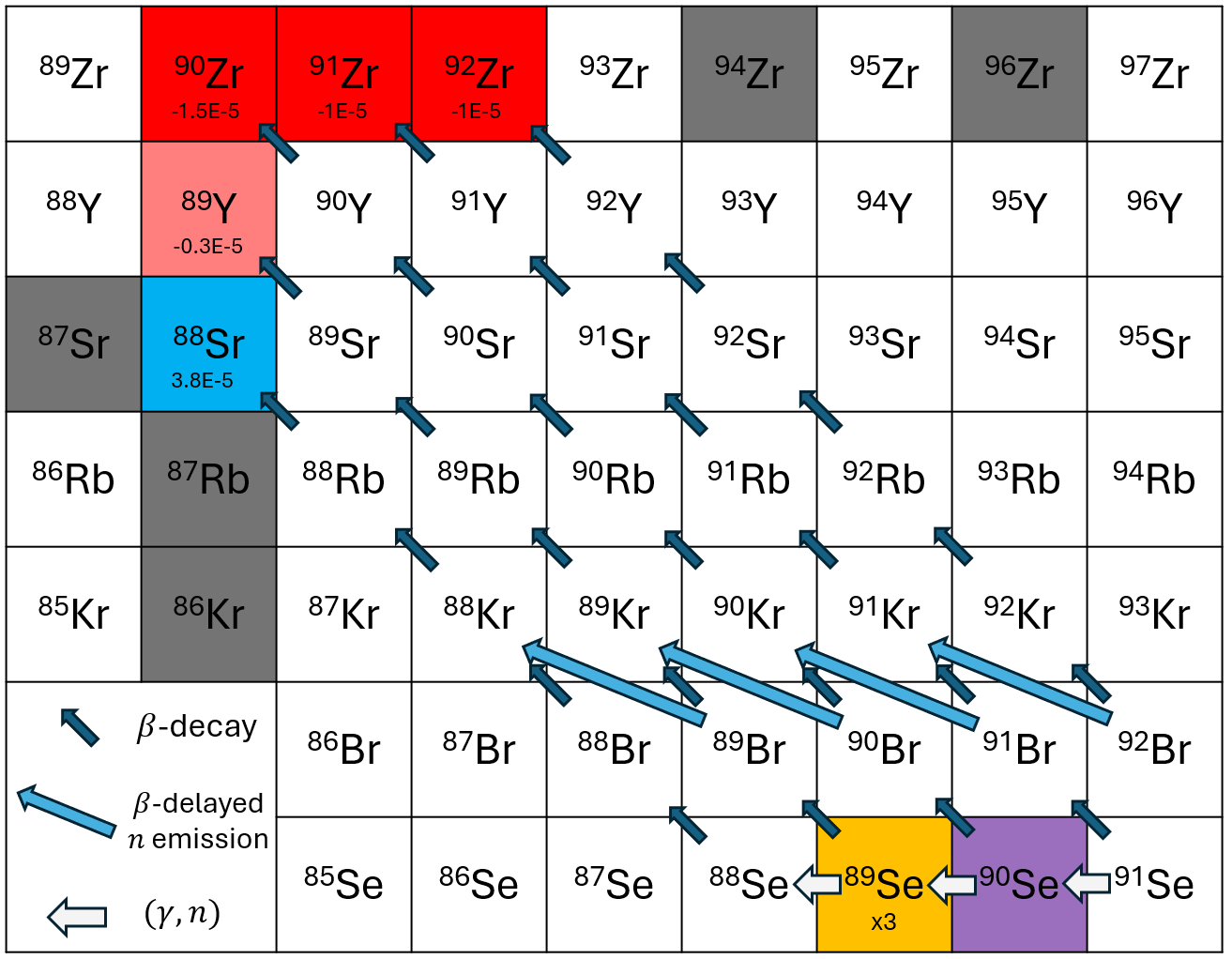}
    \caption{Same as Fig.~\ref{fig:Y-90Br} for the trajectory R+21-03 and the $^{89}$Se$(n,\gamma)$ rate. Photodissociation is represented as white arrows pointed to the left. When this rate is increased it leads to a shift in the final abundance from $^{90}$Zr, $^{91}$Zr, $^{92}$Zr and $^{89}$Y to $^{88}$Sr. }
    \label{fig:Zr-89Se}
\end{figure}

We choose $^{90}{\rm Br}(n,\gamma)^{91}{\rm Br}$, which has a Pearson correlation coefficient of $\rho \approx -0.5$ with yttrium (see the left panel of  Fig. \ref{fig:correlation-scatter}), to illustrate the underlying dynamics at play.    In Fig.~\ref{fig:Y-90Br}, we show schematically the role of $^{90}{\rm Br}(n,\gamma)^{91}{\rm Br}$ during element synthesis. The white arrow directed to the right indicates neutron capture on $^{90}{\rm Br}$, and we see that this rate competes with the $\beta$-decay and $\beta$-delayed one-neutron emission out of $^{90}{\rm Br}$. As the rate of $^{90}{\rm Br}(n,\gamma)^{91}{\rm Br}$ increases, some of the nuclei will flow to higher isobaric chains instead of making their way to $^{89}{\rm Y}$ or $^{90}{\rm Zr}$, producing a negative correlation coefficient between the final abundance of yttrium and $^{90}{\rm Br}(n,\gamma)^{91}{\rm Br}$.  A similar mechanism operates for the other neutron capture rates that are on yttrium's isobaric chain. For those nuclei that are to the left of this chain, an increase in their neutron capture rates will move nuclei into the $\beta$-decay chain that eventually reaches yttrium, so these rates exhibit a positive correlation.

Although it is most common that the cause of the correlations between neutron capture rates and final abundances is directly through the neutron capture rate, in some cases a change in a neutron capture rate has its largest effect via the associated photodissociation rate.  In the bottom panel of Fig.\ \ref{fig:correlation-all-species} we plot the Pearson correlation coefficient of each neutron capture rate with the final abundance of zirconium for the R+21, 
scenario and find that the largest correlation is with $^{89}{\rm Se}(n,\gamma)^{90}{\rm Se}$. From Table \ref{table:abundance-rate-correlation} and the right panel of Fig. \ref{fig:correlation-scatter},
we see that it has a Pearson correlation coefficient of $\rho \approx-0.6$.  Unlike the situation in the top panel of Fig.~\ref{fig:correlation-all-species}, this rate is both negatively correlated {\it and} located to the left of all of the isobaric chains that include the stable zirconium isotopes. Additionally, we see from Table \ref{table:abundance-rate-correlation} that $^{89}{\rm Se}(n,\gamma)^{90}{\rm Se}$ is also negatively correlated with the final abundance of yttrium but positively correlated with the abundance of strontium.

We illustrate the underlying dynamics behind the negative correlation between the $^{89}{\rm Se}(n,\gamma)^{90}{\rm Se}$ rate and  
the abundance of zirconium in Fig.~\ref{fig:Zr-89Se}. The pair of reactions $^{90}{\rm Se} + \gamma \leftrightarrow n + ^{89}{\rm Se}$ is one of the first in its local region to drop out of equilibrium, and for a brief time, $^{90}{\rm Se}(\gamma,n)^{89}{\rm Se}$ is larger than the reverse reaction. This is illustrated by the white arrow out of $^{90}{\rm Se}$ (i.e., from the purple colored box to yellow colored box).  This situation, where only a few rates have dropped out of equilibrium, is sometimes termed ``early freeze-out.''  The pathway directly affected by $^{90}{\rm Se} + \gamma \rightarrow n + ^{89}{\rm Se}$ is also illustrated in Fig.~\ref{fig:Zr-89Se}, where we see that an increase in this photodissociation rate produces a net flow of nuclei away from the $\beta$-decay  chain that would end in $^{90}{\rm Zr}$ and $^{89}{\rm Y}$ and toward chains that end in $^{88}{\rm Sr}$.

There are additional subdominant effects that enhance the correlation of $^{89}{\rm Se}(n,\gamma)^{90}{\rm Se}$ with the final abundance of zirconium.  When $^{89}{\rm Se} + n \leftrightarrow \gamma + ^{90}{\rm Se}$ first drops out of equilibrium, $^{90}{\rm Se} + n \leftrightarrow \gamma +  ^{91}{\rm Se}$ is still in equilibrium.  Thus, while $^{90}{\rm Se}$ abundance flows out of this isotope to the left in Fig.~\ref{fig:Zr-89Se}, abundance also flows in from the right to maintain the equilibrium ratio of $^{91}{\rm Se}$ to $^{90}{\rm Se}$.  Therefore, there is a loss of nuclei from pathways that terminate in $^{91}{\rm Zr}$. A similar dynamic occurs with $^{88}{\rm Se} + n \leftrightarrow \gamma + ^{89}{\rm Se}$, which also remains in equilibrium longer than $^{89}{\rm Se} + n \leftrightarrow \gamma + ^{90}{\rm Se}$, resulting in a flow of nuclei toward $^{88}{\rm Se}$ and eventually to $^{88}{\rm Sr}$. Hence, a singular change in $^{89}{\rm Se}$ neutron capture rate shifts the overall abundance flow from $^{92}{\rm Zr}$, $^{91}{\rm Zr}$, $^{90}{\rm Zr}$ and $^{89}{\rm Y}$ to $^{88}{\rm Sr}$.

This competition between $\beta$-decay and neutron capture/photodissociation is always present and remains the dominant underlying physical mechanisms behind the observed rate-abundance correlations, irrespective of the off‑diagonal structure of the covariance matrix. However, varying multiple rates in the freeze-out region in a correlated manner can exacerbate or mitigate the consequences of these mechanisms.  We discuss this in more detail in Sec.\ \ref{sec:correlated}.

\begin{figure}
    \centering
    \includegraphics[width=\linewidth]{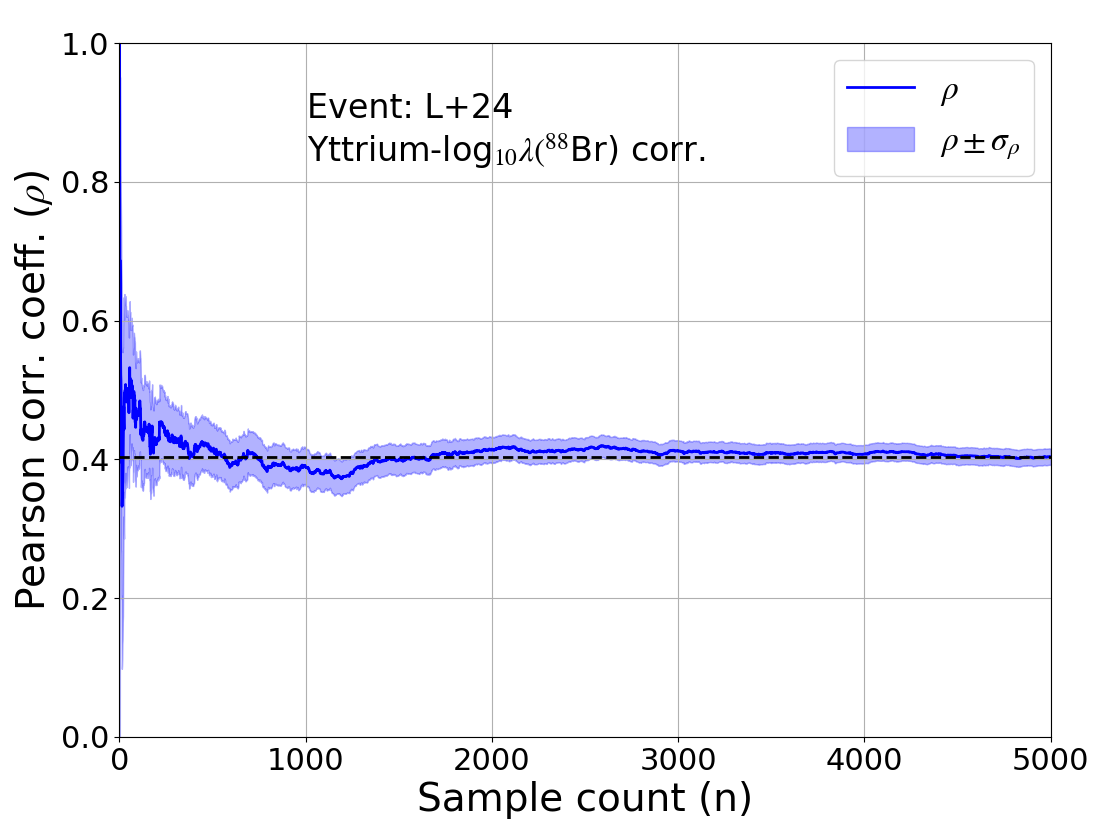}
    \caption{Pearson correlation coefficient, $\rho$ between the final abundance of the element yttrium and the rate of $^{88}{\rm Br}(n,\gamma)^{89}{\rm Br}$ as a function of the number of samples.  Also shown is the standard deviation of the correlation coefficient $\sigma_\rho$.}
    \label{fig:convergence}
\end{figure}

The standard deviation on Pearson correlation coefficients is given by $\sigma_\rho=\frac{1-\rho^2}{\sqrt{n}}$, where $n$ is the number of samples utilized to obtain the correlation. Given that we employ 5000 samples, we obtain an error $\sigma_\rho<$ 0.015 for all element-neutron capture rate correlation coefficients. We show $\sigma_\rho$ as a function of the number of samples in Fig.  \ref{fig:convergence}.

%
%

\begin{figure}
    \centering
    \includegraphics[width=\linewidth]{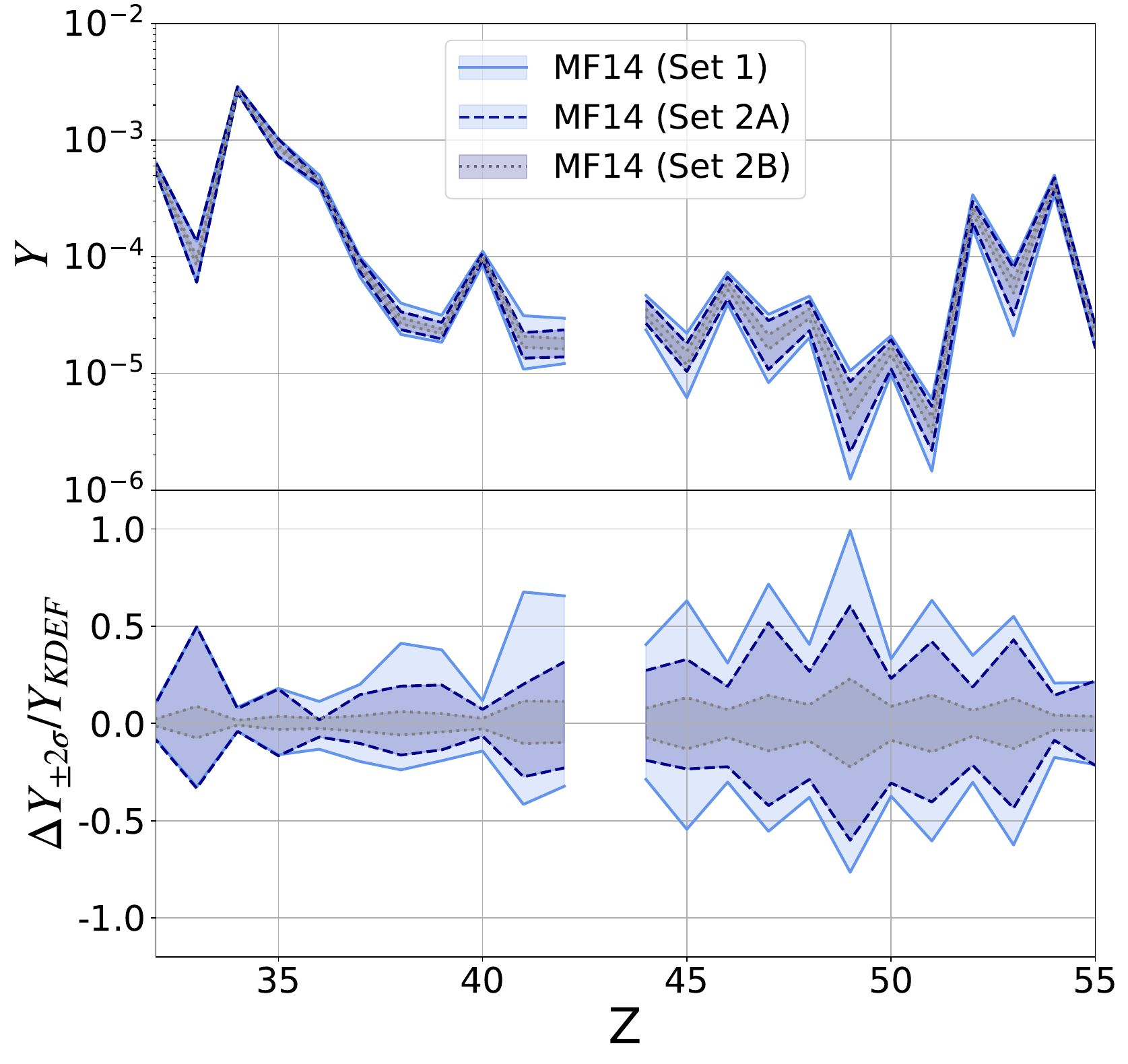}
    \caption{Top panel: Final elemental abundance distribution for the astrophysical scenario MF14. The lines and shaded distributions show the 2-$\sigma$ confidence interval of the abundance obtained in three of the Monte Carlo simulations.  Light blue (solid) shows the results from our base set of Monte Carlo simulations (set 1), dark blue (dashed) shows the results from set 2A Monte Carlo simulations, where 35 key rates are sampled within a smaller range i.e., the log normal distribution as a factor-of-5 smaller width, while the remaining rates are sampled in the same way as they are in set 1, and gray (dotted) shows the results from set 2B Monte Carlo simulations, where all rates were sampled from the smaller range (see Table \ref{tab:montecarlo-summary} for details of these Monte Carlo sets). Bottom panel: The ratio of the 2-$\sigma$ final abundance width with $Y_{\mathrm{KDEF}}$. }
    \label{fig:abundance-events-reduced}
\end{figure}

%
%

\subsection{Consequences of reduced uncertainty in selected rates}
\label{subsec:reduced}

Since in the future, improvements in the rates may come from both theory and experiment, we wish to assess the effect of obtaining reliable rates with reduced uncertainties. For this purpose, we turn to our second sets of uncorrelated Monte Carlo simulations. As described in Sec.\ \ref{subsec:MC-sampling},
we take the 35 rates identified in Table~\ref{table:rate-abundance-correlation} for MF14 for $Z=36$ to 54 and reduce the standard deviation (of the lognormal distribution) by a factor of 5 for those rates, while the remaining rates are sampled as they were in set 1. Therefore the standard deviations of the rates of the 35 isotopes are set to 0.1 while the standard deviations for the rates of the remaining isotopes are kept at the default 0.5 used in set 1. In Fig.~\ref{fig:abundance-events-reduced}, the 2-$\sigma$ result of this uncorrelated Monte Carlo simulation of set 2A is shown as the dark blue region, while the 2-$\sigma$ region of the original Monte Carlo set 1 is shown as the light blue region.  We observe that the uncertainty in the predicted abundance pattern resulting from neutron capture rates has been substantially reduced.  We also observe that further improvements in the uncertainty band can be made when the sampling width of all rates, not just the 35 key rates, is reduced by a factor of 5. This can  be seen as the gray, set 2B region in the figure.

%
%

\subsection{Relationships between neutron capture rates and {\it ratios} of elemental abundances}
\label{subsec:element-element-relationships}

The ``absolute'' abundances derived from spectra of stars are scaled arbitrarily, as it is unknown how many atoms are truly in a star's photosphere. Often, elemental ratios are used for their independence of scaling and their robustness to systematic uncertainties, making them powerful diagnostics for testing nucleosynthetic sites, as discussed in \citet{2022ApJ...935...27P} where abundance ratios and  $(\alpha,n)$ rates were considered. 
It is also the case that a single neutron capture rate can simultaneously influence the abundance of more than one element.  Therefore, a correlation between a rate and a ratio can produce a larger Pearson correlation coefficient than a correlation between a single elemental abundance and a neutron capture rate.  

%
%

\begin{figure*}
    \centering
    \includegraphics[width=0.49\linewidth]{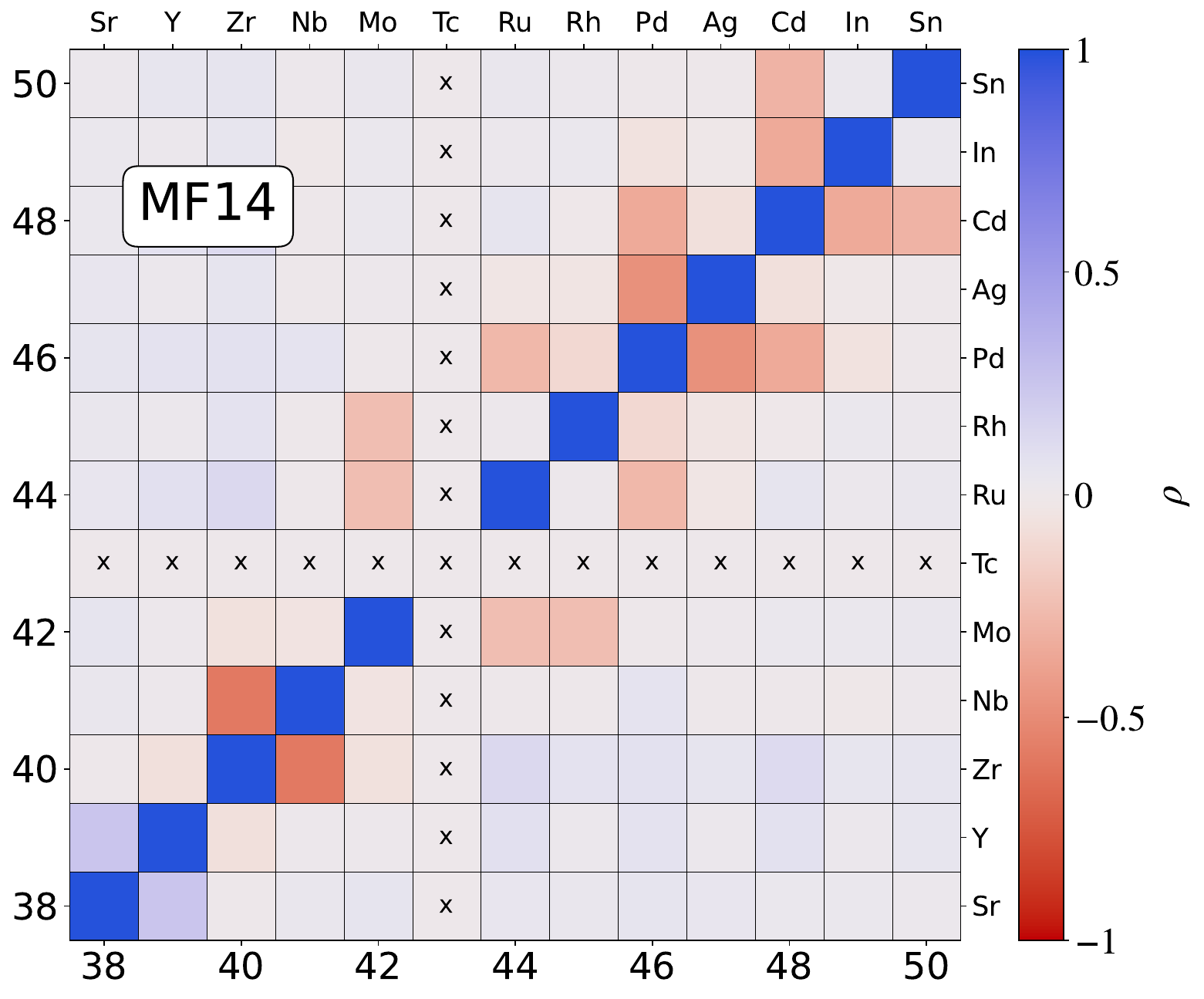}
    \includegraphics[width=0.49\linewidth]{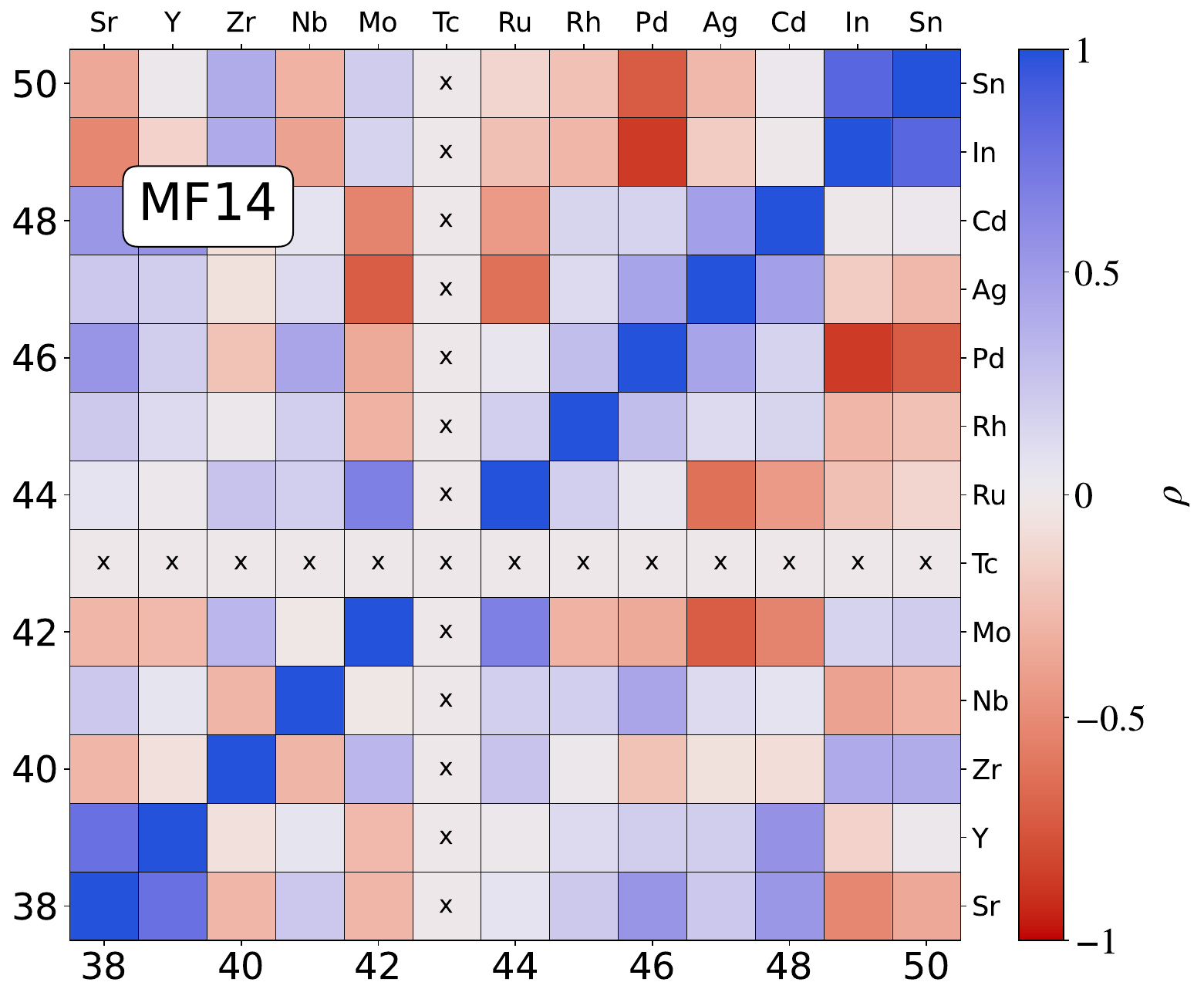}
    \caption{Correlation between elemental abundances in the astrophysical scenario MF14 for the uncorrelated Monte Carlo set 3A (left panel) and the correlated Monte Carlo set 3B (right panel). The colorbar indicates the Pearson correlation coefficient between corresponding elements.}
    \label{fig:abundance-correlations-comparison}
\end{figure*}

To explore this,  we turn to the set 3A uncorrelated Monte Carlo simulations described in Sec. \ref{subsec:MC-sampling}. Recall that in set 3A we sample using the standard deviations from KDUQEF but not the full covariance matrix. In the left panel of Fig.\ \ref{fig:abundance-correlations-comparison}, we show the final abundance-abundance correlations obtained from our set 3A Monte Carlo runs for the MF14 scenario. We see that the most common situation is that the abundance of one element is most strongly correlated with a nearby element, and that the correlation is negative, consistent with the mechanisms illustrated in Figs.\ \ref{fig:Y-90Br} and \ref{fig:Zr-89Se}. However, a glance at the right panel of Fig. \ref{fig:abundance-correlations-comparison} indicates that the abundance–abundance correlations will shift once a correlated Monte Carlo approach is adopted. Thus, we conclude that Pearson correlation coefficients between neutron capture rates and elemental abundance ratios will be more sensitive to the underlying covariance matrix than abundance-rate coefficients.

\section{Correlated Monte Carlo}
\label{sec:correlated}

%
%

\begin{figure*}
    \centering
    \includegraphics[width=0.49\linewidth]{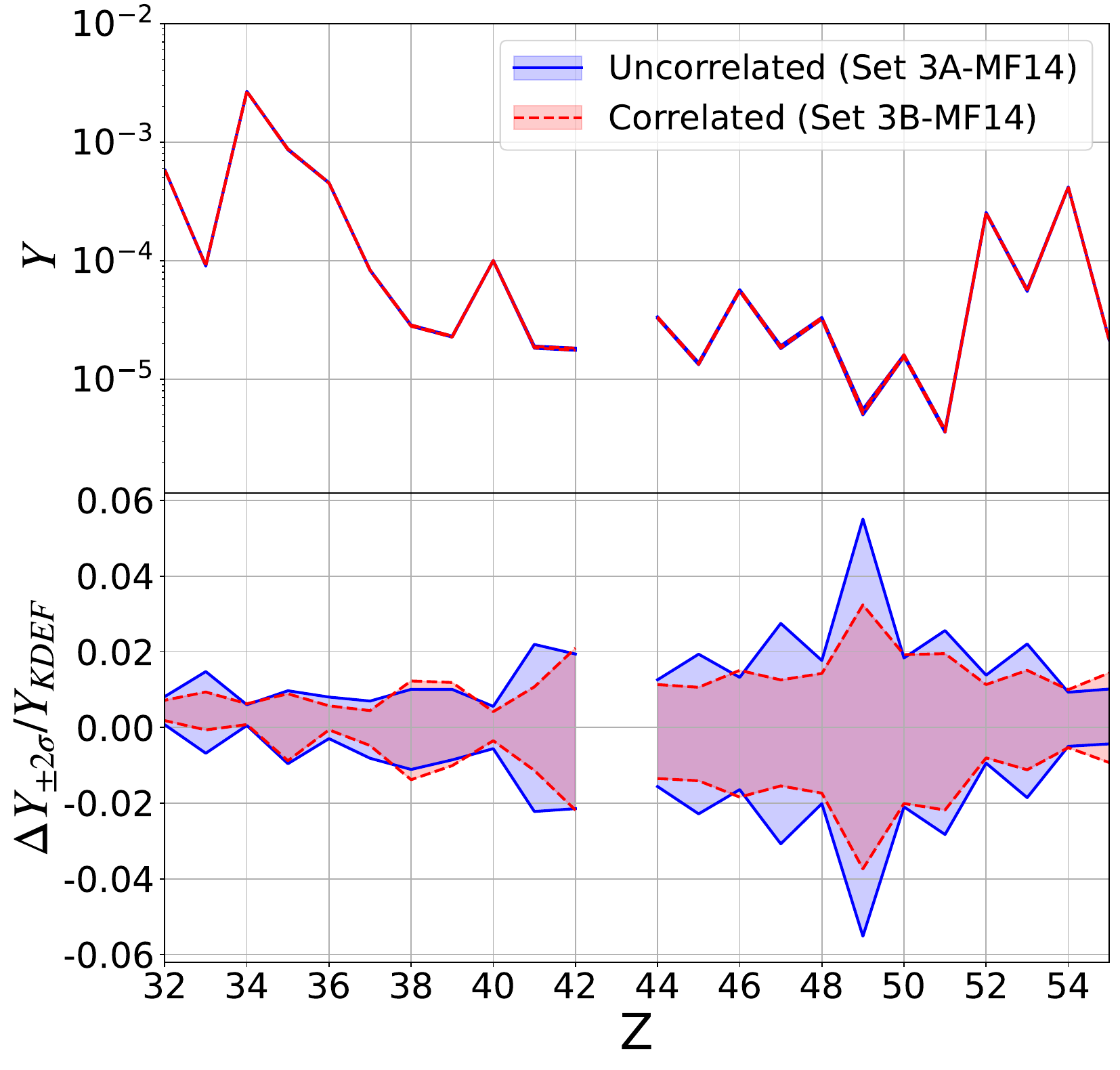}
    \includegraphics[width=0.49\linewidth]{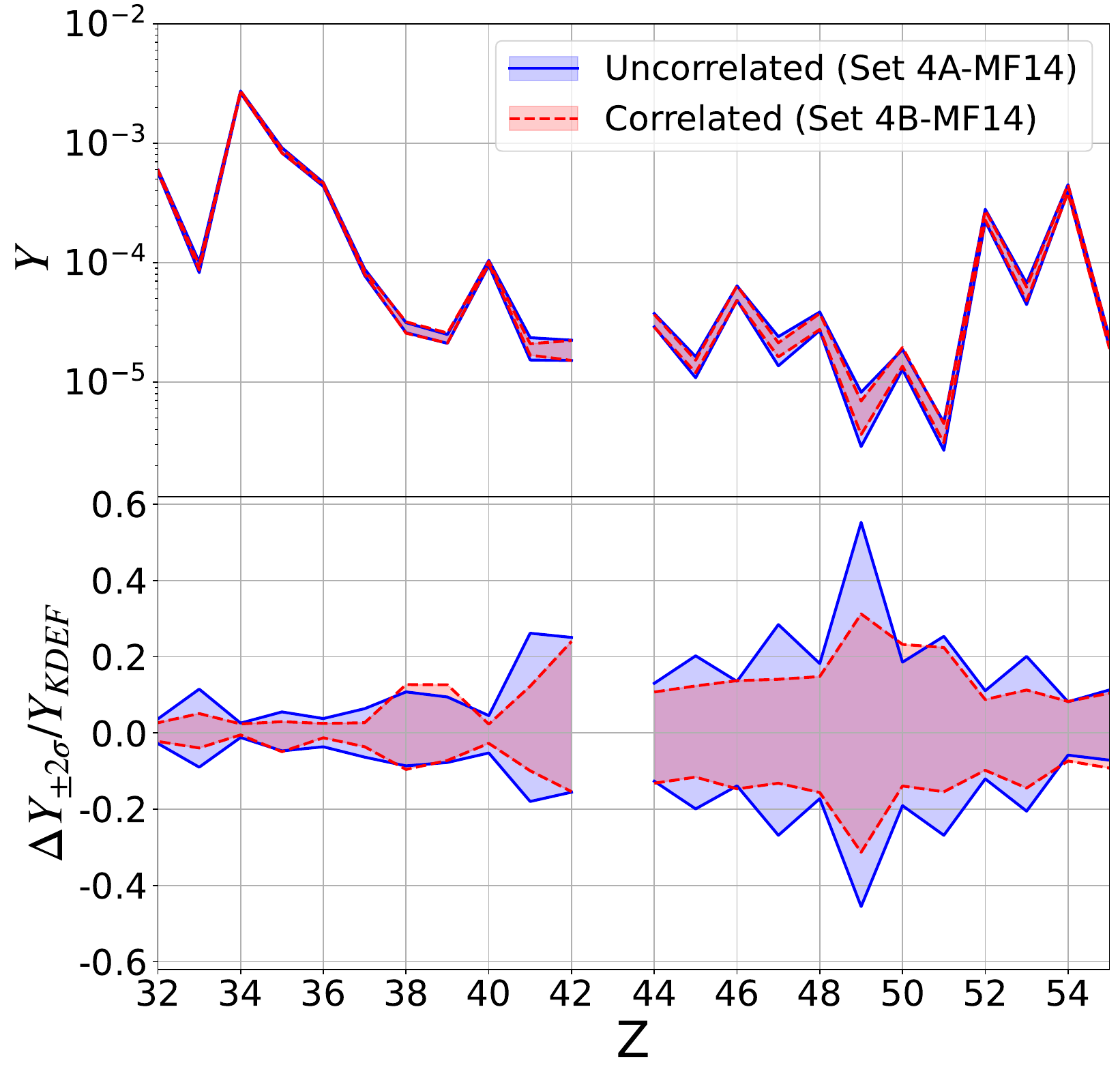}
    \caption{Top panel: Abundance distribution in the MF14 scenario obtained with and without correlations. Shown are the 2-$\sigma$ ranges for the set 3A uncorrelated Monte Carlo (solid blue) with the set 3B correlated Monte Carlo (dashed coral) (left panel) and for the set 4A uncorrelated Monte Carlo (solid blue) with the set 4B correlated Monte Carlo (dashed coral) (right panel). Bottom panel: The ratio of the 2-$\sigma$ final abundance width with $Y_{\mathrm{KDEF}}$.}
    \label{fig:corr-abundance-dist}
\end{figure*}

\begin{figure*}
    \centering
    \includegraphics[width=0.48\linewidth]{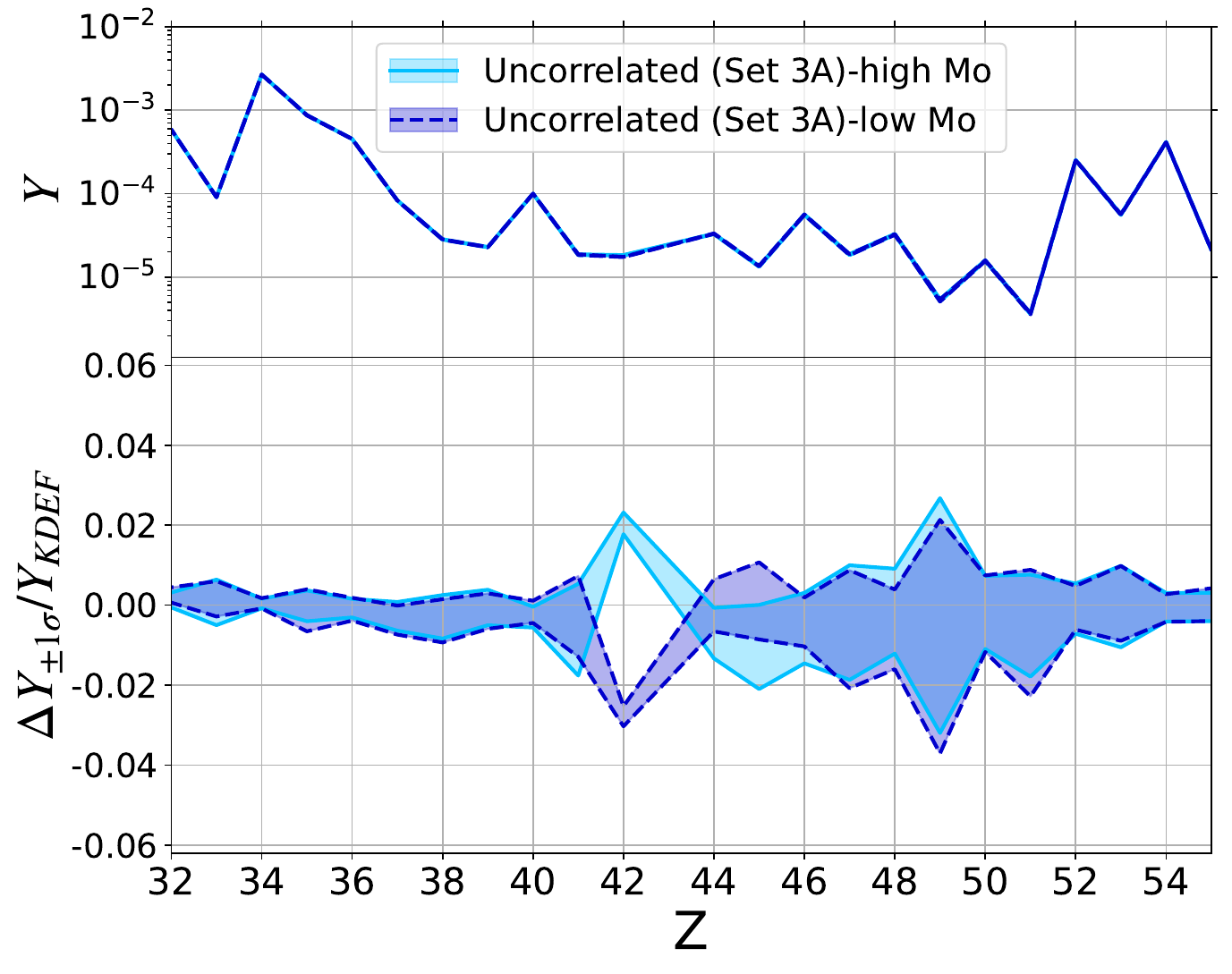}
    \includegraphics[width=0.48\linewidth]{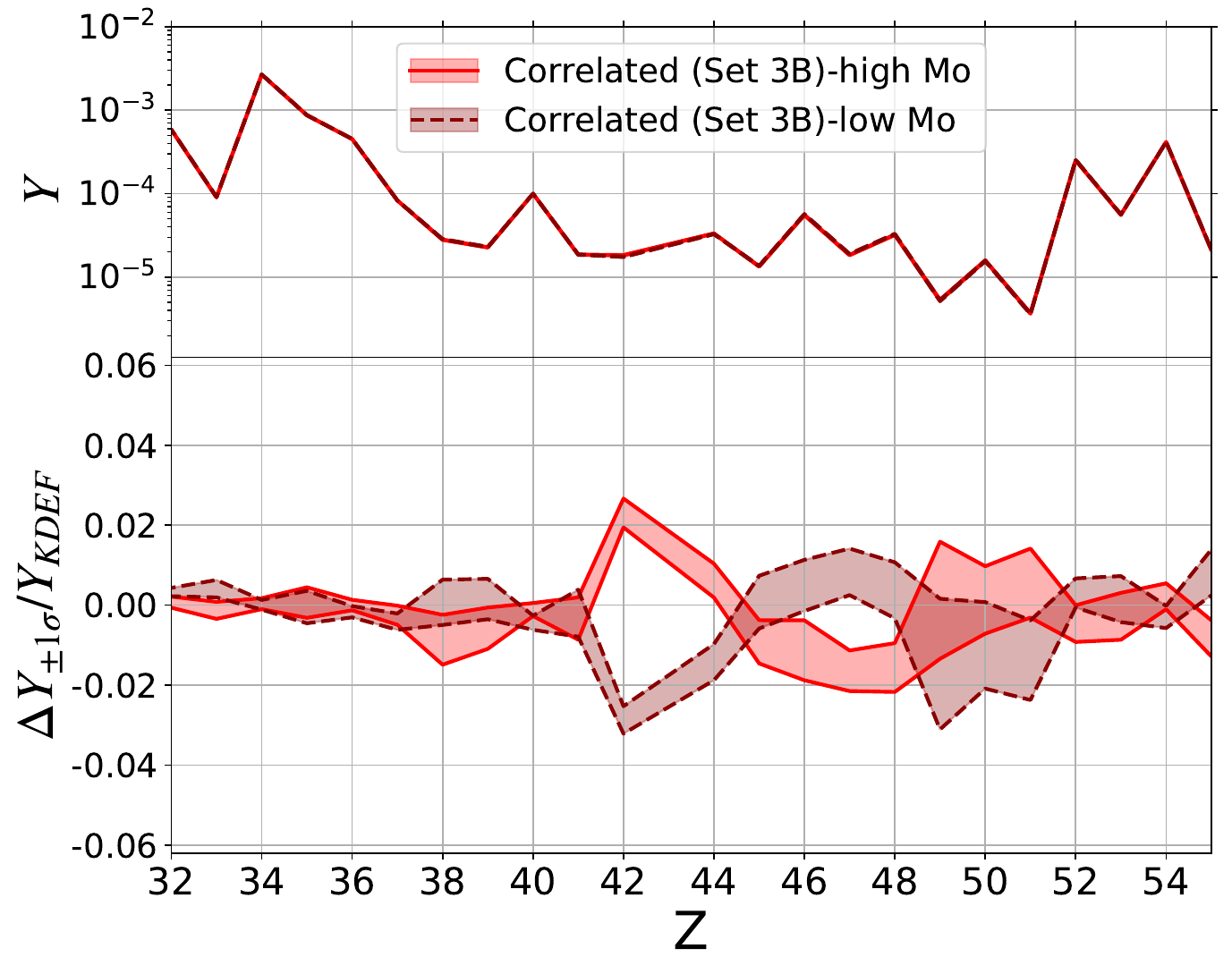}
    \caption{Abundance (1-$\sigma$ range) from the 100 MC runs which produce the largest abundance of molybdenum (solid lines) and the 100 MC runs which produce the smallest abundance (dashed lines). Left panel: Uncorrelated set 3A Right panel: Correlated set 3B }
    \label{fig:uncorr-v-corr-abundance-dist}
\end{figure*}

%
%

In this section, we compare the set 3B Monte Carlo simulations, i.e., sampling using the full covariance matrix from KDUQEF, to the set 3A Monte Carlo simulations, i.e., sampling using only the standard deviations from KDUQEF and not the full covariance matrix.  We remind the reader that the full covariance matrix (set 3B) we explore includes only the effects of the KDUQEF OMP on the cross sections and does not fully capture the nuclear physics, such as nuclear level densities and $\gamma$-strength functions, underlying a neutron capture rate. Therefore the correlations explored here represent a subset of the full model uncertainties. While in the future, we wish to probe other ingredients in $(n,\gamma)$ cross sections, such as level densities and strength functions, in this work we use our set 3B Monte Carlo simulations to begin an exploration into the role of correlated capture rates and look forward to applying the methods developed here to future studies.

The final abundance distributions for the set 3A and set 3B simulations are shown in the left panel of Fig.\ \ref{fig:corr-abundance-dist}. The upper left panel shows that the 2-$\sigma$ ranges for the final elemental distribution for these simulations are modest. The bottom panel shows these spreads in more detail and we see that these 2-$\sigma$ ranges are similar for the correlated and uncorrelated simulations. In the right panel, we show the same plot for simulations 4B and 4A, which are identical to 3B and 3A, except that the (logarithmic) covariance matrix has been multiplied by 100 in both cases. It is encouraging that we see a similar pattern of uncertainties, albeit with a larger overall magnitude, suggesting that the same underlying physics is being probed in both simulations, and the 3A and 3B simulations are not being substantially influenced by computational noise. 

Returning to simulations 3A and 3B, we examine the similarity in the size of these two uncertainty bands. Such similarity in uncertainty widths may arise when the off-diagonal elements of the covariance matrix are much smaller than the diagonal elements; however, this is not the case for the KDUQEF covariance matrix. Thus, these uncertainty bands warrant further study, and we present a subset of our runs in Fig.\ \ref{fig:uncorr-v-corr-abundance-dist}. In the left panel, we choose the 100 samples of the full 5000 that produce the most (least) molybdenum from the set of 3A simulations, and we plot the 1--$\sigma$ abundance spread as the light (dark) blue region. We see little relationship between the final abundance of molybdenum, which has $Z=42$, and that of other elements. Technetium, which has $Z=43$, does not have a stable isotope, and therefore there is no data point at $Z=43$.) We repeat the same analysis for the 3B set of simulations in the right panel, showing the high (low) molybdenum runs for the light (dark) red abundance spreads. In this case, a high molybdenum abundance is correlated with the abundance of ruthenium ($Z=44$) and is anticorrelated with the abundances of palladium, silver, and cadmium ($Z=46$, 47, and 48, respectively). Thus, we find that our full covariance matrix induces correlations between the final abundances of some elements. However, when all simulations are combined, these correlations are hidden in the full uncertainty band. Furthermore, the total uncertainty region produced by the complete covariance matrix does not differ substantially from the total uncertainty range obtained by using only the diagonal parts of the covariance matrix, as seen in Fig.\ \ref{fig:corr-abundance-dist}.

\begin{figure*}
    \centering
    \includegraphics[width=0.49\linewidth]{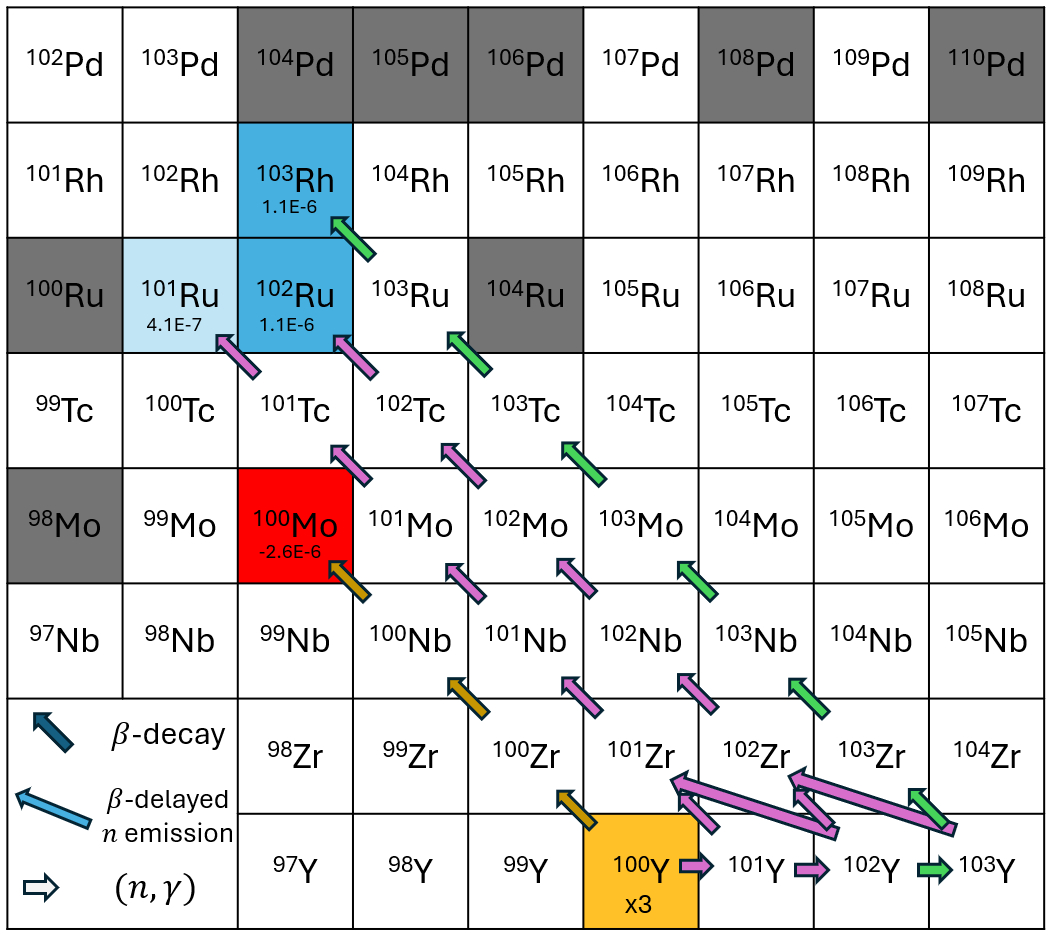}
    \includegraphics[width=0.49\linewidth]{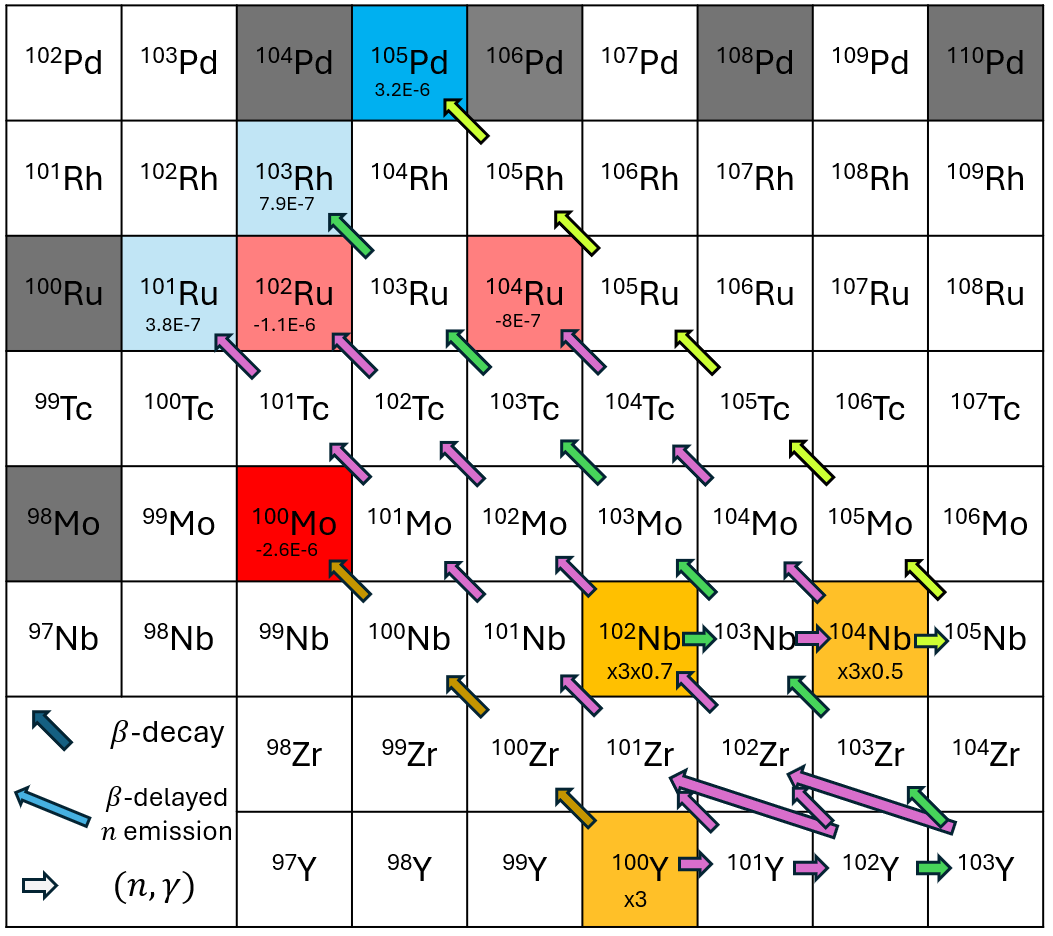}
    \caption{Similarly to Fig.\ \ref{fig:Y-90Br} for the event MF14. Left: The effect of an increase in the $^{100}$Y$(n,\gamma)$ rate. When this rate is increased, it leads to a shift in the final abundance from $^{100}$Mo to $^{101}$Ru, $^{102}$Ru, and $^{103}$Rh. Right: Effect of increasing  $^{100}$Y$(n,\gamma)$, $^{104}$Nb$(n,\gamma)$, and $^{102}$Nb$(n,\gamma)$ together. This leads to a shift in the final abundance from $^{100}$Mo, $^{102}$Ru and $^{104}$Ru to $^{101}$Ru, $^{103}$Rh, and $^{105}$Pd. A lighter blue color indicates a smaller increase in the final abundance of that nuclear species. Arrows indicate neutron capture, $\beta$ decay, and $\beta$-delayed neutron emission. Arrows of different colors lead to different stable elements: ginger for Mo, magenta for Ru, green for Rh, and lime for Pd.}
    \label{fig:Mo-Ru-100Y-104Nb-102Y}
\end{figure*}

Given the results in the right panel of Fig.\ \ref{fig:uncorr-v-corr-abundance-dist}, one might expect that the final abundance of a given element will be positively correlated with the final abundance of elements that have nearby atomic number.   Indeed, if we repeat the final abundance-abundance analysis from the left panel of Fig.\ \ref{fig:abundance-correlations-comparison} of Sec.\ \ref{subsec:element-element-relationships}, then, using the simulations with the full covariance matrix we obtain very different results as shown in the right panel. 
In Fig.\ \ref{fig:abundance-correlations-comparison}, it is particularly striking that the anticorrelation of the abundances of neighboring elements in the left panel becomes a positive correlation in the right panel.

To shed light on this behavior, we introduce an illustrative model. First, we turn to Table \ref{table:abundance-rate-correlation} to determine which neutron capture rates have the largest magnitude Pearson correlation coefficient with the final abundance of molybdenum in our set 1 Monte Carlo simulation and find that they are $^{94}{\rm Rb}(n,\gamma)$ and $^{100}{\rm Y}(n,\gamma)$. The latter reaction has the typical direct influence on the abundance pattern through the competition with the $^{100}{\rm Y}$ $\beta$-decay rate, as described in Sec.  \ref{subsec:basecorrelations}, and so we begin by focusing on this reaction.   Similarly to Fig.\ \ref{fig:Y-90Br}, in the left panel of  Fig.\ \ref{fig:Mo-Ru-100Y-104Nb-102Y} we show how nuclear flows change when only the rate of $^{100}{\rm Y}(n,\gamma)$ is increased.  The abundance of $^{100}{\rm Mo}$ decreases because there are fewer nuclei to flow along the $A=100$ isobaric chain to $^{100}{\rm Mo}$ (dark yellow arrows) but the abundances of  $^{101}{\rm Ru}$, $^{102}{\rm Ru}$ and $^{103}{\rm Rh}$ increase due to the increase in nuclei flowing down the $A=101$, $A=102$ (magenta arrows) and $A=103$ (green arrows) isobaric chains. This is consistent with the light red coloring (mild anticorrelation) of the final abundance of molybdenum with both ruthenium and rhodium seen in the left panel of  Fig.\ \ref{fig:abundance-correlations-comparison}.

The rate $^{100}{\rm Y}(n,\gamma)$ is not the only neutron capture rate that directly affects molybdenum, ruthenium, and rhodium abundances. Table \ref{table:abundance-rate-correlation} also lists $^{94}{\rm Rb}(n,\gamma)$ as positively correlated with the final abundance of molybdenum, and $^{104}{\rm Nb}(n,\gamma)$ as positively correlated with the final abundance of ruthenium.
In our KDUQEF covariance matrix, of these two, only $^{104}{\rm Nb}(n,\gamma)$ has a meaningful covariation with $^{100}{\rm Y}(n,\gamma)$. From Tables \ref{table:abundance-rate-correlation} and 
\ref{table:rate-abundance-correlation}, we see that $^{103}{\rm Nb}(n,\gamma$) $^{103}{\rm Zr}(n,\gamma)$, $^{103}{\rm Y}(n,\gamma)$, and $^{102}{\rm Nb}(n,\gamma)$ are correlated with the final abundance of rhodium. In our KDUQEF covariance matrix, of this latter set, only $^{102}{\rm Nb}(n,\gamma)$ has a meaningful co-variation with $^{100}{\rm Y}(n,\gamma)$. The covariations of $^{100}{\rm Y}(n,\gamma)$, $^{104}{\rm Nb}(n,\gamma)$, and $^{104}{\rm Nb}(n,\gamma)$ can be seen in Fig. \ref{fig:corr-MC-rates-samples-corner-only-set-3}.

Building on our illustrative model, we perform a simulation where all three neutron capture rates, i.e, $^{100}{\rm Y}(n,\gamma)$, $^{102}{\rm Nb}(n,\gamma)$, and $^{104}{\rm Nb}(n,\gamma)$, are increased. We depict the relevant pathways in the right panel of the Fig.\ \ref{fig:Mo-Ru-100Y-104Nb-102Y}. We again see that an increase $^{100}{\rm Y}(n,\gamma)$ increases the abundance of 
ruthenium because additional nuclei enter the $A=101$ isobaric decay chain (leftmost set of magenta arrows) through $^{101}{\rm Y}$. However, the associated increase in the $^{102}{\rm Nb}(n,\gamma)$ rate counteracts that effect by removing nuclei that would have continued down the $A=101$ isobaric decay chain. 
An additional effect occurs from the increase in the rate of $^{104}{\rm Nb}(n,\gamma)$ which decreases the abundance of a different isotope of ruthenium, $^{104}{\rm Ru}$ (right-most set of magenta arrows). When all three rates are increased simultaneously, the elemental abundances of molybdenum and ruthenium decrease together. Returning to set 3B of our Monte Carlo simulations, we find that molybdenum and ruthenium are positively correlated in the right panel of Fig.\ \ref{fig:abundance-correlations-comparison} as our illustrative model predicts. Also in this right panel, we observe that in the full set 3B Monte Carlo, additional elements are anticorrelated with molybdenum, such as rhodium and palladium. This is consistent with the pathways shown in the right panel of Fig. \ref{fig:Mo-Ru-100Y-104Nb-102Y}. However, we caution that to fully recover the correlation structure of the final abundances of all elements, we would need to go beyond covarying the three rates in the right panel of Fig. \ref{fig:Mo-Ru-100Y-104Nb-102Y} and use the full covariance matrix.

\section{Conclusion}
\label{sec:conclusion}

In light of the experimental advances that enable the detailed study of neutron-induced reactions on radioactive species, this work presents an updated investigation of the role of neutron capture reactions in \emph{r}-process nucleosynthesis. We focus on weak \emph{r}-process species for which ground-state information, such as masses or halflives, is already experimentally measured. Therefore the reaction rates are the most significant remaining nuclear uncertainties for these species. We quantify how uncorrelated neutron capture rate uncertainties of half an order of magnitude propagate to elemental abundance patterns in three separate astrophysical scenarios. We further utilize these studies to identify the neutron capture rates that are most strongly correlated with individual elemental abundances and confirm the mechanisms of influence from Ref. \cite{Surman:2014}. We show how reductions in the neutron capture rate uncertainties by a factor of 5  for a set of 35 most consequential rates result in $\approx$ 30 to 65 \%  reductions in abundance pattern uncertainties for a range ($Z = 36$ to 54) of weak \emph{r}-process elements. This accounts for a major portion of the improvement realized when a reduction of the same factor is applied to all $\approx 1000$ rates.

We also present a first attempt to employ a correlated Monte Carlo scheme for neutron capture rates in the weak \emph{r} process accounting for underlying optical model correlations. When we compare Monte Carlo simulations that use the full KDUQEF reaction-rate covariance matrix with those that use only its diagonal elements, we see that both approaches produce abundance-pattern variations of similar overall magnitude. At first glance, this seems inconsistent with the fact that the full-covariance Monte Carlo calculations generate stronger correlations with elemental abundances.
However, a closer examination resolves this apparent tension. Suppose that we select a subset of MC realizations in which the abundance of a particular nucleus falls within a narrower interval. In that case, the correlated calculations show reduced uncertainties in the abundances of a few other nearby nuclei but not across the full abundance pattern. This reflects the underlying physical correlations encoded in the full covariance matrix. From our uncorrelated Monte-Carlo studies, we know that the nuclear pathways that lead to the specific final elemental abundances are only directly influenced by a small number of reactions. Thus, long-range correlations of rates across the nuclear chart have less influence than they would if many rates affected the same final abundances in the same way. 
However, when we consider the entire ensemble of Monte-Carlo outcomes without conditioning on the value of any one elemental abundance, the total uncertainty range (the full abundance band) remains comparable to that of the uncorrelated case. In other words, correlations restructure how abundances covary, but they do not necessarily shrink the overall uncertainty envelope.  We consider whether we would reach the same conclusion if the covariance matrices were dramatically increased in scale, by a factor of 100, and find that we do.

We emphasize that the present analysis does not incorporate the correlations between reaction rates arising from underlying dependencies on LDs and GSFs. Nevertheless, the OMP-based study used here serves as a demonstration of our methodology, and our results suggest that physically motivated correlations between neutron-capture rates can be effectively probed with the analysis techniques developed here. However, the immediate physical conclusions from this OMP-based study are limited because, when propagated to final elemental abundances, the variation they introduce is smaller than that of other uncertainties, such as masses, $\beta$-decay rates, and the astrophysical environment. Future work that includes LD- and GSF-constrained off-diagonal covariance terms may produce
uncertainties that are of larger magnitude and could also produce additional inter-abundance correlations. Therefore, we should not draw firm conclusions until the remaining uncertainties are accounted for.

Our results show that including correlated nuclear input uncertainties does not necessarily drastically reduce the overall scale of the uncertainty of predicted abundance patterns. Thus, we expect that results from uncorrelated studies may produce reasonable uncertainty bands, but they do not provide access to the full covariance of predicted observables. However, we note that other correlation structures (our correlations between neutron capture rates are typically $\rho ~\approx ~0.43$) along with different \emph{r}-process paths should be examined in the future.

Looking to the future, we expect correlation analyses will enable improved constraints for neutron-capture rates relevant to the \emph{r} process and therefore reduced uncertainties in final abundance patterns. This future hinges on the combination of new measurements on unstable isotopes and additional theoretical developments to address model defects off-stability. Such improvements to the nuclear structure and reaction theory that underpin HF calculations will enhance predictive power in regions most relevant to the weak \emph{r} process.

\acknowledgments
We thank Konstantinos~Kravvaris, Gregory~Potel, Pranav~Nalamwar and Yanlong~Lin for stimulating discussions. This work is performed in part under the auspices of the U.S.\ Department of Energy (DOE) by Lawrence Livermore National Laboratory under Contract No. DE-AC52-107NA27344, with support from LDRD Project No. 24-ERD-023. The U.S.\ DOE partially supported this work through Contract No. DE-FG0202ER41216 (G.C.M.), No. DE-FG0295ER40934 (R.S.), and No. DE-SC00268442 (ENAF: G.C.M. and R.S.). In addition, this work was supported by the Office of Defense Nuclear Nonproliferation Research and Development (DNN R\&D), National Nuclear Security Administration, U.S.\ DOE (G.C.M. and R.S.) under Contract No. LA22-ML-DE-FOA-2440. G.C.M.\ and R.S.\ acknowledge support from the Network for Neutrinos, Nuclear Astrophysics and Symmetries (N3AS), through the National Science Foundation Physics Frontier Center Award No.\ PHY-2020275. This research was supported in part by the Notre Dame Center for Research Computing. A.K., G.C.M., R.S., and E.M.H.\ thank the Institute for Nuclear Theory at the University of Washington for its kind hospitality and stimulating research environment, U.S.\ DOE Grant No.\ DE-FG02- 00ER41132, and benefited from interactions and workshops supported by the Center for Nuclear astrophysics Across Messengers (CeNAM), which is supported by the U.S.\ DOE, Office of Science, Office of Nuclear Physics, under Award Number DE-SC0023128. J.C.G.\ acknowledges support from the GEM Fellowship supported in part by the National GEM Consortium. LLNL report number LLNL-JRNL-2015703.

Software employed is as follows: \texttt{YAHFC} (Yet Another Hauser-Feshbach Code) \citet{YAHFC} and \texttt{PRISM} (Portable Routines for Integrated nucleoSynthesis Modeling)~\cite{PRISM}.

%
%

\appendix

\section{Rate--Abundance correlations}
\label{app:tables}

In this appendix we present Table \ref{table:rate-abundance-correlation}, which lists neutron-capture rates that correlate with final elemental abundances.

\begin{table*}
    \caption{Neutron capture reactions isotopes that have a Pearson correlation coefficient $|\rho|>$0.30 with the abundances of elements in the range $Z=36$ to $Z=54$ for the three weak $r$-process conditions examined. Sorted by atomic numbers and mass numbers. }
    \resizebox{2.00\columnwidth}{!}{
    \begin{tabular}{@{\extracolsep{10pt}}c @{\hskip 7pt} !{\color{gray!50}\vrule} @{\hskip 7pt} r r r r @{\hskip 20pt} !{\color{gray!50}\vrule} r r r r @{\hskip 20pt} !{\color{gray!50}\vrule} r r r r r r@{}}
        \hline \hline
        \multirow{2}{*}{Reactant$(n,\gamma)$} & \multicolumn{4}{c}{Lund+24 (L+24)} & \multicolumn{4}{l}{Metzger \& Fern\'{a}ndez14 (MF14)} & \multicolumn{6}{c}{Reichert+21 (R+21)} \\
         & Ele. & $\rho$ & Ele. & $\rho$ & Ele. & $\rho$ & Ele. & $\rho$ & Ele. & $\rho$ & Ele. & $\rho$ & Ele. & $\rho$\\ \hline \rule{0pt}{2.5ex}
        $^{85}$Zn  & Kr & -0.40 & - & -  & - & - & - & -  & - & - & - & - & - & - \\
        $^{83}$Ga  & Kr & 0.57 & - & -  & - & - & - & -  & - & - & - & - & - & - \\
        $^{82}$Ge  & - & - & - & -  & Kr & 0.35 & - & -  & - & - & - & - & - & - \\
        $^{85}$Ge  & - & - & - & -  & Rb & -0.45 & - & -  & - & - & - & - & - & - \\
        $^{82}$As  & - & - & - & -  & Kr & 0.86 & - & -  & Kr & 0.93 & Se & -0.42 & - & - \\
        $^{87}$As  & Rb & -0.49 & Sr & 0.40  & - & - & - & -  & - & - & - & - & - & - \\
        $^{88}$As  & Y & 0.38 & Rb & -0.32  & - & - & - & -  & - & - & - & - & - & - \\
        $^{89}$As  & Sr & -0.47 & - & -  & - & - & - & -  & - & - & - & - & - & - \\
        $^{84}$Se  & - & - & - & -  & - & - & - & -  & Rb & 0.31 & - & - & - & - \\
        $^{85}$Se  & - & - & - & -  & - & - & - & -  & Rb & -0.37 & - & - & - & - \\
        $^{86}$Se  & - & - & - & -  & - & - & - & -  & Rb & 0.35 & - & - & - & - \\
        $^{87}$Se  & - & - & - & -  & Rb & -0.51 & - & -  & Sr & -0.56 & - & - & - & - \\
        $^{88}$Se  & Sr & -0.39 & Y & 0.31  & - & - & - & -  & - & - & - & - & - & - \\
        $^{89}$Se  & Y & -0.32 & - & -  & - & - & - & -  & Zr & -0.58 & Sr & 0.51 & Y & -0.38\\
        $^{87}$Br  & - & - & - & -  & Sr & 0.35 & - & -  & Rb & -0.38 & - & - & - & - \\
        $^{88}$Br  & Y & 0.40 & Sr & -0.33  & Sr & -0.58 & - & -  & Y & 0.61 & Sr & -0.49 & - & - \\
        $^{89}$Br  & - & - & - & -  & Y & -0.30 & - & -  & - & - & - & - & - & - \\
        $^{90}$Br  & Y & -0.40 & - & -  & Y & -0.54 & - & -  & Zr & -0.33 & - & - & - & - \\
        $^{92}$Br  & Nb & 0.36 & Zr & -0.34  & - & - & - & -  & - & - & - & - & - & - \\
        $^{94}$Br  & Nb & -0.51 & Mo & 0.32  & - & - & - & -  & - & - & - & - & - & - \\
        $^{92}$Kr  & - & - & - & -  & Nb & 0.50 & Zr & -0.47  & - & - & - & - & - & - \\
        $^{93}$Kr  & Nb & -0.59 & Zr & 0.34  & Nb & -0.54 & - & -  & Nb & -0.31 & - & - & - & - \\
        $^{92}$Rb  & - & - & - & -  & Nb & 0.38 & - & -  & Nb & 0.59 & - & - & - & - \\
        $^{94}$Rb  & Mo & 0.50 & Zr & -0.41  & Mo & 0.61 & Zr & -0.40  & Nb & -0.30 & - & - & - & - \\
        $^{96}$Rb  & Mo & 0.31 & Zr & -0.31  & - & - & - & -  & - & - & - & - & - & - \\
        $^{97}$Sr  & - & - & - & -  & - & - & - & -  & Mo & -0.44 & - & - & - & - \\
        $^{100}$Y  & - & - & - & -  & Mo & -0.45 & - & -  & Mo & -0.56 & - & - & - & - \\
        $^{102}$Y  & Ru & -0.38 & - & -  & - & - & - & -  & - & - & - & - & - & - \\
        $^{103}$Y  & Rh & -0.64 & Pd & 0.33  & Rh & -0.42 & - & -  & - & - & - & - & - & - \\
        $^{104}$Y  & Ru & -0.40 & - & -  & - & - & - & -  & - & - & - & - & - & - \\
        $^{103}$Zr  & Rh & -0.36 & - & -  & Rh & -0.36 & - & -  & Rh & -0.36 & - & - & - & - \\
        $^{102}$Nb  & - & - & - & -  & Rh & 0.31 & - & -  & Rh & 0.47 & Ru & -0.46 & - & - \\
        $^{103}$Nb  & - & - & - & -  & Rh & -0.45 & - & -  & Rh & -0.42 & - & - & - & - \\
        $^{104}$Nb  & - & - & - & -  & Ru & -0.57 & Pd & 0.34  & Pd & 0.60 & Ru & -0.58 & - & - \\
        $^{106}$Nb  & Ag & 0.32 & - & -  & Ag & 0.30 & - & -  & Ag & 0.38 & - & - & - & - \\
        $^{109}$Nb  & Cd & 0.38 & - & -  & - & - & - & -  & - & - & - & - & - & - \\
        $^{110}$Nb  & Cd & 0.34 & - & -  & - & - & - & -  & - & - & - & - & - & - \\
        $^{103}$Mo  & - & - & - & -  & Rh & -0.34 & - & -  & Rh & -0.50 & Ru & 0.39 & - & - \\
        $^{107}$Mo  & - & - & - & -  & - & - & - & -  & Ag & -0.41 & - & - & - & - \\
        $^{109}$Mo  & Ag & -0.37 & - & -  & Ag & -0.32 & - & -  & - & - & - & - & - & - \\
        $^{108}$Tc  & - & - & - & -  & Ag & 0.37 & - & -  & Ag & 0.58 & Pd & -0.46 & - & - \\
        $^{109}$Tc  & - & - & - & -  & Ag & -0.32 & - & -  & - & - & - & - & - & - \\
        $^{110}$Tc  & Cd & 0.41 & - & -  & Cd & 0.57 & Pd & -0.47  & Cd & 0.66 & - & - & - & - \\
        $^{109}$Ru  & - & - & - & -  & - & - & - & -  & Ag & -0.32 & - & - & - & - \\
        $^{115}$Ru  & In & -0.36 & - & -  & - & - & - & -  & - & - & - & - & - & - \\
        $^{114}$Rh  & In & 0.43 & - & -  & In & 0.53 & - & -  & In & 0.65 & Cd & -0.46 & - & - \\
        $^{115}$Rh  & In & -0.38 & - & -  & In & -0.50 & - & -  & - & - & - & - & - & - \\
        $^{116}$Rh  & - & - & - & -  & Sn & 0.46 & - & -  & Sn & 0.69 & - & - & - & - \\
        $^{115}$Pd  & - & - & - & -  & - & - & - & -  & In & -0.42 & - & - & - & - \\
        $^{120}$Ag  & - & - & - & -  & Sb & 0.33 & - & -  & Sb & 0.43 & - & - & - & - \\
        $^{123}$Ag  & Sb & -0.60 & - & -  & Sb & -0.51 & - & -  & - & - & - & - & - & - \\
        $^{124}$Ag  & Sn & -0.56 & - & -  & Sn & -0.34 & - & -  & Sn & -0.36 & Sb & -0.34 & - & - \\
        $^{128}$Ag  & Xe & 0.42 & - & -  & - & - & - & -  & I & -0.42 & - & - & - & - \\
        $^{123}$Cd  & Sb & -0.36 & - & -  & Sb & -0.44 & - & -  & Sb & -0.48 & - & - & - & - \\
        $^{127}$Cd  & I & -0.61 & Te & 0.45  & I & -0.71 & - & -  & I & -0.74 & - & - & - & - \\
        $^{129}$Cd  & Xe & -0.38 & - & -  & Xe & -0.38 & Te & 0.33  & - & - & - & - & - & - \\
        $^{124}$In  & Sn & -0.33 & - & -  & - & - & - & -  & - & - & - & - & - & - \\
        $^{126}$In  & I & 0.32 & Te & -0.31  & - & - & - & -  & - & - & - & - & - & - \\
        $^{128}$In  & Xe & 0.44 & - & -  & - & - & - & -  & Te & -0.34 & Xe & 0.33 & - & - \\
        $^{130}$In  & - & - & - & -  & Te & -0.62 & Xe & 0.62  & Xe & 0.86 & Te & -0.85 & - & - \\
        $^{129}$Sn  & - & - & - & -  & Xe & -0.32 & - & -  & - & - & - & - & - & - \\
        \hline \hline
    \end{tabular}
    }
    \label{table:rate-abundance-correlation}
\end{table*}

\bibliography{bibliography_1.bib,bibliography_2.bib,bibliography_3.bib}

\begin{thebibliography}{119}%
\makeatletter
\providecommand \@ifxundefined [1]{%
 \@ifx{#1\undefined}
}%
\providecommand \@ifnum [1]{%
 \ifnum #1\expandafter \@firstoftwo
 \else \expandafter \@secondoftwo
 \fi
}%
\providecommand \@ifx [1]{%
 \ifx #1\expandafter \@firstoftwo
 \else \expandafter \@secondoftwo
 \fi
}%
\providecommand \natexlab [1]{#1}%
\providecommand \enquote  [1]{``#1''}%
\providecommand \bibnamefont  [1]{#1}%
\providecommand \bibfnamefont [1]{#1}%
\providecommand \citenamefont [1]{#1}%
\providecommand \href@noop [0]{\@secondoftwo}%
\providecommand \href [0]{\begingroup \@sanitize@url \@href}%
\providecommand \@href[1]{\@@startlink{#1}\@@href}%
\providecommand \@@href[1]{\endgroup#1\@@endlink}%
\providecommand \@sanitize@url [0]{\catcode `\\12\catcode `\$12\catcode `\&12\catcode `\#12\catcode `\^12\catcode `\_12\catcode `\%12\relax}%
\providecommand \@@startlink[1]{}%
\providecommand \@@endlink[0]{}%
\providecommand \url  [0]{\begingroup\@sanitize@url \@url }%
\providecommand \@url [1]{\endgroup\@href {#1}{\urlprefix }}%
\providecommand \urlprefix  [0]{URL }%
\providecommand \Eprint [0]{\href }%
\providecommand \doibase [0]{https://doi.org/}%
\providecommand \selectlanguage [0]{\@gobble}%
\providecommand \bibinfo  [0]{\@secondoftwo}%
\providecommand \bibfield  [0]{\@secondoftwo}%
\providecommand \translation [1]{[#1]}%
\providecommand \BibitemOpen [0]{}%
\providecommand \bibitemStop [0]{}%
\providecommand \bibitemNoStop [0]{.\EOS\space}%
\providecommand \EOS [0]{\spacefactor3000\relax}%
\providecommand \BibitemShut  [1]{\csname bibitem#1\endcsname}%
\let\auto@bib@innerbib\@empty
\bibitem [{\citenamefont {{Burbidge}}\ \emph {et~al.}(1957)\citenamefont {{Burbidge}}, \citenamefont {{Burbidge}}, \citenamefont {{Fowler}},\ and\ \citenamefont {{Hoyle}}}]{BBFH1957}%
  \BibitemOpen
  \bibfield  {author} {\bibinfo {author} {\bibfnamefont {E.~M.}\ \bibnamefont {{Burbidge}}}, \bibinfo {author} {\bibfnamefont {G.~R.}\ \bibnamefont {{Burbidge}}}, \bibinfo {author} {\bibfnamefont {W.~A.}\ \bibnamefont {{Fowler}}},\ and\ \bibinfo {author} {\bibfnamefont {F.}~\bibnamefont {{Hoyle}}},\ }\bibfield  {title} {\bibinfo {title} {{Synthesis of the Elements in Stars}},\ }\href {https://doi.org/10.1103/RevModPhys.29.547} {\bibfield  {journal} {\bibinfo  {journal} {Rev. Mod. Phys.}\ }\textbf {\bibinfo {volume} {29}},\ \bibinfo {pages} {547} (\bibinfo {year} {1957})}\BibitemShut {NoStop}%
\bibitem [{\citenamefont {{LIGO Scientific Collaboration}}\ \emph {et~al.}(2017{\natexlab{a}})\citenamefont {{LIGO Scientific Collaboration}}, \citenamefont {{Virgo Collaboration}}, \citenamefont {{Fermi GBM}}, \citenamefont {{INTEGRAL}} \emph {et~al.}}]{LIGO-GW170817}%
  \BibitemOpen
  \bibfield  {author} {\bibinfo {author} {\bibnamefont {{LIGO Scientific Collaboration}}}, \bibinfo {author} {\bibnamefont {{Virgo Collaboration}}}, \bibinfo {author} {\bibnamefont {{Fermi GBM}}}, \bibinfo {author} {\bibnamefont {{INTEGRAL}}}, \emph {et~al.},\ }\bibfield  {title} {\bibinfo {title} {{Gravitational Waves and Gamma-Rays from a Binary Neutron Star Merger: GW170817 and GRB 170817A}},\ }\href {https://doi.org/10.3847/2041-8213/aa920c} {\bibfield  {journal} {\bibinfo  {journal} {\apjl}\ }\textbf {\bibinfo {volume} {848}},\ \bibinfo {eid} {L13} (\bibinfo {year} {2017}{\natexlab{a}})}\BibitemShut {NoStop}%
\bibitem [{\citenamefont {{LIGO Scientific Collaboration}}\ \emph {et~al.}(2017{\natexlab{b}})\citenamefont {{LIGO Scientific Collaboration}}, \citenamefont {{Virgo Collaboration}}, \citenamefont {{Fermi GBM}}, \citenamefont {{INTEGRAL}}, \citenamefont {{IceCube Collaboration}} \emph {et~al.}}]{LIGO-GW170817-MMA}%
  \BibitemOpen
  \bibfield  {author} {\bibinfo {author} {\bibnamefont {{LIGO Scientific Collaboration}}}, \bibinfo {author} {\bibnamefont {{Virgo Collaboration}}}, \bibinfo {author} {\bibnamefont {{Fermi GBM}}}, \bibinfo {author} {\bibnamefont {{INTEGRAL}}}, \bibinfo {author} {\bibnamefont {{IceCube Collaboration}}}, \emph {et~al.},\ }\bibfield  {title} {\bibinfo {title} {{Multi-messenger Observations of a Binary Neutron Star Merger}},\ }\href {https://doi.org/10.3847/2041-8213/aa91c9} {\bibfield  {journal} {\bibinfo  {journal} {\apjl}\ }\textbf {\bibinfo {volume} {848}},\ \bibinfo {eid} {L12} (\bibinfo {year} {2017}{\natexlab{b}})}\BibitemShut {NoStop}%
\bibitem [{\citenamefont {{Hansen}}\ \emph {et~al.}(2018)\citenamefont {{Hansen}}, \citenamefont {{Holmbeck}}, \citenamefont {{Beers}}, \citenamefont {{Placco}}, \citenamefont {{Roederer}}, \citenamefont {{Frebel}}, \citenamefont {{Sakari}}, \citenamefont {{Simon}},\ and\ \citenamefont {{Thompson}}}]{Hansen+2018}%
  \BibitemOpen
  \bibfield  {author} {\bibinfo {author} {\bibfnamefont {T.~T.}\ \bibnamefont {{Hansen}}}, \bibinfo {author} {\bibfnamefont {E.~M.}\ \bibnamefont {{Holmbeck}}}, \bibinfo {author} {\bibfnamefont {T.~C.}\ \bibnamefont {{Beers}}}, \bibinfo {author} {\bibfnamefont {V.~M.}\ \bibnamefont {{Placco}}}, \bibinfo {author} {\bibfnamefont {I.~U.}\ \bibnamefont {{Roederer}}}, \bibinfo {author} {\bibfnamefont {A.}~\bibnamefont {{Frebel}}}, \bibinfo {author} {\bibfnamefont {C.~M.}\ \bibnamefont {{Sakari}}}, \bibinfo {author} {\bibfnamefont {J.~D.}\ \bibnamefont {{Simon}}},\ and\ \bibinfo {author} {\bibfnamefont {I.~B.}\ \bibnamefont {{Thompson}}},\ }\bibfield  {title} {\bibinfo {title} {{The R-process Alliance: First Release from the Southern Search for R-process-enhanced Stars in the Galactic Halo}},\ }\href {https://doi.org/10.3847/1538-4357/aabacc} {\bibfield  {journal} {\bibinfo  {journal} {\apj}\ }\textbf {\bibinfo {volume} {858}},\ \bibinfo {eid} {92} (\bibinfo {year} {2018})}\BibitemShut {NoStop}%
\bibitem [{\citenamefont {{Sakari}}\ \emph {et~al.}(2018)\citenamefont {{Sakari}}, \citenamefont {{Placco}}, \citenamefont {{Farrell}}, \citenamefont {{Roederer}}, \citenamefont {{Wallerstein}}, \citenamefont {{Beers}}, \citenamefont {{Ezzeddine}}, \citenamefont {{Frebel}}, \citenamefont {{Hansen}}, \citenamefont {{Holmbeck}} \emph {et~al.}}]{Sakari+2018}%
  \BibitemOpen
  \bibfield  {author} {\bibinfo {author} {\bibfnamefont {C.~M.}\ \bibnamefont {{Sakari}}}, \bibinfo {author} {\bibfnamefont {V.~M.}\ \bibnamefont {{Placco}}}, \bibinfo {author} {\bibfnamefont {E.~M.}\ \bibnamefont {{Farrell}}}, \bibinfo {author} {\bibfnamefont {I.~U.}\ \bibnamefont {{Roederer}}}, \bibinfo {author} {\bibfnamefont {G.}~\bibnamefont {{Wallerstein}}}, \bibinfo {author} {\bibfnamefont {T.~C.}\ \bibnamefont {{Beers}}}, \bibinfo {author} {\bibfnamefont {R.}~\bibnamefont {{Ezzeddine}}}, \bibinfo {author} {\bibfnamefont {A.}~\bibnamefont {{Frebel}}}, \bibinfo {author} {\bibfnamefont {T.}~\bibnamefont {{Hansen}}}, \bibinfo {author} {\bibfnamefont {E.~M.}\ \bibnamefont {{Holmbeck}}}, \emph {et~al.},\ }\bibfield  {title} {\bibinfo {title} {{The R-Process Alliance: First Release from the Northern Search for r-process-enhanced Metal-poor Stars in the Galactic Halo}},\ }\href {https://doi.org/10.3847/1538-4357/aae9df} {\bibfield  {journal} {\bibinfo  {journal} {\apj}\ }\textbf {\bibinfo {volume} {868}},\
  \bibinfo {eid} {110} (\bibinfo {year} {2018})}\BibitemShut {NoStop}%
\bibitem [{\citenamefont {{Holmbeck}}\ \emph {et~al.}(2020)\citenamefont {{Holmbeck}}, \citenamefont {{Hansen}}, \citenamefont {{Beers}}, \citenamefont {{Placco}}, \citenamefont {{Whitten}}, \citenamefont {{Rasmussen}}, \citenamefont {{Roederer}}, \citenamefont {{Ezzeddine}}, \citenamefont {{Sakari}}, \citenamefont {{Frebel}}, \citenamefont {{Drout}}, \citenamefont {{Simon}}, \citenamefont {{Thompson}} \emph {et~al.}}]{Holmbeck+2020}%
  \BibitemOpen
  \bibfield  {author} {\bibinfo {author} {\bibfnamefont {E.~M.}\ \bibnamefont {{Holmbeck}}}, \bibinfo {author} {\bibfnamefont {T.~T.}\ \bibnamefont {{Hansen}}}, \bibinfo {author} {\bibfnamefont {T.~C.}\ \bibnamefont {{Beers}}}, \bibinfo {author} {\bibfnamefont {V.~M.}\ \bibnamefont {{Placco}}}, \bibinfo {author} {\bibfnamefont {D.~D.}\ \bibnamefont {{Whitten}}}, \bibinfo {author} {\bibfnamefont {K.~C.}\ \bibnamefont {{Rasmussen}}}, \bibinfo {author} {\bibfnamefont {I.~U.}\ \bibnamefont {{Roederer}}}, \bibinfo {author} {\bibfnamefont {R.}~\bibnamefont {{Ezzeddine}}}, \bibinfo {author} {\bibfnamefont {C.~M.}\ \bibnamefont {{Sakari}}}, \bibinfo {author} {\bibfnamefont {A.}~\bibnamefont {{Frebel}}}, \bibinfo {author} {\bibfnamefont {M.~R.}\ \bibnamefont {{Drout}}}, \bibinfo {author} {\bibfnamefont {J.~D.}\ \bibnamefont {{Simon}}}, \bibinfo {author} {\bibfnamefont {I.~B.}\ \bibnamefont {{Thompson}}}, \emph {et~al.},\ }\bibfield  {title} {\bibinfo {title} {{The R-Process Alliance: Fourth Data Release from the
  Search for R-process-enhanced Stars in the Galactic Halo}},\ }\href {https://doi.org/10.3847/1538-4365/ab9c19} {\bibfield  {journal} {\bibinfo  {journal} {\apjs}\ }\textbf {\bibinfo {volume} {249}},\ \bibinfo {eid} {30} (\bibinfo {year} {2020})}\BibitemShut {NoStop}%
\bibitem [{\citenamefont {{Bandyopadhyay}}\ \emph {et~al.}(2024)\citenamefont {{Bandyopadhyay}}, \citenamefont {{Ezzeddine}}, \citenamefont {{Allende Prieto}}, \citenamefont {{Aria}}, \citenamefont {{Shah}}, \citenamefont {{Beers}}, \citenamefont {{Frebel}}, \citenamefont {{Hansen}}, \citenamefont {{Holmbeck}}, \citenamefont {{Placco}} \emph {et~al.}}]{Bandyopadhyay+2024}%
  \BibitemOpen
  \bibfield  {author} {\bibinfo {author} {\bibfnamefont {A.}~\bibnamefont {{Bandyopadhyay}}}, \bibinfo {author} {\bibfnamefont {R.}~\bibnamefont {{Ezzeddine}}}, \bibinfo {author} {\bibfnamefont {C.}~\bibnamefont {{Allende Prieto}}}, \bibinfo {author} {\bibfnamefont {N.}~\bibnamefont {{Aria}}}, \bibinfo {author} {\bibfnamefont {S.~P.}\ \bibnamefont {{Shah}}}, \bibinfo {author} {\bibfnamefont {T.~C.}\ \bibnamefont {{Beers}}}, \bibinfo {author} {\bibfnamefont {A.}~\bibnamefont {{Frebel}}}, \bibinfo {author} {\bibfnamefont {T.~T.}\ \bibnamefont {{Hansen}}}, \bibinfo {author} {\bibfnamefont {E.~M.}\ \bibnamefont {{Holmbeck}}}, \bibinfo {author} {\bibfnamefont {V.~M.}\ \bibnamefont {{Placco}}}, \emph {et~al.},\ }\bibfield  {title} {\bibinfo {title} {{The R-process Alliance: Fifth Data Release from the Search for R-process-enhanced Metal-poor Stars in the Galactic Halo with the GTC}},\ }\href {https://doi.org/10.3847/1538-4365/ad6f0f} {\bibfield  {journal} {\bibinfo  {journal} {\apjs}\ }\textbf {\bibinfo {volume}
  {274}},\ \bibinfo {eid} {39} (\bibinfo {year} {2024})}\BibitemShut {NoStop}%
\bibitem [{\citenamefont {{Meyer}}(1989)}]{Meyer1989}%
  \BibitemOpen
  \bibfield  {author} {\bibinfo {author} {\bibfnamefont {B.~S.}\ \bibnamefont {{Meyer}}},\ }\bibfield  {title} {\bibinfo {title} {{Decompression of Initially Cold Neutron Star Matter: A Mechanism for the r-Process?}},\ }\href {https://doi.org/10.1086/167702} {\bibfield  {journal} {\bibinfo  {journal} {\apj}\ }\textbf {\bibinfo {volume} {343}},\ \bibinfo {pages} {254} (\bibinfo {year} {1989})}\BibitemShut {NoStop}%
\bibitem [{\citenamefont {{Just}}\ \emph {et~al.}(2015)\citenamefont {{Just}}, \citenamefont {{Bauswein}}, \citenamefont {{Ardevol Pulpillo}}, \citenamefont {{Goriely}},\ and\ \citenamefont {{Janka}}}]{Just+2015}%
  \BibitemOpen
  \bibfield  {author} {\bibinfo {author} {\bibfnamefont {O.}~\bibnamefont {{Just}}}, \bibinfo {author} {\bibfnamefont {A.}~\bibnamefont {{Bauswein}}}, \bibinfo {author} {\bibfnamefont {R.}~\bibnamefont {{Ardevol Pulpillo}}}, \bibinfo {author} {\bibfnamefont {S.}~\bibnamefont {{Goriely}}},\ and\ \bibinfo {author} {\bibfnamefont {H.-T.}\ \bibnamefont {{Janka}}},\ }\bibfield  {title} {\bibinfo {title} {{Comprehensive nucleosynthesis analysis for ejecta of compact binary mergers}},\ }\href {https://doi.org/10.1093/mnras/stv009} {\bibfield  {journal} {\bibinfo  {journal} {\mnras}\ }\textbf {\bibinfo {volume} {448}},\ \bibinfo {pages} {541} (\bibinfo {year} {2015})}\BibitemShut {NoStop}%
\bibitem [{\citenamefont {{Bovard}}\ \emph {et~al.}(2017)\citenamefont {{Bovard}}, \citenamefont {{Martin}}, \citenamefont {{Guercilena}}, \citenamefont {{Arcones}}, \citenamefont {{Rezzolla}},\ and\ \citenamefont {{Korobkin}}}]{Bovard+2017}%
  \BibitemOpen
  \bibfield  {author} {\bibinfo {author} {\bibfnamefont {L.}~\bibnamefont {{Bovard}}}, \bibinfo {author} {\bibfnamefont {D.}~\bibnamefont {{Martin}}}, \bibinfo {author} {\bibfnamefont {F.}~\bibnamefont {{Guercilena}}}, \bibinfo {author} {\bibfnamefont {A.}~\bibnamefont {{Arcones}}}, \bibinfo {author} {\bibfnamefont {L.}~\bibnamefont {{Rezzolla}}},\ and\ \bibinfo {author} {\bibfnamefont {O.}~\bibnamefont {{Korobkin}}},\ }\bibfield  {title} {\bibinfo {title} {{r -process nucleosynthesis from matter ejected in binary neutron star mergers}},\ }\href {https://doi.org/10.1103/PhysRevD.96.124005} {\bibfield  {journal} {\bibinfo  {journal} {\prd}\ }\textbf {\bibinfo {volume} {96}},\ \bibinfo {eid} {124005} (\bibinfo {year} {2017})}\BibitemShut {NoStop}%
\bibitem [{\citenamefont {Radice}\ \emph {et~al.}(2018)\citenamefont {Radice}, \citenamefont {Perego}, \citenamefont {Hotokezaka}, \citenamefont {Bernuzzi}, \citenamefont {Fromm},\ and\ \citenamefont {Roberts}}]{Radice:2018ghv}%
  \BibitemOpen
  \bibfield  {author} {\bibinfo {author} {\bibfnamefont {D.}~\bibnamefont {Radice}}, \bibinfo {author} {\bibfnamefont {A.}~\bibnamefont {Perego}}, \bibinfo {author} {\bibfnamefont {K.}~\bibnamefont {Hotokezaka}}, \bibinfo {author} {\bibfnamefont {S.}~\bibnamefont {Bernuzzi}}, \bibinfo {author} {\bibfnamefont {S.~A.}\ \bibnamefont {Fromm}},\ and\ \bibinfo {author} {\bibfnamefont {L.~F.}\ \bibnamefont {Roberts}},\ }\bibfield  {title} {\bibinfo {title} {{Viscous-Dynamical Ejecta from Binary Neutron Star Merger}},\ }\href {https://doi.org/10.3847/2041-8213/aaf053} {\bibfield  {journal} {\bibinfo  {journal} {Astrophys. J. Lett.}\ }\textbf {\bibinfo {volume} {869}},\ \bibinfo {pages} {L35} (\bibinfo {year} {2018})}\BibitemShut {NoStop}%
\bibitem [{\citenamefont {{Fujibayashi}}\ \emph {et~al.}(2023)\citenamefont {{Fujibayashi}}, \citenamefont {{Kiuchi}}, \citenamefont {{Wanajo}}, \citenamefont {{Kyutoku}}, \citenamefont {{Sekiguchi}},\ and\ \citenamefont {{Shibata}}}]{Fujibayashi+2023}%
  \BibitemOpen
  \bibfield  {author} {\bibinfo {author} {\bibfnamefont {S.}~\bibnamefont {{Fujibayashi}}}, \bibinfo {author} {\bibfnamefont {K.}~\bibnamefont {{Kiuchi}}}, \bibinfo {author} {\bibfnamefont {S.}~\bibnamefont {{Wanajo}}}, \bibinfo {author} {\bibfnamefont {K.}~\bibnamefont {{Kyutoku}}}, \bibinfo {author} {\bibfnamefont {Y.}~\bibnamefont {{Sekiguchi}}},\ and\ \bibinfo {author} {\bibfnamefont {M.}~\bibnamefont {{Shibata}}},\ }\bibfield  {title} {\bibinfo {title} {{Comprehensive Study of Mass Ejection and Nucleosynthesis in Binary Neutron Star Mergers Leaving Short-lived Massive Neutron Stars}},\ }\href {https://doi.org/10.3847/1538-4357/ac9ce0} {\bibfield  {journal} {\bibinfo  {journal} {\apj}\ }\textbf {\bibinfo {volume} {942}},\ \bibinfo {eid} {39} (\bibinfo {year} {2023})}\BibitemShut {NoStop}%
\bibitem [{\citenamefont {Foucart}\ \emph {et~al.}(2024)\citenamefont {Foucart}, \citenamefont {Duez}, \citenamefont {Kidder}, \citenamefont {Pfeiffer},\ and\ \citenamefont {Scheel}}]{Foucart:2024kci}%
  \BibitemOpen
  \bibfield  {author} {\bibinfo {author} {\bibfnamefont {F.}~\bibnamefont {Foucart}}, \bibinfo {author} {\bibfnamefont {M.~D.}\ \bibnamefont {Duez}}, \bibinfo {author} {\bibfnamefont {L.~E.}\ \bibnamefont {Kidder}}, \bibinfo {author} {\bibfnamefont {H.~P.}\ \bibnamefont {Pfeiffer}},\ and\ \bibinfo {author} {\bibfnamefont {M.~A.}\ \bibnamefont {Scheel}},\ }\bibfield  {title} {\bibinfo {title} {{Dynamical ejecta from binary neutron star mergers: Impact of a small residual eccentricity and of the equation of state implementation}},\ }\href {https://doi.org/10.1103/PhysRevD.110.024003} {\bibfield  {journal} {\bibinfo  {journal} {Phys. Rev. D}\ }\textbf {\bibinfo {volume} {110}},\ \bibinfo {pages} {024003} (\bibinfo {year} {2024})}\BibitemShut {NoStop}%
\bibitem [{\citenamefont {Kedia}\ \emph {et~al.}(2023)\citenamefont {Kedia}, \citenamefont {Ristic}, \citenamefont {O'Shaughnessy}, \citenamefont {Yelikar}, \citenamefont {Wollaeger}, \citenamefont {Korobkin}, \citenamefont {Chase}, \citenamefont {Fryer},\ and\ \citenamefont {Fontes}}]{PhysRevResearch.5.013168}%
  \BibitemOpen
  \bibfield  {author} {\bibinfo {author} {\bibfnamefont {A.}~\bibnamefont {Kedia}}, \bibinfo {author} {\bibfnamefont {M.}~\bibnamefont {Ristic}}, \bibinfo {author} {\bibfnamefont {R.}~\bibnamefont {O'Shaughnessy}}, \bibinfo {author} {\bibfnamefont {A.~B.}\ \bibnamefont {Yelikar}}, \bibinfo {author} {\bibfnamefont {R.~T.}\ \bibnamefont {Wollaeger}}, \bibinfo {author} {\bibfnamefont {O.}~\bibnamefont {Korobkin}}, \bibinfo {author} {\bibfnamefont {E.~A.}\ \bibnamefont {Chase}}, \bibinfo {author} {\bibfnamefont {C.~L.}\ \bibnamefont {Fryer}},\ and\ \bibinfo {author} {\bibfnamefont {C.~J.}\ \bibnamefont {Fontes}},\ }\bibfield  {title} {\bibinfo {title} {Surrogate light curve models for kilonovae with comprehensive wind ejecta outflows and parameter estimation for {AT2017gfo}},\ }\href {https://doi.org/10.1103/PhysRevResearch.5.013168} {\bibfield  {journal} {\bibinfo  {journal} {Phys. Rev. Res.}\ }\textbf {\bibinfo {volume} {5}},\ \bibinfo {pages} {013168} (\bibinfo {year} {2023})}\BibitemShut {NoStop}%
\bibitem [{\citenamefont {Peng}\ \emph {et~al.}(2024)\citenamefont {Peng}, \citenamefont {Risti\ifmmode~\acute{c}\else \'{c}\fi{}}, \citenamefont {Kedia}, \citenamefont {O'Shaughnessy}, \citenamefont {Fontes}, \citenamefont {Fryer}, \citenamefont {Korobkin}, \citenamefont {Mumpower}, \citenamefont {Villar},\ and\ \citenamefont {Wollaeger}}]{PhysRevResearch.6.033078}%
  \BibitemOpen
  \bibfield  {author} {\bibinfo {author} {\bibfnamefont {Y.}~\bibnamefont {Peng}}, \bibinfo {author} {\bibfnamefont {M.}~\bibnamefont {Risti\ifmmode~\acute{c}\else \'{c}\fi{}}}, \bibinfo {author} {\bibfnamefont {A.}~\bibnamefont {Kedia}}, \bibinfo {author} {\bibfnamefont {R.}~\bibnamefont {O'Shaughnessy}}, \bibinfo {author} {\bibfnamefont {C.~J.}\ \bibnamefont {Fontes}}, \bibinfo {author} {\bibfnamefont {C.~L.}\ \bibnamefont {Fryer}}, \bibinfo {author} {\bibfnamefont {O.}~\bibnamefont {Korobkin}}, \bibinfo {author} {\bibfnamefont {M.~R.}\ \bibnamefont {Mumpower}}, \bibinfo {author} {\bibfnamefont {V.~A.}\ \bibnamefont {Villar}},\ and\ \bibinfo {author} {\bibfnamefont {R.~T.}\ \bibnamefont {Wollaeger}},\ }\bibfield  {title} {\bibinfo {title} {Kilonova light-curve interpolation with neural networks},\ }\href {https://doi.org/10.1103/PhysRevResearch.6.033078} {\bibfield  {journal} {\bibinfo  {journal} {Phys. Rev. Res.}\ }\textbf {\bibinfo {volume} {6}},\ \bibinfo {pages} {033078} (\bibinfo {year}
  {2024})}\BibitemShut {NoStop}%
\bibitem [{\citenamefont {Fryer}\ \emph {et~al.}(2024)\citenamefont {Fryer}, \citenamefont {Hungerford}, \citenamefont {Wollaeger}, \citenamefont {Miller}, \citenamefont {De}, \citenamefont {Fontes}, \citenamefont {Korobkin}, \citenamefont {Kedia}, \citenamefont {Ristic},\ and\ \citenamefont {O’Shaughnessy}}]{Fryer_2024}%
  \BibitemOpen
  \bibfield  {author} {\bibinfo {author} {\bibfnamefont {C.~L.}\ \bibnamefont {Fryer}}, \bibinfo {author} {\bibfnamefont {A.~L.}\ \bibnamefont {Hungerford}}, \bibinfo {author} {\bibfnamefont {R.~T.}\ \bibnamefont {Wollaeger}}, \bibinfo {author} {\bibfnamefont {J.~M.}\ \bibnamefont {Miller}}, \bibinfo {author} {\bibfnamefont {S.}~\bibnamefont {De}}, \bibinfo {author} {\bibfnamefont {C.~J.}\ \bibnamefont {Fontes}}, \bibinfo {author} {\bibfnamefont {O.}~\bibnamefont {Korobkin}}, \bibinfo {author} {\bibfnamefont {A.}~\bibnamefont {Kedia}}, \bibinfo {author} {\bibfnamefont {M.}~\bibnamefont {Ristic}},\ and\ \bibinfo {author} {\bibfnamefont {R.}~\bibnamefont {O’Shaughnessy}},\ }\bibfield  {title} {\bibinfo {title} {The effect of the velocity distribution on kilonova emission},\ }\href {https://doi.org/10.3847/1538-4357/ad1036} {\bibfield  {journal} {\bibinfo  {journal} {\apj}\ }\textbf {\bibinfo {volume} {961}},\ \bibinfo {pages} {9} (\bibinfo {year} {2024})}\BibitemShut {NoStop}%
\bibitem [{\citenamefont {Risti\ifmmode~\acute{c}\else \'{c}\fi{}}\ \emph {et~al.}(2023)\citenamefont {Risti\ifmmode~\acute{c}\else \'{c}\fi{}}, \citenamefont {O'Shaughnessy}, \citenamefont {Villar}, \citenamefont {Wollaeger}, \citenamefont {Korobkin}, \citenamefont {Fryer}, \citenamefont {Fontes},\ and\ \citenamefont {Kedia}}]{PhysRevResearch.5.043106}%
  \BibitemOpen
  \bibfield  {author} {\bibinfo {author} {\bibfnamefont {M.}~\bibnamefont {Risti\ifmmode~\acute{c}\else \'{c}\fi{}}}, \bibinfo {author} {\bibfnamefont {R.}~\bibnamefont {O'Shaughnessy}}, \bibinfo {author} {\bibfnamefont {V.~A.}\ \bibnamefont {Villar}}, \bibinfo {author} {\bibfnamefont {R.~T.}\ \bibnamefont {Wollaeger}}, \bibinfo {author} {\bibfnamefont {O.}~\bibnamefont {Korobkin}}, \bibinfo {author} {\bibfnamefont {C.~L.}\ \bibnamefont {Fryer}}, \bibinfo {author} {\bibfnamefont {C.~J.}\ \bibnamefont {Fontes}},\ and\ \bibinfo {author} {\bibfnamefont {A.}~\bibnamefont {Kedia}},\ }\bibfield  {title} {\bibinfo {title} {Interpolated kilonova spectra models: Examining the effects of a phenomenological, blue component in the fitting of {AT2017gfo} spectra},\ }\href {https://doi.org/10.1103/PhysRevResearch.5.043106} {\bibfield  {journal} {\bibinfo  {journal} {Phys. Rev. Res.}\ }\textbf {\bibinfo {volume} {5}},\ \bibinfo {pages} {043106} (\bibinfo {year} {2023})}\BibitemShut {NoStop}%
\bibitem [{\citenamefont {Surman}\ \emph {et~al.}(2008)\citenamefont {Surman}, \citenamefont {McLaughlin}, \citenamefont {Ruffert}, \citenamefont {Janka},\ and\ \citenamefont {Hix}}]{Surman:2008qf}%
  \BibitemOpen
  \bibfield  {author} {\bibinfo {author} {\bibfnamefont {R.}~\bibnamefont {Surman}}, \bibinfo {author} {\bibfnamefont {G.~C.}\ \bibnamefont {McLaughlin}}, \bibinfo {author} {\bibfnamefont {M.}~\bibnamefont {Ruffert}}, \bibinfo {author} {\bibfnamefont {H.~T.}\ \bibnamefont {Janka}},\ and\ \bibinfo {author} {\bibfnamefont {W.~R.}\ \bibnamefont {Hix}},\ }\bibfield  {title} {\bibinfo {title} {{r-Process Nucleosynthesis in Hot Accretion Disk Flows from Black Hole - Neutron Star Mergers}},\ }\href {https://doi.org/10.1086/589507} {\bibfield  {journal} {\bibinfo  {journal} {Astrophys. J. Lett.}\ }\textbf {\bibinfo {volume} {679}},\ \bibinfo {pages} {L117} (\bibinfo {year} {2008})}\BibitemShut {NoStop}%
\bibitem [{\citenamefont {{Darbha}}\ \emph {et~al.}(2021)\citenamefont {{Darbha}}, \citenamefont {{Kasen}}, \citenamefont {{Foucart}},\ and\ \citenamefont {{Price}}}]{Darbha+2021}%
  \BibitemOpen
  \bibfield  {author} {\bibinfo {author} {\bibfnamefont {S.}~\bibnamefont {{Darbha}}}, \bibinfo {author} {\bibfnamefont {D.}~\bibnamefont {{Kasen}}}, \bibinfo {author} {\bibfnamefont {F.}~\bibnamefont {{Foucart}}},\ and\ \bibinfo {author} {\bibfnamefont {D.~J.}\ \bibnamefont {{Price}}},\ }\bibfield  {title} {\bibinfo {title} {{Electromagnetic Signatures from the Tidal Tail of a Black Hole-Neutron Star Merger}},\ }\href {https://doi.org/10.3847/1538-4357/abff5d} {\bibfield  {journal} {\bibinfo  {journal} {\apj}\ }\textbf {\bibinfo {volume} {915}},\ \bibinfo {eid} {69} (\bibinfo {year} {2021})}\BibitemShut {NoStop}%
\bibitem [{\citenamefont {{Curtis}}\ \emph {et~al.}(2023)\citenamefont {{Curtis}}, \citenamefont {{Miller}}, \citenamefont {{Fr{\"o}hlich}}, \citenamefont {{Sprouse}}, \citenamefont {{Lloyd-Ronning}},\ and\ \citenamefont {{Mumpower}}}]{Curtis+2023}%
  \BibitemOpen
  \bibfield  {author} {\bibinfo {author} {\bibfnamefont {S.}~\bibnamefont {{Curtis}}}, \bibinfo {author} {\bibfnamefont {J.~M.}\ \bibnamefont {{Miller}}}, \bibinfo {author} {\bibfnamefont {C.}~\bibnamefont {{Fr{\"o}hlich}}}, \bibinfo {author} {\bibfnamefont {T.}~\bibnamefont {{Sprouse}}}, \bibinfo {author} {\bibfnamefont {N.}~\bibnamefont {{Lloyd-Ronning}}},\ and\ \bibinfo {author} {\bibfnamefont {M.}~\bibnamefont {{Mumpower}}},\ }\bibfield  {title} {\bibinfo {title} {{Nucleosynthesis in Outflows from Black Hole-Neutron Star Merger Disks with Full GR({\ensuremath{\nu}})RMHD}},\ }\href {https://doi.org/10.3847/2041-8213/acba16} {\bibfield  {journal} {\bibinfo  {journal} {\apjl}\ }\textbf {\bibinfo {volume} {945}},\ \bibinfo {eid} {L13} (\bibinfo {year} {2023})}\BibitemShut {NoStop}%
\bibitem [{\citenamefont {{Wanajo}}\ \emph {et~al.}(2024)\citenamefont {{Wanajo}}, \citenamefont {{Fujibayashi}}, \citenamefont {{Hayashi}}, \citenamefont {{Kiuchi}}, \citenamefont {{Sekiguchi}},\ and\ \citenamefont {{Shibata}}}]{Wanajo+2024}%
  \BibitemOpen
  \bibfield  {author} {\bibinfo {author} {\bibfnamefont {S.}~\bibnamefont {{Wanajo}}}, \bibinfo {author} {\bibfnamefont {S.}~\bibnamefont {{Fujibayashi}}}, \bibinfo {author} {\bibfnamefont {K.}~\bibnamefont {{Hayashi}}}, \bibinfo {author} {\bibfnamefont {K.}~\bibnamefont {{Kiuchi}}}, \bibinfo {author} {\bibfnamefont {Y.}~\bibnamefont {{Sekiguchi}}},\ and\ \bibinfo {author} {\bibfnamefont {M.}~\bibnamefont {{Shibata}}},\ }\bibfield  {title} {\bibinfo {title} {{Actinide-Boosting r Process in Black-Hole{\textendash}Neutron-Star Merger Ejecta}},\ }\href {https://doi.org/10.1103/PhysRevLett.133.241201} {\bibfield  {journal} {\bibinfo  {journal} {\prl}\ }\textbf {\bibinfo {volume} {133}},\ \bibinfo {eid} {241201} (\bibinfo {year} {2024})}\BibitemShut {NoStop}%
\bibitem [{\citenamefont {{Surman}}\ \emph {et~al.}(2006)\citenamefont {{Surman}}, \citenamefont {{McLaughlin}},\ and\ \citenamefont {{Hix}}}]{Surman+2006}%
  \BibitemOpen
  \bibfield  {author} {\bibinfo {author} {\bibfnamefont {R.}~\bibnamefont {{Surman}}}, \bibinfo {author} {\bibfnamefont {G.~C.}\ \bibnamefont {{McLaughlin}}},\ and\ \bibinfo {author} {\bibfnamefont {W.~R.}\ \bibnamefont {{Hix}}},\ }\bibfield  {title} {\bibinfo {title} {{Nucleosynthesis in the Outflow from Gamma-Ray Burst Accretion Disks}},\ }\href {https://doi.org/10.1086/501116} {\bibfield  {journal} {\bibinfo  {journal} {\apj}\ }\textbf {\bibinfo {volume} {643}},\ \bibinfo {pages} {1057} (\bibinfo {year} {2006})}\BibitemShut {NoStop}%
\bibitem [{\citenamefont {{Siegel}}\ \emph {et~al.}(2019)\citenamefont {{Siegel}}, \citenamefont {{Barnes}},\ and\ \citenamefont {{Metzger}}}]{Siegel+2019}%
  \BibitemOpen
  \bibfield  {author} {\bibinfo {author} {\bibfnamefont {D.~M.}\ \bibnamefont {{Siegel}}}, \bibinfo {author} {\bibfnamefont {J.}~\bibnamefont {{Barnes}}},\ and\ \bibinfo {author} {\bibfnamefont {B.~D.}\ \bibnamefont {{Metzger}}},\ }\bibfield  {title} {\bibinfo {title} {{Collapsars as a major source of r-process elements}},\ }\href {https://doi.org/10.1038/s41586-019-1136-0} {\bibfield  {journal} {\bibinfo  {journal} {\nat}\ }\textbf {\bibinfo {volume} {569}},\ \bibinfo {pages} {241} (\bibinfo {year} {2019})}\BibitemShut {NoStop}%
\bibitem [{\citenamefont {{Miller}}\ \emph {et~al.}(2020)\citenamefont {{Miller}}, \citenamefont {{Sprouse}}, \citenamefont {{Fryer}}, \citenamefont {{Ryan}}, \citenamefont {{Dolence}}, \citenamefont {{Mumpower}},\ and\ \citenamefont {{Surman}}}]{Miller+2020}%
  \BibitemOpen
  \bibfield  {author} {\bibinfo {author} {\bibfnamefont {J.~M.}\ \bibnamefont {{Miller}}}, \bibinfo {author} {\bibfnamefont {T.~M.}\ \bibnamefont {{Sprouse}}}, \bibinfo {author} {\bibfnamefont {C.~L.}\ \bibnamefont {{Fryer}}}, \bibinfo {author} {\bibfnamefont {B.~R.}\ \bibnamefont {{Ryan}}}, \bibinfo {author} {\bibfnamefont {J.~C.}\ \bibnamefont {{Dolence}}}, \bibinfo {author} {\bibfnamefont {M.~R.}\ \bibnamefont {{Mumpower}}},\ and\ \bibinfo {author} {\bibfnamefont {R.}~\bibnamefont {{Surman}}},\ }\bibfield  {title} {\bibinfo {title} {{Full Transport General Relativistic Radiation Magnetohydrodynamics for Nucleosynthesis in Collapsars}},\ }\href {https://doi.org/10.3847/1538-4357/abb4e3} {\bibfield  {journal} {\bibinfo  {journal} {\apj}\ }\textbf {\bibinfo {volume} {902}},\ \bibinfo {eid} {66} (\bibinfo {year} {2020})}\BibitemShut {NoStop}%
\bibitem [{\citenamefont {{Nishimura}}\ \emph {et~al.}(2015)\citenamefont {{Nishimura}}, \citenamefont {{Takiwaki}},\ and\ \citenamefont {{Thielemann}}}]{Nishimura+2015}%
  \BibitemOpen
  \bibfield  {author} {\bibinfo {author} {\bibfnamefont {N.}~\bibnamefont {{Nishimura}}}, \bibinfo {author} {\bibfnamefont {T.}~\bibnamefont {{Takiwaki}}},\ and\ \bibinfo {author} {\bibfnamefont {F.-K.}\ \bibnamefont {{Thielemann}}},\ }\bibfield  {title} {\bibinfo {title} {{The r-process Nucleosynthesis in the Various Jet-like Explosions of Magnetorotational Core-collapse Supernovae}},\ }\href {https://doi.org/10.1088/0004-637X/810/2/109} {\bibfield  {journal} {\bibinfo  {journal} {\apj}\ }\textbf {\bibinfo {volume} {810}},\ \bibinfo {eid} {109} (\bibinfo {year} {2015})}\BibitemShut {NoStop}%
\bibitem [{\citenamefont {{M{\"o}sta}}\ \emph {et~al.}(2018)\citenamefont {{M{\"o}sta}}, \citenamefont {{Roberts}}, \citenamefont {{Halevi}}, \citenamefont {{Ott}}, \citenamefont {{Lippuner}}, \citenamefont {{Haas}},\ and\ \citenamefont {{Schnetter}}}]{Mosta+2018}%
  \BibitemOpen
  \bibfield  {author} {\bibinfo {author} {\bibfnamefont {P.}~\bibnamefont {{M{\"o}sta}}}, \bibinfo {author} {\bibfnamefont {L.~F.}\ \bibnamefont {{Roberts}}}, \bibinfo {author} {\bibfnamefont {G.}~\bibnamefont {{Halevi}}}, \bibinfo {author} {\bibfnamefont {C.~D.}\ \bibnamefont {{Ott}}}, \bibinfo {author} {\bibfnamefont {J.}~\bibnamefont {{Lippuner}}}, \bibinfo {author} {\bibfnamefont {R.}~\bibnamefont {{Haas}}},\ and\ \bibinfo {author} {\bibfnamefont {E.}~\bibnamefont {{Schnetter}}},\ }\bibfield  {title} {\bibinfo {title} {{r-process Nucleosynthesis from Three-dimensional Magnetorotational Core-collapse Supernovae}},\ }\href {https://doi.org/10.3847/1538-4357/aad6ec} {\bibfield  {journal} {\bibinfo  {journal} {\apj}\ }\textbf {\bibinfo {volume} {864}},\ \bibinfo {eid} {171} (\bibinfo {year} {2018})}\BibitemShut {NoStop}%
\bibitem [{\citenamefont {Reichert}\ \emph {et~al.}(2021)\citenamefont {Reichert}, \citenamefont {Obergaulinger}, \citenamefont {Eichler}, \citenamefont {Aloy},\ and\ \citenamefont {Arcones}}]{Reichert:2020mjo}%
  \BibitemOpen
  \bibfield  {author} {\bibinfo {author} {\bibfnamefont {M.}~\bibnamefont {Reichert}}, \bibinfo {author} {\bibfnamefont {M.}~\bibnamefont {Obergaulinger}}, \bibinfo {author} {\bibfnamefont {M.}~\bibnamefont {Eichler}}, \bibinfo {author} {\bibfnamefont {M.-{\'A}.}\ \bibnamefont {Aloy}},\ and\ \bibinfo {author} {\bibfnamefont {A.}~\bibnamefont {Arcones}},\ }\bibfield  {title} {\bibinfo {title} {{Nucleosynthesis in magneto-rotational supernovae}},\ }\href {https://doi.org/10.1093/mnras/stab029} {\bibfield  {journal} {\bibinfo  {journal} {Mon. Not. Roy. Astron. Soc.}\ }\textbf {\bibinfo {volume} {501}},\ \bibinfo {pages} {5733} (\bibinfo {year} {2021})}\BibitemShut {NoStop}%
\bibitem [{\citenamefont {{Zha}}\ \emph {et~al.}(2024)\citenamefont {{Zha}}, \citenamefont {{M{\"u}ller}},\ and\ \citenamefont {{Powell}}}]{Zha+2024}%
  \BibitemOpen
  \bibfield  {author} {\bibinfo {author} {\bibfnamefont {S.}~\bibnamefont {{Zha}}}, \bibinfo {author} {\bibfnamefont {B.}~\bibnamefont {{M{\"u}ller}}},\ and\ \bibinfo {author} {\bibfnamefont {J.}~\bibnamefont {{Powell}}},\ }\bibfield  {title} {\bibinfo {title} {{Nucleosynthesis in the Innermost Ejecta of Magnetorotational Supernova Explosions in Three Dimensions}},\ }\href {https://doi.org/10.3847/1538-4357/ad4ae7} {\bibfield  {journal} {\bibinfo  {journal} {\apj}\ }\textbf {\bibinfo {volume} {969}},\ \bibinfo {eid} {141} (\bibinfo {year} {2024})}\BibitemShut {NoStop}%
\bibitem [{\citenamefont {Liu}\ \emph {et~al.}(2025)\citenamefont {Liu}, \citenamefont {Grohs}, \citenamefont {Lund}, \citenamefont {McLaughlin}, \citenamefont {Reichert}, \citenamefont {Roederer}, \citenamefont {Surman},\ and\ \citenamefont {Wang}}]{Liu:2025auu}%
  \BibitemOpen
  \bibfield  {author} {\bibinfo {author} {\bibfnamefont {Z.}~\bibnamefont {Liu}}, \bibinfo {author} {\bibfnamefont {E.}~\bibnamefont {Grohs}}, \bibinfo {author} {\bibfnamefont {K.~A.}\ \bibnamefont {Lund}}, \bibinfo {author} {\bibfnamefont {G.~C.}\ \bibnamefont {McLaughlin}}, \bibinfo {author} {\bibfnamefont {M.}~\bibnamefont {Reichert}}, \bibinfo {author} {\bibfnamefont {I.~U.}\ \bibnamefont {Roederer}}, \bibinfo {author} {\bibfnamefont {R.}~\bibnamefont {Surman}},\ and\ \bibinfo {author} {\bibfnamefont {X.}~\bibnamefont {Wang}},\ }\bibfield  {title} {\bibinfo {title} {{Gamma Rays as a Signature of r-process Producing Supernovae: Remnants and Future Galactic Explosions}},\ }\href {https://doi.org/10.3847/1538-4357/ae1298} {\bibfield  {journal} {\bibinfo  {journal} {Astrophys. J.}\ }\textbf {\bibinfo {volume} {995}},\ \bibinfo {pages} {34} (\bibinfo {year} {2025})}\BibitemShut {NoStop}%
\bibitem [{\citenamefont {{Patel}}\ \emph {et~al.}(2025)\citenamefont {{Patel}}, \citenamefont {{Metzger}}, \citenamefont {{Goldberg}}, \citenamefont {{Cehula}}, \citenamefont {{Thompson}},\ and\ \citenamefont {{Renzo}}}]{Patel+2025}%
  \BibitemOpen
  \bibfield  {author} {\bibinfo {author} {\bibfnamefont {A.}~\bibnamefont {{Patel}}}, \bibinfo {author} {\bibfnamefont {B.~D.}\ \bibnamefont {{Metzger}}}, \bibinfo {author} {\bibfnamefont {J.~A.}\ \bibnamefont {{Goldberg}}}, \bibinfo {author} {\bibfnamefont {J.}~\bibnamefont {{Cehula}}}, \bibinfo {author} {\bibfnamefont {T.~A.}\ \bibnamefont {{Thompson}}},\ and\ \bibinfo {author} {\bibfnamefont {M.}~\bibnamefont {{Renzo}}},\ }\bibfield  {title} {\bibinfo {title} {{r-process Nucleosynthesis and Radioactively Powered Transients from Magnetar Giant Flares}},\ }\href {https://doi.org/10.3847/1538-4357/adceb7} {\bibfield  {journal} {\bibinfo  {journal} {\apj}\ }\textbf {\bibinfo {volume} {985}},\ \bibinfo {eid} {234} (\bibinfo {year} {2025})}\BibitemShut {NoStop}%
\bibitem [{\citenamefont {{Fuller}}\ \emph {et~al.}(2017)\citenamefont {{Fuller}}, \citenamefont {{Kusenko}},\ and\ \citenamefont {{Takhistov}}}]{Fuller+2017}%
  \BibitemOpen
  \bibfield  {author} {\bibinfo {author} {\bibfnamefont {G.~M.}\ \bibnamefont {{Fuller}}}, \bibinfo {author} {\bibfnamefont {A.}~\bibnamefont {{Kusenko}}},\ and\ \bibinfo {author} {\bibfnamefont {V.}~\bibnamefont {{Takhistov}}},\ }\bibfield  {title} {\bibinfo {title} {{Primordial Black Holes and \textit{r}-Process Nucleosynthesis}},\ }\href {https://doi.org/10.1103/PhysRevLett.119.061101} {\bibfield  {journal} {\bibinfo  {journal} {\prl}\ }\textbf {\bibinfo {volume} {119}},\ \bibinfo {eid} {061101} (\bibinfo {year} {2017})}\BibitemShut {NoStop}%
\bibitem [{\citenamefont {{Fischer}}\ \emph {et~al.}(2020)\citenamefont {{Fischer}}, \citenamefont {{Wu}}, \citenamefont {{Wehmeyer}}, \citenamefont {{Bastian}}, \citenamefont {{Mart{\'\i}nez-Pinedo}},\ and\ \citenamefont {{Thielemann}}}]{Fischer+2020}%
  \BibitemOpen
  \bibfield  {author} {\bibinfo {author} {\bibfnamefont {T.}~\bibnamefont {{Fischer}}}, \bibinfo {author} {\bibfnamefont {M.-R.}\ \bibnamefont {{Wu}}}, \bibinfo {author} {\bibfnamefont {B.}~\bibnamefont {{Wehmeyer}}}, \bibinfo {author} {\bibfnamefont {N.-U.~F.}\ \bibnamefont {{Bastian}}}, \bibinfo {author} {\bibfnamefont {G.}~\bibnamefont {{Mart{\'\i}nez-Pinedo}}},\ and\ \bibinfo {author} {\bibfnamefont {F.-K.}\ \bibnamefont {{Thielemann}}},\ }\bibfield  {title} {\bibinfo {title} {{Core-collapse Supernova Explosions Driven by the Hadron-quark Phase Transition as a Rare r-process Site}},\ }\href {https://doi.org/10.3847/1538-4357/ab86b0} {\bibfield  {journal} {\bibinfo  {journal} {\apj}\ }\textbf {\bibinfo {volume} {894}},\ \bibinfo {eid} {9} (\bibinfo {year} {2020})}\BibitemShut {NoStop}%
\bibitem [{\citenamefont {Zhu}\ \emph {et~al.}(2021)\citenamefont {Zhu}, \citenamefont {Lund}, \citenamefont {Barnes}, \citenamefont {Sprouse}, \citenamefont {Vassh}, \citenamefont {McLaughlin}, \citenamefont {Mumpower},\ and\ \citenamefont {Surman}}]{Zhu:2020eyk}%
  \BibitemOpen
  \bibfield  {author} {\bibinfo {author} {\bibfnamefont {Y.~L.}\ \bibnamefont {Zhu}}, \bibinfo {author} {\bibfnamefont {K.}~\bibnamefont {Lund}}, \bibinfo {author} {\bibfnamefont {J.}~\bibnamefont {Barnes}}, \bibinfo {author} {\bibfnamefont {T.~M.}\ \bibnamefont {Sprouse}}, \bibinfo {author} {\bibfnamefont {N.}~\bibnamefont {Vassh}}, \bibinfo {author} {\bibfnamefont {G.~C.}\ \bibnamefont {McLaughlin}}, \bibinfo {author} {\bibfnamefont {M.~R.}\ \bibnamefont {Mumpower}},\ and\ \bibinfo {author} {\bibfnamefont {R.}~\bibnamefont {Surman}},\ }\bibfield  {title} {\bibinfo {title} {{Modeling Kilonova Light Curves: Dependence on Nuclear Inputs}},\ }\href {https://doi.org/10.3847/1538-4357/abc69e} {\bibfield  {journal} {\bibinfo  {journal} {Astrophys. J.}\ }\textbf {\bibinfo {volume} {906}},\ \bibinfo {pages} {94} (\bibinfo {year} {2021})}\BibitemShut {NoStop}%
\bibitem [{\citenamefont {Barnes}\ \emph {et~al.}(2021)\citenamefont {Barnes}, \citenamefont {Zhu}, \citenamefont {Lund}, \citenamefont {Sprouse}, \citenamefont {Vassh}, \citenamefont {McLaughlin}, \citenamefont {Mumpower},\ and\ \citenamefont {Surman}}]{Barnes:2020nfi}%
  \BibitemOpen
  \bibfield  {author} {\bibinfo {author} {\bibfnamefont {J.}~\bibnamefont {Barnes}}, \bibinfo {author} {\bibfnamefont {Y.~L.}\ \bibnamefont {Zhu}}, \bibinfo {author} {\bibfnamefont {K.~A.}\ \bibnamefont {Lund}}, \bibinfo {author} {\bibfnamefont {T.~M.}\ \bibnamefont {Sprouse}}, \bibinfo {author} {\bibfnamefont {N.}~\bibnamefont {Vassh}}, \bibinfo {author} {\bibfnamefont {G.~C.}\ \bibnamefont {McLaughlin}}, \bibinfo {author} {\bibfnamefont {M.~R.}\ \bibnamefont {Mumpower}},\ and\ \bibinfo {author} {\bibfnamefont {R.}~\bibnamefont {Surman}},\ }\bibfield  {title} {\bibinfo {title} {{Kilonovae Across the Nuclear Physics Landscape: The Impact of Nuclear Physics Uncertainties on r-process-powered Emission}},\ }\href {https://doi.org/10.3847/1538-4357/ac0aec} {\bibfield  {journal} {\bibinfo  {journal} {Astrophys. J.}\ }\textbf {\bibinfo {volume} {918}},\ \bibinfo {pages} {44} (\bibinfo {year} {2021})}\BibitemShut {NoStop}%
\bibitem [{\citenamefont {Lund}\ \emph {et~al.}(2023)\citenamefont {Lund}, \citenamefont {Engel}, \citenamefont {McLaughlin}, \citenamefont {Mumpower}, \citenamefont {Ney},\ and\ \citenamefont {Surman}}]{Lund:2022bsr}%
  \BibitemOpen
  \bibfield  {author} {\bibinfo {author} {\bibfnamefont {K.~A.}\ \bibnamefont {Lund}}, \bibinfo {author} {\bibfnamefont {J.}~\bibnamefont {Engel}}, \bibinfo {author} {\bibfnamefont {G.~C.}\ \bibnamefont {McLaughlin}}, \bibinfo {author} {\bibfnamefont {M.~R.}\ \bibnamefont {Mumpower}}, \bibinfo {author} {\bibfnamefont {E.~M.}\ \bibnamefont {Ney}},\ and\ \bibinfo {author} {\bibfnamefont {R.}~\bibnamefont {Surman}},\ }\bibfield  {title} {\bibinfo {title} {{The Influence of {\ensuremath{\beta}}-decay Rates on r-process Observables}},\ }\href {https://doi.org/10.3847/1538-4357/acaf56} {\bibfield  {journal} {\bibinfo  {journal} {Astrophys. J.}\ }\textbf {\bibinfo {volume} {944}},\ \bibinfo {pages} {144} (\bibinfo {year} {2023})}\BibitemShut {NoStop}%
\bibitem [{\citenamefont {Holmbeck}\ \emph {et~al.}(2023)\citenamefont {Holmbeck}, \citenamefont {Surman}, \citenamefont {Roederer}, \citenamefont {McLaughlin},\ and\ \citenamefont {Frebel}}]{Holmbeck:2022mog}%
  \BibitemOpen
  \bibfield  {author} {\bibinfo {author} {\bibfnamefont {E.~M.}\ \bibnamefont {Holmbeck}}, \bibinfo {author} {\bibfnamefont {R.}~\bibnamefont {Surman}}, \bibinfo {author} {\bibfnamefont {I.~U.}\ \bibnamefont {Roederer}}, \bibinfo {author} {\bibfnamefont {G.~C.}\ \bibnamefont {McLaughlin}},\ and\ \bibinfo {author} {\bibfnamefont {A.}~\bibnamefont {Frebel}},\ }\bibfield  {title} {\bibinfo {title} {{HD 222925: A New Opportunity to Explore the Astrophysical and Nuclear Conditions of r-process Sites}},\ }\href {https://doi.org/10.3847/1538-4357/acccf3} {\bibfield  {journal} {\bibinfo  {journal} {Astrophys. J.}\ }\textbf {\bibinfo {volume} {951}},\ \bibinfo {pages} {30} (\bibinfo {year} {2023})}\BibitemShut {NoStop}%
\bibitem [{\citenamefont {{Reiter}}\ \emph {et~al.}(2020)\citenamefont {{Reiter}}, \citenamefont {{Ayet San Andr{\'e}s}}, \citenamefont {{Nikas}}, \citenamefont {{Lippuner}}, \citenamefont {{Andreoiu}}, \citenamefont {{Babcock}}, \citenamefont {{Barquest}}, \citenamefont {{Bollig}}, \citenamefont {{Brunner}}, \citenamefont {{Dickel}} \emph {et~al.}}]{Reiter+2020}%
  \BibitemOpen
  \bibfield  {author} {\bibinfo {author} {\bibfnamefont {M.~P.}\ \bibnamefont {{Reiter}}}, \bibinfo {author} {\bibfnamefont {S.}~\bibnamefont {{Ayet San Andr{\'e}s}}}, \bibinfo {author} {\bibfnamefont {S.}~\bibnamefont {{Nikas}}}, \bibinfo {author} {\bibfnamefont {J.}~\bibnamefont {{Lippuner}}}, \bibinfo {author} {\bibfnamefont {C.}~\bibnamefont {{Andreoiu}}}, \bibinfo {author} {\bibfnamefont {C.}~\bibnamefont {{Babcock}}}, \bibinfo {author} {\bibfnamefont {B.~R.}\ \bibnamefont {{Barquest}}}, \bibinfo {author} {\bibfnamefont {J.}~\bibnamefont {{Bollig}}}, \bibinfo {author} {\bibfnamefont {T.}~\bibnamefont {{Brunner}}}, \bibinfo {author} {\bibfnamefont {T.}~\bibnamefont {{Dickel}}}, \emph {et~al.},\ }\bibfield  {title} {\bibinfo {title} {{Mass measurements of neutron-rich gallium isotopes refine production of nuclei of the first r -process abundance peak in neutron-star merger calculations}},\ }\href {https://doi.org/10.1103/PhysRevC.101.025803} {\bibfield  {journal} {\bibinfo  {journal} {\prc}\ }\textbf
  {\bibinfo {volume} {101}},\ \bibinfo {eid} {025803} (\bibinfo {year} {2020})}\BibitemShut {NoStop}%
\bibitem [{\citenamefont {{Li}}\ \emph {et~al.}(2022)\citenamefont {{Li}}, \citenamefont {{Naimi}}, \citenamefont {{Sprouse}}, \citenamefont {{Mumpower}}, \citenamefont {{Abe}}, \citenamefont {{Yamaguchi}}, \citenamefont {{Nagae}}, \citenamefont {{Suzaki}}, \citenamefont {{Wakasugi}}, \citenamefont {{Arakawa}} \emph {et~al.}}]{Li+2022}%
  \BibitemOpen
  \bibfield  {author} {\bibinfo {author} {\bibfnamefont {H.~F.}\ \bibnamefont {{Li}}}, \bibinfo {author} {\bibfnamefont {S.}~\bibnamefont {{Naimi}}}, \bibinfo {author} {\bibfnamefont {T.~M.}\ \bibnamefont {{Sprouse}}}, \bibinfo {author} {\bibfnamefont {M.~R.}\ \bibnamefont {{Mumpower}}}, \bibinfo {author} {\bibfnamefont {Y.}~\bibnamefont {{Abe}}}, \bibinfo {author} {\bibfnamefont {Y.}~\bibnamefont {{Yamaguchi}}}, \bibinfo {author} {\bibfnamefont {D.}~\bibnamefont {{Nagae}}}, \bibinfo {author} {\bibfnamefont {F.}~\bibnamefont {{Suzaki}}}, \bibinfo {author} {\bibfnamefont {M.}~\bibnamefont {{Wakasugi}}}, \bibinfo {author} {\bibfnamefont {H.}~\bibnamefont {{Arakawa}}}, \emph {et~al.},\ }\bibfield  {title} {\bibinfo {title} {{First Application of Mass Measurements with the Rare-RI Ring Reveals the Solar r -Process Abundance Trend at A =122 and A =123}},\ }\href {https://doi.org/10.1103/PhysRevLett.128.152701} {\bibfield  {journal} {\bibinfo  {journal} {\prl}\ }\textbf {\bibinfo {volume} {128}},\ \bibinfo {eid}
  {152701} (\bibinfo {year} {2022})}\BibitemShut {NoStop}%
\bibitem [{\citenamefont {{Yokoyama}}\ \emph {et~al.}(2019)\citenamefont {{Yokoyama}}, \citenamefont {{Grzywacz}}, \citenamefont {{Rasco}}, \citenamefont {{Brewer}}, \citenamefont {{Rykaczewski}}, \citenamefont {{Dillmann}}, \citenamefont {{Tain}}, \citenamefont {{Nishimura}}, \citenamefont {{Ahn}}, \citenamefont {{Algora}} \emph {et~al.}}]{Yokoyama+2019}%
  \BibitemOpen
  \bibfield  {author} {\bibinfo {author} {\bibfnamefont {R.}~\bibnamefont {{Yokoyama}}}, \bibinfo {author} {\bibfnamefont {R.}~\bibnamefont {{Grzywacz}}}, \bibinfo {author} {\bibfnamefont {B.~C.}\ \bibnamefont {{Rasco}}}, \bibinfo {author} {\bibfnamefont {N.}~\bibnamefont {{Brewer}}}, \bibinfo {author} {\bibfnamefont {K.~P.}\ \bibnamefont {{Rykaczewski}}}, \bibinfo {author} {\bibfnamefont {I.}~\bibnamefont {{Dillmann}}}, \bibinfo {author} {\bibfnamefont {J.~L.}\ \bibnamefont {{Tain}}}, \bibinfo {author} {\bibfnamefont {S.}~\bibnamefont {{Nishimura}}}, \bibinfo {author} {\bibfnamefont {D.~S.}\ \bibnamefont {{Ahn}}}, \bibinfo {author} {\bibfnamefont {A.}~\bibnamefont {{Algora}}}, \emph {et~al.},\ }\bibfield  {title} {\bibinfo {title} {{Strong one-neutron emission from two-neutron unbound states in {\ensuremath{\beta}} decays of the \textit{r}-process nuclei $^{87,86}$Ga}},\ }\href {https://doi.org/10.1103/PhysRevC.100.031302} {\bibfield  {journal} {\bibinfo  {journal} {\prc}\ }\textbf {\bibinfo {volume}
  {100}},\ \bibinfo {eid} {031302} (\bibinfo {year} {2019})}\BibitemShut {NoStop}%
\bibitem [{\citenamefont {{Xu}}\ \emph {et~al.}(2023)\citenamefont {{Xu}}, \citenamefont {{Madurga}}, \citenamefont {{Grzywacz}}, \citenamefont {{King}}, \citenamefont {{Algora}}, \citenamefont {{Andreyev}}, \citenamefont {{Benito}}, \citenamefont {{Berry}}, \citenamefont {{Borge}}, \citenamefont {{Costache}} \emph {et~al.}}]{Xu+2023}%
  \BibitemOpen
  \bibfield  {author} {\bibinfo {author} {\bibfnamefont {Z.~Y.}\ \bibnamefont {{Xu}}}, \bibinfo {author} {\bibfnamefont {M.}~\bibnamefont {{Madurga}}}, \bibinfo {author} {\bibfnamefont {R.}~\bibnamefont {{Grzywacz}}}, \bibinfo {author} {\bibfnamefont {T.~T.}\ \bibnamefont {{King}}}, \bibinfo {author} {\bibfnamefont {A.}~\bibnamefont {{Algora}}}, \bibinfo {author} {\bibfnamefont {A.~N.}\ \bibnamefont {{Andreyev}}}, \bibinfo {author} {\bibfnamefont {J.}~\bibnamefont {{Benito}}}, \bibinfo {author} {\bibfnamefont {T.}~\bibnamefont {{Berry}}}, \bibinfo {author} {\bibfnamefont {M.~J.~G.}\ \bibnamefont {{Borge}}}, \bibinfo {author} {\bibfnamefont {C.}~\bibnamefont {{Costache}}}, \emph {et~al.},\ }\bibfield  {title} {\bibinfo {title} {{$^{133}$In: A Rosetta Stone for Decays of r -Process Nuclei}},\ }\href {https://doi.org/10.1103/PhysRevLett.131.022501} {\bibfield  {journal} {\bibinfo  {journal} {\prl}\ }\textbf {\bibinfo {volume} {131}},\ \bibinfo {eid} {022501} (\bibinfo {year} {2023})}\BibitemShut {NoStop}%
\bibitem [{\citenamefont {{Reifarth}}\ \emph {et~al.}(2017)\citenamefont {{Reifarth}}, \citenamefont {{G{\"o}bel}}, \citenamefont {{Heftrich}}, \citenamefont {{Weigand}}, \citenamefont {{Jurado}}, \citenamefont {{K{\"a}ppeler}},\ and\ \citenamefont {{Litvinov}}}]{Reifarth+2017}%
  \BibitemOpen
  \bibfield  {author} {\bibinfo {author} {\bibfnamefont {R.}~\bibnamefont {{Reifarth}}}, \bibinfo {author} {\bibfnamefont {K.}~\bibnamefont {{G{\"o}bel}}}, \bibinfo {author} {\bibfnamefont {T.}~\bibnamefont {{Heftrich}}}, \bibinfo {author} {\bibfnamefont {M.}~\bibnamefont {{Weigand}}}, \bibinfo {author} {\bibfnamefont {B.}~\bibnamefont {{Jurado}}}, \bibinfo {author} {\bibfnamefont {F.}~\bibnamefont {{K{\"a}ppeler}}},\ and\ \bibinfo {author} {\bibfnamefont {Y.~A.}\ \bibnamefont {{Litvinov}}},\ }\bibfield  {title} {\bibinfo {title} {{Spallation-based neutron target for direct studies of neutron-induced reactions in inverse kinematics}},\ }\href {https://doi.org/10.1103/PhysRevAccelBeams.20.044701} {\bibfield  {journal} {\bibinfo  {journal} {Phys. Rev. Accel. Beams}\ }\textbf {\bibinfo {volume} {20}},\ \bibinfo {eid} {044701} (\bibinfo {year} {2017})}\BibitemShut {NoStop}%
\bibitem [{\citenamefont {Capote}\ \emph {et~al.}(2009)\citenamefont {Capote}, \citenamefont {Herman}, \citenamefont {Oblozinsk{\'y}}, \citenamefont {Young}, \citenamefont {Goriely}, \citenamefont {Belgya}, \citenamefont {Ignatyuk}, \citenamefont {Koning}, \citenamefont {Hilaire}, \citenamefont {Plujko} \emph {et~al.}}]{Capote:09}%
  \BibitemOpen
  \bibfield  {author} {\bibinfo {author} {\bibfnamefont {R.}~\bibnamefont {Capote}}, \bibinfo {author} {\bibfnamefont {M.}~\bibnamefont {Herman}}, \bibinfo {author} {\bibfnamefont {P.}~\bibnamefont {Oblozinsk{\'y}}}, \bibinfo {author} {\bibfnamefont {P.}~\bibnamefont {Young}}, \bibinfo {author} {\bibfnamefont {S.}~\bibnamefont {Goriely}}, \bibinfo {author} {\bibfnamefont {T.}~\bibnamefont {Belgya}}, \bibinfo {author} {\bibfnamefont {A.}~\bibnamefont {Ignatyuk}}, \bibinfo {author} {\bibfnamefont {A.}~\bibnamefont {Koning}}, \bibinfo {author} {\bibfnamefont {S.}~\bibnamefont {Hilaire}}, \bibinfo {author} {\bibfnamefont {V.}~\bibnamefont {Plujko}}, \emph {et~al.},\ }\bibfield  {title} {\bibinfo {title} {{RIPL} - reference input parameter library for calculation of nuclear reactions and nuclear data evaluations},\ }\href {https://doi.org/DOI: 10.1016/j.nds.2009.10.004} {\bibfield  {journal} {\bibinfo  {journal} {Nuclear Data Sheets}\ }\textbf {\bibinfo {volume} {110}},\ \bibinfo {pages} {3107 } (\bibinfo {year}
  {2009})}\BibitemShut {NoStop}%
\bibitem [{\citenamefont {Hilaire}\ and\ \citenamefont {Goriely}(2021)}]{Hilaire:21}%
  \BibitemOpen
  \bibfield  {author} {\bibinfo {author} {\bibfnamefont {S.}~\bibnamefont {Hilaire}}\ and\ \bibinfo {author} {\bibfnamefont {S.}~\bibnamefont {Goriely}},\ }\bibfield  {title} {\bibinfo {title} {Towards more predictive nuclear reaction modelling},\ }in\ \href@noop {} {\emph {\bibinfo {booktitle} {Compound-Nuclear Reactions}}}\ (\bibinfo  {publisher} {Springer International Publishing},\ \bibinfo {address} {Cham},\ \bibinfo {year} {2021})\ pp.\ \bibinfo {pages} {3--15}\BibitemShut {NoStop}%
\bibitem [{\citenamefont {Alhassid}\ \emph {et~al.}(2008)\citenamefont {Alhassid}, \citenamefont {Fang},\ and\ \citenamefont {Nakada}}]{Alhassid:08}%
  \BibitemOpen
  \bibfield  {author} {\bibinfo {author} {\bibfnamefont {Y.}~\bibnamefont {Alhassid}}, \bibinfo {author} {\bibfnamefont {L.}~\bibnamefont {Fang}},\ and\ \bibinfo {author} {\bibfnamefont {H.}~\bibnamefont {Nakada}},\ }\bibfield  {title} {\bibinfo {title} {Heavy deformed nuclei in the shell model monte carlo method},\ }\href {https://doi.org/10.1103/PhysRevLett.101.082501} {\bibfield  {journal} {\bibinfo  {journal} {Phys. Rev. Lett.}\ }\textbf {\bibinfo {volume} {101}},\ \bibinfo {pages} {082501} (\bibinfo {year} {2008})}\BibitemShut {NoStop}%
\bibitem [{\citenamefont {Alhassid}\ \emph {et~al.}(2015)\citenamefont {Alhassid}, \citenamefont {Bonett-Matiz}, \citenamefont {Liu},\ and\ \citenamefont {Nakada}}]{Alhassid:2015a}%
  \BibitemOpen
  \bibfield  {author} {\bibinfo {author} {\bibfnamefont {Y.}~\bibnamefont {Alhassid}}, \bibinfo {author} {\bibfnamefont {M.}~\bibnamefont {Bonett-Matiz}}, \bibinfo {author} {\bibfnamefont {S.}~\bibnamefont {Liu}},\ and\ \bibinfo {author} {\bibfnamefont {H.}~\bibnamefont {Nakada}},\ }\bibfield  {title} {\bibinfo {title} {Direct microscopic calculation of nuclear level densities in the shell model monte carlo approach},\ }\href {https://doi.org/10.1103/PhysRevC.92.024307} {\bibfield  {journal} {\bibinfo  {journal} {Phys. Rev. C}\ }\textbf {\bibinfo {volume} {92}},\ \bibinfo {pages} {024307} (\bibinfo {year} {2015})}\BibitemShut {NoStop}%
\bibitem [{\citenamefont {Shimizu}\ \emph {et~al.}(2016)\citenamefont {Shimizu}, \citenamefont {Utsuno}, \citenamefont {Futamura}, \citenamefont {Sakurai}, \citenamefont {Mizusaki},\ and\ \citenamefont {Otsuka}}]{Shimizu:16}%
  \BibitemOpen
  \bibfield  {author} {\bibinfo {author} {\bibfnamefont {N.}~\bibnamefont {Shimizu}}, \bibinfo {author} {\bibfnamefont {Y.}~\bibnamefont {Utsuno}}, \bibinfo {author} {\bibfnamefont {Y.}~\bibnamefont {Futamura}}, \bibinfo {author} {\bibfnamefont {T.}~\bibnamefont {Sakurai}}, \bibinfo {author} {\bibfnamefont {T.}~\bibnamefont {Mizusaki}},\ and\ \bibinfo {author} {\bibfnamefont {T.}~\bibnamefont {Otsuka}},\ }\bibfield  {title} {\bibinfo {title} {Stochastic estimation of nuclear level density in the nuclear shell model: An application to parity-dependent level density in $^{58}${Ni}},\ }\href {https://doi.org/10.1016/j.physletb.2015.12.005} {\bibfield  {journal} {\bibinfo  {journal} {\plb}\ }\textbf {\bibinfo {volume} {753}},\ \bibinfo {pages} {13} (\bibinfo {year} {2016})}\BibitemShut {NoStop}%
\bibitem [{\citenamefont {Karampagia}\ and\ \citenamefont {Zelevinsky}(2020)}]{Karampagia:20}%
  \BibitemOpen
  \bibfield  {author} {\bibinfo {author} {\bibfnamefont {S.}~\bibnamefont {Karampagia}}\ and\ \bibinfo {author} {\bibfnamefont {V.}~\bibnamefont {Zelevinsky}},\ }\bibfield  {title} {\bibinfo {title} {Nuclear shell model and level density},\ }\href {https://doi.org/10.1142/S0218301320300052} {\bibfield  {journal} {\bibinfo  {journal} {Int. J. Mod. Phys. E}\ }\textbf {\bibinfo {volume} {29}},\ \bibinfo {pages} {2030005} (\bibinfo {year} {2020})}\BibitemShut {NoStop}%
\bibitem [{\citenamefont {Ormand}\ and\ \citenamefont {Brown}(2020)}]{Ormand:20}%
  \BibitemOpen
  \bibfield  {author} {\bibinfo {author} {\bibfnamefont {W.~E.}\ \bibnamefont {Ormand}}\ and\ \bibinfo {author} {\bibfnamefont {B.~A.}\ \bibnamefont {Brown}},\ }\bibfield  {title} {\bibinfo {title} {Microscopic calculations of nuclear level densities with the lanczos method},\ }\href {https://doi.org/10.1103/PhysRevC.102.014315} {\bibfield  {journal} {\bibinfo  {journal} {Phys. Rev. C}\ }\textbf {\bibinfo {volume} {102}},\ \bibinfo {pages} {014315} (\bibinfo {year} {2020})}\BibitemShut {NoStop}%
\bibitem [{\citenamefont {Tsoneva}\ and\ \citenamefont {Lenske}(2016)}]{Tsoneva:2016aa}%
  \BibitemOpen
  \bibfield  {author} {\bibinfo {author} {\bibfnamefont {N.}~\bibnamefont {Tsoneva}}\ and\ \bibinfo {author} {\bibfnamefont {H.}~\bibnamefont {Lenske}},\ }\bibfield  {title} {\bibinfo {title} {Energy--density functional plus quasiparticle--phonon model theory as a powerful tool for nuclear structure and astrophysics},\ }\href {https://doi.org/10.1134/S1063778816060247} {\bibfield  {journal} {\bibinfo  {journal} {Phys. At. Nucl.}\ }\textbf {\bibinfo {volume} {79}},\ \bibinfo {pages} {885} (\bibinfo {year} {2016})}\BibitemShut {NoStop}%
\bibitem [{\citenamefont {Goriely}\ \emph {et~al.}(2018)\citenamefont {Goriely}, \citenamefont {Hilaire}, \citenamefont {P\'eru},\ and\ \citenamefont {Sieja}}]{Goriely:18a}%
  \BibitemOpen
  \bibfield  {author} {\bibinfo {author} {\bibfnamefont {S.}~\bibnamefont {Goriely}}, \bibinfo {author} {\bibfnamefont {S.}~\bibnamefont {Hilaire}}, \bibinfo {author} {\bibfnamefont {S.}~\bibnamefont {P\'eru}},\ and\ \bibinfo {author} {\bibfnamefont {K.}~\bibnamefont {Sieja}},\ }\bibfield  {title} {\bibinfo {title} {Gogny-hfb+qrpa dipole strength function and its application to radiative nucleon capture cross section},\ }\href {https://doi.org/10.1103/PhysRevC.98.014327} {\bibfield  {journal} {\bibinfo  {journal} {Phys. Rev. C}\ }\textbf {\bibinfo {volume} {98}},\ \bibinfo {pages} {014327} (\bibinfo {year} {2018})}\BibitemShut {NoStop}%
\bibitem [{\citenamefont {Goriely}\ and\ \citenamefont {Plujko}(2019)}]{Goriely:19a}%
  \BibitemOpen
  \bibfield  {author} {\bibinfo {author} {\bibfnamefont {S.}~\bibnamefont {Goriely}}\ and\ \bibinfo {author} {\bibfnamefont {V.}~\bibnamefont {Plujko}},\ }\bibfield  {title} {\bibinfo {title} {Simple empirical $e1$ and $m1$ strength functions for practical applications},\ }\href {https://doi.org/10.1103/PhysRevC.99.014303} {\bibfield  {journal} {\bibinfo  {journal} {Phys. Rev. C}\ }\textbf {\bibinfo {volume} {99}},\ \bibinfo {pages} {014303} (\bibinfo {year} {2019})}\BibitemShut {NoStop}%
\bibitem [{\citenamefont {Goriely}\ \emph {et~al.}(2019)\citenamefont {Goriely}, \citenamefont {Dimitriou}, \citenamefont {Wiedeking}, \citenamefont {Belgya}, \citenamefont {Firestone}, \citenamefont {Kopecky}, \citenamefont {Krti{\v c}ka}, \citenamefont {Plujko}, \citenamefont {Schwengner}, \citenamefont {Siem} \emph {et~al.}}]{Goriely:2019aa}%
  \BibitemOpen
  \bibfield  {author} {\bibinfo {author} {\bibfnamefont {S.}~\bibnamefont {Goriely}}, \bibinfo {author} {\bibfnamefont {P.}~\bibnamefont {Dimitriou}}, \bibinfo {author} {\bibfnamefont {M.}~\bibnamefont {Wiedeking}}, \bibinfo {author} {\bibfnamefont {T.}~\bibnamefont {Belgya}}, \bibinfo {author} {\bibfnamefont {R.}~\bibnamefont {Firestone}}, \bibinfo {author} {\bibfnamefont {J.}~\bibnamefont {Kopecky}}, \bibinfo {author} {\bibfnamefont {M.}~\bibnamefont {Krti{\v c}ka}}, \bibinfo {author} {\bibfnamefont {V.}~\bibnamefont {Plujko}}, \bibinfo {author} {\bibfnamefont {R.}~\bibnamefont {Schwengner}}, \bibinfo {author} {\bibfnamefont {S.}~\bibnamefont {Siem}}, \emph {et~al.},\ }\bibfield  {title} {\bibinfo {title} {Reference database for photon strength functions},\ }\href {https://doi.org/10.1140/epja/i2019-12840-1} {\bibfield  {journal} {\bibinfo  {journal} {\epja}\ }\textbf {\bibinfo {volume} {55}},\ \bibinfo {pages} {172} (\bibinfo {year} {2019})}\BibitemShut {NoStop}%
\bibitem [{\citenamefont {Hebborn}\ \emph {et~al.}(2023)\citenamefont {Hebborn}, \citenamefont {Nunes}, \citenamefont {Potel}, \citenamefont {Dickhoff}, \citenamefont {Holt}, \citenamefont {Atkinson}, \citenamefont {Baker}, \citenamefont {Barbieri}, \citenamefont {Blanchon}, \citenamefont {Burrows} \emph {et~al.}}]{HebbornNPDH2022}%
  \BibitemOpen
  \bibfield  {author} {\bibinfo {author} {\bibfnamefont {C.}~\bibnamefont {Hebborn}}, \bibinfo {author} {\bibfnamefont {F.~M.}\ \bibnamefont {Nunes}}, \bibinfo {author} {\bibfnamefont {G.}~\bibnamefont {Potel}}, \bibinfo {author} {\bibfnamefont {W.~H.}\ \bibnamefont {Dickhoff}}, \bibinfo {author} {\bibfnamefont {J.~W.}\ \bibnamefont {Holt}}, \bibinfo {author} {\bibfnamefont {M.~C.}\ \bibnamefont {Atkinson}}, \bibinfo {author} {\bibfnamefont {R.~B.}\ \bibnamefont {Baker}}, \bibinfo {author} {\bibfnamefont {C.}~\bibnamefont {Barbieri}}, \bibinfo {author} {\bibfnamefont {G.}~\bibnamefont {Blanchon}}, \bibinfo {author} {\bibfnamefont {M.}~\bibnamefont {Burrows}}, \emph {et~al.},\ }\bibfield  {title} {\bibinfo {title} {Optical potentials for the rare-isotope beam era},\ }\href {https://dx.doi.org/10.1088/1361-6471/acc348} {\bibfield  {journal} {\bibinfo  {journal} {\jpg}\ }\textbf {\bibinfo {volume} {50}},\ \bibinfo {pages} {060501} (\bibinfo {year} {2023})}\BibitemShut {NoStop}%
\bibitem [{\citenamefont {{Pruitt, Cole D.}}\ \emph {et~al.}(2025)\citenamefont {{Pruitt, Cole D.}}, \citenamefont {{Perrotta, Salvatore Simone}}, \citenamefont {{Escher, Jutta}},\ and\ \citenamefont {{Gorton, Oliver}}}]{Pruitt:25cnr}%
  \BibitemOpen
  \bibfield  {author} {\bibinfo {author} {\bibnamefont {{Pruitt, Cole D.}}}, \bibinfo {author} {\bibnamefont {{Perrotta, Salvatore Simone}}}, \bibinfo {author} {\bibnamefont {{Escher, Jutta}}},\ and\ \bibinfo {author} {\bibnamefont {{Gorton, Oliver}}},\ }\bibfield  {title} {\bibinfo {title} {Extending the dispersive optical model to $\beta$-unstable systems},\ }\href {https://doi.org/10.1051/epjconf/202532205001} {\bibfield  {journal} {\bibinfo  {journal} {EPJ Web Conf.}\ }\textbf {\bibinfo {volume} {322}},\ \bibinfo {pages} {05001} (\bibinfo {year} {2025})}\BibitemShut {NoStop}%
\bibitem [{\citenamefont {Dickhoff}\ and\ \citenamefont {Charity}(2019)}]{Dickhoff:19}%
  \BibitemOpen
  \bibfield  {author} {\bibinfo {author} {\bibfnamefont {W.}~\bibnamefont {Dickhoff}}\ and\ \bibinfo {author} {\bibfnamefont {R.}~\bibnamefont {Charity}},\ }\bibfield  {title} {\bibinfo {title} {Recent developments for the optical model of nuclei},\ }\href {https://doi.org/https://doi.org/10.1016/j.ppnp.2018.11.002} {\bibfield  {journal} {\bibinfo  {journal} {\ppnp}\ }\textbf {\bibinfo {volume} {105}},\ \bibinfo {pages} {252} (\bibinfo {year} {2019})}\BibitemShut {NoStop}%
\bibitem [{\citenamefont {Bauge}\ \emph {et~al.}(2000)\citenamefont {Bauge}, \citenamefont {Delaroche}, \citenamefont {Girod}, \citenamefont {Haouat}, \citenamefont {Lachkar}, \citenamefont {Patin}, \citenamefont {Sigaud},\ and\ \citenamefont {Chardine}}]{Bauge:00a}%
  \BibitemOpen
  \bibfield  {author} {\bibinfo {author} {\bibfnamefont {E.}~\bibnamefont {Bauge}}, \bibinfo {author} {\bibfnamefont {J.~P.}\ \bibnamefont {Delaroche}}, \bibinfo {author} {\bibfnamefont {M.}~\bibnamefont {Girod}}, \bibinfo {author} {\bibfnamefont {G.}~\bibnamefont {Haouat}}, \bibinfo {author} {\bibfnamefont {J.}~\bibnamefont {Lachkar}}, \bibinfo {author} {\bibfnamefont {Y.}~\bibnamefont {Patin}}, \bibinfo {author} {\bibfnamefont {J.}~\bibnamefont {Sigaud}},\ and\ \bibinfo {author} {\bibfnamefont {J.}~\bibnamefont {Chardine}},\ }\bibfield  {title} {\bibinfo {title} {Neutron scattering from the ${}^{155,156,157,158,160}\mathrm{Gd}$ isotopes: Measurements and analyses with a deformed, semimicroscopic optical model},\ }\href {https://doi.org/10.1103/PhysRevC.61.034306} {\bibfield  {journal} {\bibinfo  {journal} {Phys. Rev. C}\ }\textbf {\bibinfo {volume} {61}},\ \bibinfo {pages} {034306} (\bibinfo {year} {2000})}\BibitemShut {NoStop}%
\bibitem [{\citenamefont {Blanchon}\ \emph {et~al.}(2015)\citenamefont {Blanchon}, \citenamefont {Dupuis}, \citenamefont {Arellano},\ and\ \citenamefont {Vinh~Mau}}]{Blanchon:15}%
  \BibitemOpen
  \bibfield  {author} {\bibinfo {author} {\bibfnamefont {G.}~\bibnamefont {Blanchon}}, \bibinfo {author} {\bibfnamefont {M.}~\bibnamefont {Dupuis}}, \bibinfo {author} {\bibfnamefont {H.~F.}\ \bibnamefont {Arellano}},\ and\ \bibinfo {author} {\bibfnamefont {N.}~\bibnamefont {Vinh~Mau}},\ }\bibfield  {title} {\bibinfo {title} {Microscopic positive-energy potential based on the gogny interaction},\ }\href {https://doi.org/10.1103/PhysRevC.91.014612} {\bibfield  {journal} {\bibinfo  {journal} {Phys. Rev. C}\ }\textbf {\bibinfo {volume} {91}},\ \bibinfo {pages} {014612} (\bibinfo {year} {2015})}\BibitemShut {NoStop}%
\bibitem [{\citenamefont {Bostr\"om}\ \emph {et~al.}(2025)\citenamefont {Bostr\"om}, \citenamefont {Rotureau}, \citenamefont {Carlsson},\ and\ \citenamefont {Idini}}]{Bostrom:25}%
  \BibitemOpen
  \bibfield  {author} {\bibinfo {author} {\bibfnamefont {J.}~\bibnamefont {Bostr\"om}}, \bibinfo {author} {\bibfnamefont {J.}~\bibnamefont {Rotureau}}, \bibinfo {author} {\bibfnamefont {B.~G.}\ \bibnamefont {Carlsson}},\ and\ \bibinfo {author} {\bibfnamefont {A.}~\bibnamefont {Idini}},\ }\bibfield  {title} {\bibinfo {title} {Nuclear cross sections from low-energy interactions},\ }\href {https://doi.org/10.1103/hfnm-ytj5} {\bibfield  {journal} {\bibinfo  {journal} {Phys. Rev. C}\ }\textbf {\bibinfo {volume} {112}},\ \bibinfo {pages} {L051602} (\bibinfo {year} {2025})}\BibitemShut {NoStop}%
\bibitem [{\citenamefont {Catacora-Rios}\ \emph {et~al.}(2021)\citenamefont {Catacora-Rios}, \citenamefont {King}, \citenamefont {Lovell},\ and\ \citenamefont {Nunes}}]{CatacoraRios:21}%
  \BibitemOpen
  \bibfield  {author} {\bibinfo {author} {\bibfnamefont {M.}~\bibnamefont {Catacora-Rios}}, \bibinfo {author} {\bibfnamefont {G.~B.}\ \bibnamefont {King}}, \bibinfo {author} {\bibfnamefont {A.~E.}\ \bibnamefont {Lovell}},\ and\ \bibinfo {author} {\bibfnamefont {F.~M.}\ \bibnamefont {Nunes}},\ }\bibfield  {title} {\bibinfo {title} {Statistical tools for a better optical model},\ }\href {https://doi.org/10.1103/PhysRevC.104.064611} {\bibfield  {journal} {\bibinfo  {journal} {Phys. Rev. C}\ }\textbf {\bibinfo {volume} {104}},\ \bibinfo {pages} {064611} (\bibinfo {year} {2021})}\BibitemShut {NoStop}%
\bibitem [{\citenamefont {{Pruitt}}\ \emph {et~al.}(2023)\citenamefont {{Pruitt}}, \citenamefont {{Escher}},\ and\ \citenamefont {{Rahman}}}]{2023PhRvC.107a4602P}%
  \BibitemOpen
  \bibfield  {author} {\bibinfo {author} {\bibfnamefont {C.~D.}\ \bibnamefont {{Pruitt}}}, \bibinfo {author} {\bibfnamefont {J.~E.}\ \bibnamefont {{Escher}}},\ and\ \bibinfo {author} {\bibfnamefont {R.}~\bibnamefont {{Rahman}}},\ }\bibfield  {title} {\bibinfo {title} {{Uncertainty-quantified phenomenological optical potentials for single-nucleon scattering}},\ }\href {https://doi.org/10.1103/PhysRevC.107.014602} {\bibfield  {journal} {\bibinfo  {journal} {\prc}\ }\textbf {\bibinfo {volume} {107}},\ \bibinfo {eid} {014602} (\bibinfo {year} {2023})}\BibitemShut {NoStop}%
\bibitem [{\citenamefont {Beyer}\ \emph {et~al.}(2025)\citenamefont {Beyer}, \citenamefont {Lovell}, \citenamefont {Pruitt}, \citenamefont {Giha},\ and\ \citenamefont {Kiedrowski}}]{Beyer-PRC-2025}%
  \BibitemOpen
  \bibfield  {author} {\bibinfo {author} {\bibfnamefont {K.~A.}\ \bibnamefont {Beyer}}, \bibinfo {author} {\bibfnamefont {A.~E.}\ \bibnamefont {Lovell}}, \bibinfo {author} {\bibfnamefont {C.~D.}\ \bibnamefont {Pruitt}}, \bibinfo {author} {\bibfnamefont {N.~P.}\ \bibnamefont {Giha}},\ and\ \bibinfo {author} {\bibfnamefont {B.~C.}\ \bibnamefont {Kiedrowski}},\ }\bibfield  {title} {\bibinfo {title} {Uncertainty quantification of optical models in fission fragment deexcitation},\ }\href {https://doi.org/10.1103/4kny-mj7g} {\bibfield  {journal} {\bibinfo  {journal} {Phys. Rev. C}\ }\textbf {\bibinfo {volume} {112}},\ \bibinfo {pages} {024604} (\bibinfo {year} {2025})}\BibitemShut {NoStop}%
\bibitem [{\citenamefont {{Dimitrakopoulos}}\ \emph {et~al.}(2025)\citenamefont {{Dimitrakopoulos}}, \citenamefont {{Perdikakis}}, \citenamefont {{Montes}}, \citenamefont {{Gastis}}, \citenamefont {{Kuvin}}, \citenamefont {{Lee}}, \citenamefont {{Tsintari}},\ and\ \citenamefont {{Voinov}}}]{Dimitrakopoulos:2025}%
  \BibitemOpen
  \bibfield  {author} {\bibinfo {author} {\bibfnamefont {N.}~\bibnamefont {{Dimitrakopoulos}}}, \bibinfo {author} {\bibfnamefont {G.}~\bibnamefont {{Perdikakis}}}, \bibinfo {author} {\bibfnamefont {F.}~\bibnamefont {{Montes}}}, \bibinfo {author} {\bibfnamefont {P.}~\bibnamefont {{Gastis}}}, \bibinfo {author} {\bibfnamefont {S.~A.}\ \bibnamefont {{Kuvin}}}, \bibinfo {author} {\bibfnamefont {H.~Y.}\ \bibnamefont {{Lee}}}, \bibinfo {author} {\bibfnamefont {P.}~\bibnamefont {{Tsintari}}},\ and\ \bibinfo {author} {\bibfnamefont {A.~V.}\ \bibnamefont {{Voinov}}},\ }\bibfield  {title} {\bibinfo {title} {{New approach for the quantification of uncertainties in reaction modeling via data-driven multi-objective optimization}},\ }\href {https://doi.org/10.48550/arXiv.2507.06370} {\bibfield  {journal} {\bibinfo  {journal} {arXiv e-prints}\ ,\ \bibinfo {eid} {arXiv:2507.06370}} (\bibinfo {year} {2025})}\BibitemShut {NoStop}%
\bibitem [{\citenamefont {Escher}\ \emph {et~al.}(2012)\citenamefont {Escher}, \citenamefont {Burke}, \citenamefont {Dietrich}, \citenamefont {Scielzo}, \citenamefont {Thompson},\ and\ \citenamefont {Younes}}]{Escher:12rmp}%
  \BibitemOpen
  \bibfield  {author} {\bibinfo {author} {\bibfnamefont {J.~E.}\ \bibnamefont {Escher}}, \bibinfo {author} {\bibfnamefont {J.~T.}\ \bibnamefont {Burke}}, \bibinfo {author} {\bibfnamefont {F.~S.}\ \bibnamefont {Dietrich}}, \bibinfo {author} {\bibfnamefont {N.~D.}\ \bibnamefont {Scielzo}}, \bibinfo {author} {\bibfnamefont {I.~J.}\ \bibnamefont {Thompson}},\ and\ \bibinfo {author} {\bibfnamefont {W.}~\bibnamefont {Younes}},\ }\bibfield  {title} {\bibinfo {title} {Compound-nuclear reaction cross sections from surrogate measurements},\ }\href {https://doi.org/10.1103/RevModPhys.84.353} {\bibfield  {journal} {\bibinfo  {journal} {Rev. Mod. Phys.}\ }\textbf {\bibinfo {volume} {84}},\ \bibinfo {pages} {353} (\bibinfo {year} {2012})}\BibitemShut {NoStop}%
\bibitem [{\citenamefont {{Escher, J. E.}}\ \emph {et~al.}(2016)\citenamefont {{Escher, J. E.}}, \citenamefont {{Tonchev, A. P.}}, \citenamefont {{Burke, J. T.}}, \citenamefont {{Bedrossian, P.}}, \citenamefont {{Casperson, R. J.}}, \citenamefont {{Cooper, N.}}, \citenamefont {{Hughes, R. O.}}, \citenamefont {{Humby, P.}}, \citenamefont {{Ilieva, R. S.}}, \citenamefont {{Ota, S.}} \emph {et~al.}}]{Escher:16a}%
  \BibitemOpen
  \bibfield  {author} {\bibinfo {author} {\bibnamefont {{Escher, J. E.}}}, \bibinfo {author} {\bibnamefont {{Tonchev, A. P.}}}, \bibinfo {author} {\bibnamefont {{Burke, J. T.}}}, \bibinfo {author} {\bibnamefont {{Bedrossian, P.}}}, \bibinfo {author} {\bibnamefont {{Casperson, R. J.}}}, \bibinfo {author} {\bibnamefont {{Cooper, N.}}}, \bibinfo {author} {\bibnamefont {{Hughes, R. O.}}}, \bibinfo {author} {\bibnamefont {{Humby, P.}}}, \bibinfo {author} {\bibnamefont {{Ilieva, R. S.}}}, \bibinfo {author} {\bibnamefont {{Ota, S.}}}, \emph {et~al.},\ }\bibfield  {title} {\bibinfo {title} {Compound-nuclear reactions with unstable nuclei: Constraining theory through innovative experimental approaches},\ }\href {https://doi.org/10.1051/epjconf/201612212001} {\bibfield  {journal} {\bibinfo  {journal} {EPJ Web of Conferences}\ }\textbf {\bibinfo {volume} {122}},\ \bibinfo {pages} {12001} (\bibinfo {year} {2016})}\BibitemShut {NoStop}%
\bibitem [{\citenamefont {Escher}(2025)}]{Escher:2025puf}%
  \BibitemOpen
  \bibfield  {author} {\bibinfo {author} {\bibfnamefont {J.~E.}\ \bibnamefont {Escher}},\ }\bibfield  {title} {\bibinfo {title} {{The Surrogate Nuclear Reaction Method: Concept, recent advances, and new opportunities}},\ }\href {https://doi.org/10.1051/epjconf/202532203001} {\bibfield  {journal} {\bibinfo  {journal} {EPJ Web Conf.}\ }\textbf {\bibinfo {volume} {322}},\ \bibinfo {pages} {03001} (\bibinfo {year} {2025})}\BibitemShut {NoStop}%
\bibitem [{\citenamefont {Larsen}\ \emph {et~al.}(2019)\citenamefont {Larsen}, \citenamefont {Spyrou}, \citenamefont {Liddick},\ and\ \citenamefont {Guttormsen}}]{Larsen:19}%
  \BibitemOpen
  \bibfield  {author} {\bibinfo {author} {\bibfnamefont {A.}~\bibnamefont {Larsen}}, \bibinfo {author} {\bibfnamefont {A.}~\bibnamefont {Spyrou}}, \bibinfo {author} {\bibfnamefont {S.}~\bibnamefont {Liddick}},\ and\ \bibinfo {author} {\bibfnamefont {M.}~\bibnamefont {Guttormsen}},\ }\bibfield  {title} {\bibinfo {title} {Novel techniques for constraining neutron-capture rates relevant for r-process heavy-element nucleosynthesis},\ }\href {https://doi.org/10.1016/j.ppnp.2019.04.002} {\bibfield  {journal} {\bibinfo  {journal} {\ppnp}\ }\textbf {\bibinfo {volume} {107}},\ \bibinfo {pages} {69} (\bibinfo {year} {2019})}\BibitemShut {NoStop}%
\bibitem [{\citenamefont {Ingeberg}\ \emph {et~al.}(2020)\citenamefont {Ingeberg}, \citenamefont {Siem}, \citenamefont {Wiedeking}, \citenamefont {Sieja}, \citenamefont {Bleuel}, \citenamefont {Brits}, \citenamefont {Bucher}, \citenamefont {Dinoko}, \citenamefont {Easton}, \citenamefont {G{\"o}rgen} \emph {et~al.}}]{Ingeberg:2020aa}%
  \BibitemOpen
  \bibfield  {author} {\bibinfo {author} {\bibfnamefont {V.~W.}\ \bibnamefont {Ingeberg}}, \bibinfo {author} {\bibfnamefont {S.}~\bibnamefont {Siem}}, \bibinfo {author} {\bibfnamefont {M.}~\bibnamefont {Wiedeking}}, \bibinfo {author} {\bibfnamefont {K.}~\bibnamefont {Sieja}}, \bibinfo {author} {\bibfnamefont {D.~L.}\ \bibnamefont {Bleuel}}, \bibinfo {author} {\bibfnamefont {C.~P.}\ \bibnamefont {Brits}}, \bibinfo {author} {\bibfnamefont {T.~D.}\ \bibnamefont {Bucher}}, \bibinfo {author} {\bibfnamefont {T.~S.}\ \bibnamefont {Dinoko}}, \bibinfo {author} {\bibfnamefont {J.~L.}\ \bibnamefont {Easton}}, \bibinfo {author} {\bibfnamefont {A.}~\bibnamefont {G{\"o}rgen}}, \emph {et~al.},\ }\bibfield  {title} {\bibinfo {title} {First application of the oslo method in inverse kinematics},\ }\href {https://doi.org/10.1140/epja/s10050-020-00070-7} {\bibfield  {journal} {\bibinfo  {journal} {\epja}\ }\textbf {\bibinfo {volume} {56}},\ \bibinfo {pages} {68} (\bibinfo {year} {2020})}\BibitemShut {NoStop}%
\bibitem [{\citenamefont {Wiedeking}\ \emph {et~al.}(2021)\citenamefont {Wiedeking}, \citenamefont {Guttormsen}, \citenamefont {Larsen}, \citenamefont {Zeiser}, \citenamefont {G\"orgen}, \citenamefont {Liddick}, \citenamefont {M\"ucher}, \citenamefont {Siem},\ and\ \citenamefont {Spyrou}}]{Wiedeking:21}%
  \BibitemOpen
  \bibfield  {author} {\bibinfo {author} {\bibfnamefont {M.}~\bibnamefont {Wiedeking}}, \bibinfo {author} {\bibfnamefont {M.}~\bibnamefont {Guttormsen}}, \bibinfo {author} {\bibfnamefont {A.~C.}\ \bibnamefont {Larsen}}, \bibinfo {author} {\bibfnamefont {F.}~\bibnamefont {Zeiser}}, \bibinfo {author} {\bibfnamefont {A.}~\bibnamefont {G\"orgen}}, \bibinfo {author} {\bibfnamefont {S.~N.}\ \bibnamefont {Liddick}}, \bibinfo {author} {\bibfnamefont {D.}~\bibnamefont {M\"ucher}}, \bibinfo {author} {\bibfnamefont {S.}~\bibnamefont {Siem}},\ and\ \bibinfo {author} {\bibfnamefont {A.}~\bibnamefont {Spyrou}},\ }\bibfield  {title} {\bibinfo {title} {Independent normalization for $\ensuremath{\gamma}$-ray strength functions: The shape method},\ }\href {https://doi.org/10.1103/PhysRevC.104.014311} {\bibfield  {journal} {\bibinfo  {journal} {Phys. Rev. C}\ }\textbf {\bibinfo {volume} {104}},\ \bibinfo {pages} {014311} (\bibinfo {year} {2021})}\BibitemShut {NoStop}%
\bibitem [{\citenamefont {Escher}\ \emph {et~al.}(2018)\citenamefont {Escher}, \citenamefont {Harke}, \citenamefont {Hughes}, \citenamefont {Scielzo}, \citenamefont {Casperson}, \citenamefont {Ota}, \citenamefont {Park}, \citenamefont {Saastamoinen},\ and\ \citenamefont {Ross}}]{Escher:18prl}%
  \BibitemOpen
  \bibfield  {author} {\bibinfo {author} {\bibfnamefont {J.~E.}\ \bibnamefont {Escher}}, \bibinfo {author} {\bibfnamefont {J.~T.}\ \bibnamefont {Harke}}, \bibinfo {author} {\bibfnamefont {R.~O.}\ \bibnamefont {Hughes}}, \bibinfo {author} {\bibfnamefont {N.~D.}\ \bibnamefont {Scielzo}}, \bibinfo {author} {\bibfnamefont {R.~J.}\ \bibnamefont {Casperson}}, \bibinfo {author} {\bibfnamefont {S.}~\bibnamefont {Ota}}, \bibinfo {author} {\bibfnamefont {H.~I.}\ \bibnamefont {Park}}, \bibinfo {author} {\bibfnamefont {A.}~\bibnamefont {Saastamoinen}},\ and\ \bibinfo {author} {\bibfnamefont {T.~J.}\ \bibnamefont {Ross}},\ }\bibfield  {title} {\bibinfo {title} {Constraining neutron capture cross sections for unstable nuclei with surrogate reaction data and theory},\ }\href {https://doi.org/10.1103/PhysRevLett.121.052501} {\bibfield  {journal} {\bibinfo  {journal} {Phys. Rev. Lett.}\ }\textbf {\bibinfo {volume} {121}},\ \bibinfo {pages} {052501} (\bibinfo {year} {2018})}\BibitemShut {NoStop}%
\bibitem [{\citenamefont {Ratkiewicz}\ \emph {et~al.}(2019)\citenamefont {Ratkiewicz}, \citenamefont {Cizewski}, \citenamefont {Escher}, \citenamefont {Potel}, \citenamefont {Burke}, \citenamefont {Casperson}, \citenamefont {McCleskey}, \citenamefont {Austin}, \citenamefont {Burcher}, \citenamefont {Hughes} \emph {et~al.}}]{Ratkiewicz:19prl}%
  \BibitemOpen
  \bibfield  {author} {\bibinfo {author} {\bibfnamefont {A.}~\bibnamefont {Ratkiewicz}}, \bibinfo {author} {\bibfnamefont {J.~A.}\ \bibnamefont {Cizewski}}, \bibinfo {author} {\bibfnamefont {J.~E.}\ \bibnamefont {Escher}}, \bibinfo {author} {\bibfnamefont {G.}~\bibnamefont {Potel}}, \bibinfo {author} {\bibfnamefont {J.~T.}\ \bibnamefont {Burke}}, \bibinfo {author} {\bibfnamefont {R.~J.}\ \bibnamefont {Casperson}}, \bibinfo {author} {\bibfnamefont {M.}~\bibnamefont {McCleskey}}, \bibinfo {author} {\bibfnamefont {R.~A.~E.}\ \bibnamefont {Austin}}, \bibinfo {author} {\bibfnamefont {S.}~\bibnamefont {Burcher}}, \bibinfo {author} {\bibfnamefont {R.~O.}\ \bibnamefont {Hughes}}, \emph {et~al.},\ }\bibfield  {title} {\bibinfo {title} {Towards neutron capture on exotic nuclei: Demonstrating $(d,p\ensuremath{\gamma})$ as a surrogate reaction for $(n,\ensuremath{\gamma})$},\ }\href {https://doi.org/10.1103/PhysRevLett.122.052502} {\bibfield  {journal} {\bibinfo  {journal} {Phys. Rev. Lett.}\ }\textbf {\bibinfo {volume}
  {122}},\ \bibinfo {pages} {052502} (\bibinfo {year} {2019})}\BibitemShut {NoStop}%
\bibitem [{\citenamefont {P\'erez~S\'anchez}\ \emph {et~al.}(2020)\citenamefont {P\'erez~S\'anchez}, \citenamefont {Jurado}, \citenamefont {M\'eot}, \citenamefont {Roig}, \citenamefont {Dupuis}, \citenamefont {Bouland}, \citenamefont {Denis-Petit}, \citenamefont {Marini}, \citenamefont {Mathieu}, \citenamefont {Tsekhanovich} \emph {et~al.}}]{PerezSanchez:20}%
  \BibitemOpen
  \bibfield  {author} {\bibinfo {author} {\bibfnamefont {R.}~\bibnamefont {P\'erez~S\'anchez}}, \bibinfo {author} {\bibfnamefont {B.}~\bibnamefont {Jurado}}, \bibinfo {author} {\bibfnamefont {V.}~\bibnamefont {M\'eot}}, \bibinfo {author} {\bibfnamefont {O.}~\bibnamefont {Roig}}, \bibinfo {author} {\bibfnamefont {M.}~\bibnamefont {Dupuis}}, \bibinfo {author} {\bibfnamefont {O.}~\bibnamefont {Bouland}}, \bibinfo {author} {\bibfnamefont {D.}~\bibnamefont {Denis-Petit}}, \bibinfo {author} {\bibfnamefont {P.}~\bibnamefont {Marini}}, \bibinfo {author} {\bibfnamefont {L.}~\bibnamefont {Mathieu}}, \bibinfo {author} {\bibfnamefont {I.}~\bibnamefont {Tsekhanovich}}, \emph {et~al.},\ }\bibfield  {title} {\bibinfo {title} {Simultaneous determination of neutron-induced fission and radiative capture cross sections from decay probabilities obtained with a surrogate reaction},\ }\href {https://doi.org/10.1103/PhysRevLett.125.122502} {\bibfield  {journal} {\bibinfo  {journal} {Phys. Rev. Lett.}\ }\textbf {\bibinfo {volume}
  {125}},\ \bibinfo {pages} {122502} (\bibinfo {year} {2020})}\BibitemShut {NoStop}%
\bibitem [{\citenamefont {Sguazzin}\ \emph {et~al.}(2025)\citenamefont {Sguazzin}, \citenamefont {Jurado}, \citenamefont {Pibernat}, \citenamefont {Swartz}, \citenamefont {Grieser}, \citenamefont {Glorius}, \citenamefont {Litvinov}, \citenamefont {Adamczewski-Musch}, \citenamefont {Alfaurt}, \citenamefont {Ascher} \emph {et~al.}}]{Sguazzin:25prl}%
  \BibitemOpen
  \bibfield  {author} {\bibinfo {author} {\bibfnamefont {M.}~\bibnamefont {Sguazzin}}, \bibinfo {author} {\bibfnamefont {B.}~\bibnamefont {Jurado}}, \bibinfo {author} {\bibfnamefont {J.}~\bibnamefont {Pibernat}}, \bibinfo {author} {\bibfnamefont {J.~A.}\ \bibnamefont {Swartz}}, \bibinfo {author} {\bibfnamefont {M.}~\bibnamefont {Grieser}}, \bibinfo {author} {\bibfnamefont {J.}~\bibnamefont {Glorius}}, \bibinfo {author} {\bibfnamefont {Y.~A.}\ \bibnamefont {Litvinov}}, \bibinfo {author} {\bibfnamefont {J.}~\bibnamefont {Adamczewski-Musch}}, \bibinfo {author} {\bibfnamefont {P.}~\bibnamefont {Alfaurt}}, \bibinfo {author} {\bibfnamefont {P.}~\bibnamefont {Ascher}}, \emph {et~al.},\ }\bibfield  {title} {\bibinfo {title} {First measurement of the neutron-emission probability with a surrogate reaction in inverse kinematics at a heavy-ion storage ring},\ }\href {https://doi.org/10.1103/PhysRevLett.134.072501} {\bibfield  {journal} {\bibinfo  {journal} {Phys. Rev. Lett.}\ }\textbf {\bibinfo {volume} {134}},\ \bibinfo
  {pages} {072501} (\bibinfo {year} {2025})}\BibitemShut {NoStop}%
\bibitem [{\citenamefont {{Thapa}}\ \emph {et~al.}(2026)\citenamefont {{Thapa}}, \citenamefont {{Escher}}, \citenamefont {{Chimanski}}, \citenamefont {{Gorton}}, \citenamefont {{Dupuis}}, \citenamefont {{In}}, \citenamefont {{Ota}}, \citenamefont {{P{\'e}ru}},\ and\ \citenamefont {{Younes}}}]{2026PhLB..87640433T}%
  \BibitemOpen
  \bibfield  {author} {\bibinfo {author} {\bibfnamefont {A.}~\bibnamefont {{Thapa}}}, \bibinfo {author} {\bibfnamefont {J.}~\bibnamefont {{Escher}}}, \bibinfo {author} {\bibfnamefont {E.}~\bibnamefont {{Chimanski}}}, \bibinfo {author} {\bibfnamefont {O.}~\bibnamefont {{Gorton}}}, \bibinfo {author} {\bibfnamefont {M.}~\bibnamefont {{Dupuis}}}, \bibinfo {author} {\bibfnamefont {E.~J.}\ \bibnamefont {{In}}}, \bibinfo {author} {\bibfnamefont {S.}~\bibnamefont {{Ota}}}, \bibinfo {author} {\bibfnamefont {S.}~\bibnamefont {{P{\'e}ru}}},\ and\ \bibinfo {author} {\bibfnamefont {W.}~\bibnamefont {{Younes}}},\ }\bibfield  {title} {\bibinfo {title} {{Constraining capture cross sections using proton inelastic scattering as a surrogate reaction}},\ }\href {https://doi.org/10.1016/j.physletb.2026.140433} {\bibfield  {journal} {\bibinfo  {journal} {\plb}\ }\textbf {\bibinfo {volume} {876}},\ \bibinfo {eid} {140433} (\bibinfo {year} {2026})}\BibitemShut {NoStop}%
\bibitem [{\citenamefont {{Spyrou}}\ \emph {et~al.}(2014)\citenamefont {{Spyrou}}, \citenamefont {{Liddick}}, \citenamefont {{Larsen}}, \citenamefont {{Guttormsen}}, \citenamefont {{Cooper}}, \citenamefont {{Dombos}}, \citenamefont {{Morrissey}}, \citenamefont {{Naqvi}}, \citenamefont {{Perdikakis}}, \citenamefont {{Quinn}} \emph {et~al.}}]{Spyrou:2014}%
  \BibitemOpen
  \bibfield  {author} {\bibinfo {author} {\bibfnamefont {A.}~\bibnamefont {{Spyrou}}}, \bibinfo {author} {\bibfnamefont {S.~N.}\ \bibnamefont {{Liddick}}}, \bibinfo {author} {\bibfnamefont {A.~C.}\ \bibnamefont {{Larsen}}}, \bibinfo {author} {\bibfnamefont {M.}~\bibnamefont {{Guttormsen}}}, \bibinfo {author} {\bibfnamefont {K.}~\bibnamefont {{Cooper}}}, \bibinfo {author} {\bibfnamefont {A.~C.}\ \bibnamefont {{Dombos}}}, \bibinfo {author} {\bibfnamefont {D.~J.}\ \bibnamefont {{Morrissey}}}, \bibinfo {author} {\bibfnamefont {F.}~\bibnamefont {{Naqvi}}}, \bibinfo {author} {\bibfnamefont {G.}~\bibnamefont {{Perdikakis}}}, \bibinfo {author} {\bibfnamefont {S.~J.}\ \bibnamefont {{Quinn}}}, \emph {et~al.},\ }\bibfield  {title} {\bibinfo {title} {{Novel technique for Constraining r -Process (n , {\ensuremath{\gamma}} ) Reaction Rates}},\ }\href {https://doi.org/10.1103/PhysRevLett.113.232502} {\bibfield  {journal} {\bibinfo  {journal} {\prl}\ }\textbf {\bibinfo {volume} {113}},\ \bibinfo {eid} {232502} (\bibinfo
  {year} {2014})}\BibitemShut {NoStop}%
\bibitem [{\citenamefont {{Liddick}}\ \emph {et~al.}(2016)\citenamefont {{Liddick}}, \citenamefont {{Spyrou}}, \citenamefont {{Crider}}, \citenamefont {{Naqvi}}, \citenamefont {{Larsen}}, \citenamefont {{Guttormsen}}, \citenamefont {{Mumpower}}, \citenamefont {{Surman}}, \citenamefont {{Perdikakis}}, \citenamefont {{Bleuel}} \emph {et~al.}}]{Liddick:2016}%
  \BibitemOpen
  \bibfield  {author} {\bibinfo {author} {\bibfnamefont {S.~N.}\ \bibnamefont {{Liddick}}}, \bibinfo {author} {\bibfnamefont {A.}~\bibnamefont {{Spyrou}}}, \bibinfo {author} {\bibfnamefont {B.~P.}\ \bibnamefont {{Crider}}}, \bibinfo {author} {\bibfnamefont {F.}~\bibnamefont {{Naqvi}}}, \bibinfo {author} {\bibfnamefont {A.~C.}\ \bibnamefont {{Larsen}}}, \bibinfo {author} {\bibfnamefont {M.}~\bibnamefont {{Guttormsen}}}, \bibinfo {author} {\bibfnamefont {M.}~\bibnamefont {{Mumpower}}}, \bibinfo {author} {\bibfnamefont {R.}~\bibnamefont {{Surman}}}, \bibinfo {author} {\bibfnamefont {G.}~\bibnamefont {{Perdikakis}}}, \bibinfo {author} {\bibfnamefont {D.~L.}\ \bibnamefont {{Bleuel}}}, \emph {et~al.},\ }\bibfield  {title} {\bibinfo {title} {{Experimental Neutron Capture Rate Constraint Far from Stability}},\ }\href {https://doi.org/10.1103/PhysRevLett.116.242502} {\bibfield  {journal} {\bibinfo  {journal} {\prl}\ }\textbf {\bibinfo {volume} {116}},\ \bibinfo {eid} {242502} (\bibinfo {year} {2016})}\BibitemShut
  {NoStop}%
\bibitem [{\citenamefont {{Spyrou}}\ \emph {et~al.}(2017)\citenamefont {{Spyrou}}, \citenamefont {{Larsen}}, \citenamefont {{Liddick}}, \citenamefont {{Naqvi}}, \citenamefont {{Crider}}, \citenamefont {{Dombos}}, \citenamefont {{Guttormsen}}, \citenamefont {{Bleuel}}, \citenamefont {{Couture}}, \citenamefont {{Crespo Campo}} \emph {et~al.}}]{Spyrou:2017}%
  \BibitemOpen
  \bibfield  {author} {\bibinfo {author} {\bibfnamefont {A.}~\bibnamefont {{Spyrou}}}, \bibinfo {author} {\bibfnamefont {A.~C.}\ \bibnamefont {{Larsen}}}, \bibinfo {author} {\bibfnamefont {S.~N.}\ \bibnamefont {{Liddick}}}, \bibinfo {author} {\bibfnamefont {F.}~\bibnamefont {{Naqvi}}}, \bibinfo {author} {\bibfnamefont {B.~P.}\ \bibnamefont {{Crider}}}, \bibinfo {author} {\bibfnamefont {A.~C.}\ \bibnamefont {{Dombos}}}, \bibinfo {author} {\bibfnamefont {M.}~\bibnamefont {{Guttormsen}}}, \bibinfo {author} {\bibfnamefont {D.~L.}\ \bibnamefont {{Bleuel}}}, \bibinfo {author} {\bibfnamefont {A.}~\bibnamefont {{Couture}}}, \bibinfo {author} {\bibfnamefont {L.}~\bibnamefont {{Crespo Campo}}}, \emph {et~al.},\ }\bibfield  {title} {\bibinfo {title} {{Neutron-capture rates for explosive nucleosynthesis: the case of $^{68}$Ni(n, {\ensuremath{\gamma}})$^{69}$Ni}},\ }\href {https://doi.org/10.1088/1361-6471/aa5ae7} {\bibfield  {journal} {\bibinfo  {journal} {\jpg}\ }\textbf {\bibinfo {volume} {44}},\ \bibinfo {eid}
  {044002} (\bibinfo {year} {2017})}\BibitemShut {NoStop}%
\bibitem [{\citenamefont {{M{\"u}cher}}\ \emph {et~al.}(2023)\citenamefont {{M{\"u}cher}}, \citenamefont {{Spyrou}}, \citenamefont {{Wiedeking}}, \citenamefont {{Guttormsen}}, \citenamefont {{Larsen}}, \citenamefont {{Zeiser}}, \citenamefont {{Harris}}, \citenamefont {{Richard}}, \citenamefont {{Smith}}, \citenamefont {{G{\"o}rgen}} \emph {et~al.}}]{Mucher:2023}%
  \BibitemOpen
  \bibfield  {author} {\bibinfo {author} {\bibfnamefont {D.}~\bibnamefont {{M{\"u}cher}}}, \bibinfo {author} {\bibfnamefont {A.}~\bibnamefont {{Spyrou}}}, \bibinfo {author} {\bibfnamefont {M.}~\bibnamefont {{Wiedeking}}}, \bibinfo {author} {\bibfnamefont {M.}~\bibnamefont {{Guttormsen}}}, \bibinfo {author} {\bibfnamefont {A.~C.}\ \bibnamefont {{Larsen}}}, \bibinfo {author} {\bibfnamefont {F.}~\bibnamefont {{Zeiser}}}, \bibinfo {author} {\bibfnamefont {C.}~\bibnamefont {{Harris}}}, \bibinfo {author} {\bibfnamefont {A.~L.}\ \bibnamefont {{Richard}}}, \bibinfo {author} {\bibfnamefont {M.~K.}\ \bibnamefont {{Smith}}}, \bibinfo {author} {\bibfnamefont {A.}~\bibnamefont {{G{\"o}rgen}}}, \emph {et~al.},\ }\bibfield  {title} {\bibinfo {title} {{Extracting model-independent nuclear level densities away from stability}},\ }\href {https://doi.org/10.1103/PhysRevC.107.L011602} {\bibfield  {journal} {\bibinfo  {journal} {\prc}\ }\textbf {\bibinfo {volume} {107}},\ \bibinfo {eid} {L011602} (\bibinfo {year}
  {2023})}\BibitemShut {NoStop}%
\bibitem [{\citenamefont {{Horowitz}}\ \emph {et~al.}(2019)\citenamefont {{Horowitz}}, \citenamefont {{Arcones}}, \citenamefont {{C{\^o}t{\'e}}}, \citenamefont {{Dillmann}}, \citenamefont {{Nazarewicz}}, \citenamefont {{Roederer}}, \citenamefont {{Schatz}}, \citenamefont {{Aprahamian}}, \citenamefont {{Atanasov}}, \citenamefont {{Bauswein}} \emph {et~al.}}]{Horowitz+2019}%
  \BibitemOpen
  \bibfield  {author} {\bibinfo {author} {\bibfnamefont {C.~J.}\ \bibnamefont {{Horowitz}}}, \bibinfo {author} {\bibfnamefont {A.}~\bibnamefont {{Arcones}}}, \bibinfo {author} {\bibfnamefont {B.}~\bibnamefont {{C{\^o}t{\'e}}}}, \bibinfo {author} {\bibfnamefont {I.}~\bibnamefont {{Dillmann}}}, \bibinfo {author} {\bibfnamefont {W.}~\bibnamefont {{Nazarewicz}}}, \bibinfo {author} {\bibfnamefont {I.~U.}\ \bibnamefont {{Roederer}}}, \bibinfo {author} {\bibfnamefont {H.}~\bibnamefont {{Schatz}}}, \bibinfo {author} {\bibfnamefont {A.}~\bibnamefont {{Aprahamian}}}, \bibinfo {author} {\bibfnamefont {D.}~\bibnamefont {{Atanasov}}}, \bibinfo {author} {\bibfnamefont {A.}~\bibnamefont {{Bauswein}}}, \emph {et~al.},\ }\bibfield  {title} {\bibinfo {title} {{r-process nucleosynthesis: connecting rare-isotope beam facilities with the cosmos}},\ }\href {https://doi.org/10.1088/1361-6471/ab0849} {\bibfield  {journal} {\bibinfo  {journal} {\jpg}\ }\textbf {\bibinfo {volume} {46}},\ \bibinfo {pages} {083001} (\bibinfo {year}
  {2019})}\BibitemShut {NoStop}%
\bibitem [{\citenamefont {{Brown}}\ \emph {et~al.}(2025)\citenamefont {{Brown}}, \citenamefont {{Gade}}, \citenamefont {{Stroberg}}, \citenamefont {{Escher}}, \citenamefont {{Fossez}}, \citenamefont {{Giuliani}}, \citenamefont {{Hoffman}}, \citenamefont {{Nazarewicz}}, \citenamefont {{Seng}}, \citenamefont {{Sorensen}} \emph {et~al.}}]{FRIB-motivation-2024}%
  \BibitemOpen
  \bibfield  {author} {\bibinfo {author} {\bibfnamefont {B.~A.}\ \bibnamefont {{Brown}}}, \bibinfo {author} {\bibfnamefont {A.}~\bibnamefont {{Gade}}}, \bibinfo {author} {\bibfnamefont {S.~R.}\ \bibnamefont {{Stroberg}}}, \bibinfo {author} {\bibfnamefont {J.~E.}\ \bibnamefont {{Escher}}}, \bibinfo {author} {\bibfnamefont {K.}~\bibnamefont {{Fossez}}}, \bibinfo {author} {\bibfnamefont {P.}~\bibnamefont {{Giuliani}}}, \bibinfo {author} {\bibfnamefont {C.~R.}\ \bibnamefont {{Hoffman}}}, \bibinfo {author} {\bibfnamefont {W.}~\bibnamefont {{Nazarewicz}}}, \bibinfo {author} {\bibfnamefont {C.-Y.}\ \bibnamefont {{Seng}}}, \bibinfo {author} {\bibfnamefont {A.}~\bibnamefont {{Sorensen}}}, \emph {et~al.},\ }\bibfield  {title} {\bibinfo {title} {{Motivations for early high-profile FRIB experiments}},\ }\href {https://doi.org/10.1088/1361-6471/adb449} {\bibfield  {journal} {\bibinfo  {journal} {\jpg}\ }\textbf {\bibinfo {volume} {52}},\ \bibinfo {eid} {050501} (\bibinfo {year} {2025})}\BibitemShut {NoStop}%
\bibitem [{\citenamefont {{Surman}}\ and\ \citenamefont {{Engel}}(2001)}]{Surman+2001}%
  \BibitemOpen
  \bibfield  {author} {\bibinfo {author} {\bibfnamefont {R.}~\bibnamefont {{Surman}}}\ and\ \bibinfo {author} {\bibfnamefont {J.}~\bibnamefont {{Engel}}},\ }\bibfield  {title} {\bibinfo {title} {{Changes in r-process abundances at late times}},\ }\href {https://doi.org/10.1103/PhysRevC.64.035801} {\bibfield  {journal} {\bibinfo  {journal} {\prc}\ }\textbf {\bibinfo {volume} {64}},\ \bibinfo {eid} {035801} (\bibinfo {year} {2001})}\BibitemShut {NoStop}%
\bibitem [{\citenamefont {{Surman}}\ \emph {et~al.}(2009)\citenamefont {{Surman}}, \citenamefont {{Beun}}, \citenamefont {{McLaughlin}},\ and\ \citenamefont {{Hix}}}]{Surman+2009}%
  \BibitemOpen
  \bibfield  {author} {\bibinfo {author} {\bibfnamefont {R.}~\bibnamefont {{Surman}}}, \bibinfo {author} {\bibfnamefont {J.}~\bibnamefont {{Beun}}}, \bibinfo {author} {\bibfnamefont {G.~C.}\ \bibnamefont {{McLaughlin}}},\ and\ \bibinfo {author} {\bibfnamefont {W.~R.}\ \bibnamefont {{Hix}}},\ }\bibfield  {title} {\bibinfo {title} {{Neutron capture rates near A=130 that effect a global change to the r-process abundance distribution}},\ }\href {https://doi.org/10.1103/PhysRevC.79.045809} {\bibfield  {journal} {\bibinfo  {journal} {\prc}\ }\textbf {\bibinfo {volume} {79}},\ \bibinfo {eid} {045809} (\bibinfo {year} {2009})}\BibitemShut {NoStop}%
\bibitem [{\citenamefont {{Arcones}}\ and\ \citenamefont {{Mart{\'\i}nez-Pinedo}}(2011)}]{Arcones+2011}%
  \BibitemOpen
  \bibfield  {author} {\bibinfo {author} {\bibfnamefont {A.}~\bibnamefont {{Arcones}}}\ and\ \bibinfo {author} {\bibfnamefont {G.}~\bibnamefont {{Mart{\'\i}nez-Pinedo}}},\ }\bibfield  {title} {\bibinfo {title} {{Dynamical r-process studies within the neutrino-driven wind scenario and its sensitivity to the nuclear physics input}},\ }\href {https://doi.org/10.1103/PhysRevC.83.045809} {\bibfield  {journal} {\bibinfo  {journal} {\prc}\ }\textbf {\bibinfo {volume} {83}},\ \bibinfo {eid} {045809} (\bibinfo {year} {2011})}\BibitemShut {NoStop}%
\bibitem [{\citenamefont {{Mumpower}}\ \emph {et~al.}(2012)\citenamefont {{Mumpower}}, \citenamefont {{McLaughlin}},\ and\ \citenamefont {{Surman}}}]{Mumpower+2012}%
  \BibitemOpen
  \bibfield  {author} {\bibinfo {author} {\bibfnamefont {M.~R.}\ \bibnamefont {{Mumpower}}}, \bibinfo {author} {\bibfnamefont {G.~C.}\ \bibnamefont {{McLaughlin}}},\ and\ \bibinfo {author} {\bibfnamefont {R.}~\bibnamefont {{Surman}}},\ }\bibfield  {title} {\bibinfo {title} {{Influence of neutron capture rates in the rare earth region on the r-process abundance pattern}},\ }\href {https://doi.org/10.1103/PhysRevC.86.035803} {\bibfield  {journal} {\bibinfo  {journal} {\prc}\ }\textbf {\bibinfo {volume} {86}},\ \bibinfo {eid} {035803} (\bibinfo {year} {2012})}\BibitemShut {NoStop}%
\bibitem [{\citenamefont {{Mumpower}}\ \emph {et~al.}(2015)\citenamefont {{Mumpower}}, \citenamefont {{Surman}},\ and\ \citenamefont {{Aprahamian}}}]{Mumpower+2015}%
  \BibitemOpen
  \bibfield  {author} {\bibinfo {author} {\bibfnamefont {M.}~\bibnamefont {{Mumpower}}}, \bibinfo {author} {\bibfnamefont {R.}~\bibnamefont {{Surman}}},\ and\ \bibinfo {author} {\bibfnamefont {A.}~\bibnamefont {{Aprahamian}}},\ }\bibfield  {title} {\bibinfo {title} {{Variances in r-process predictions from uncertain nuclear rates}},\ }in\ \href {https://doi.org/10.1088/1742-6596/599/1/012031} {\emph {\bibinfo {booktitle} {J. Phys.: Conf. Ser.}}},\ Vol.\ \bibinfo {volume} {599}\ (\bibinfo  {publisher} {IOP},\ \bibinfo {year} {2015})\ p.\ \bibinfo {pages} {012031}\BibitemShut {NoStop}%
\bibitem [{\citenamefont {{Martinet}}\ \emph {et~al.}(2025)\citenamefont {{Martinet}}, \citenamefont {{Goriely}}, \citenamefont {{Choplin}},\ and\ \citenamefont {{Siess}}}]{2025EPJA...61...48M}%
  \BibitemOpen
  \bibfield  {author} {\bibinfo {author} {\bibfnamefont {S.}~\bibnamefont {{Martinet}}}, \bibinfo {author} {\bibfnamefont {S.}~\bibnamefont {{Goriely}}}, \bibinfo {author} {\bibfnamefont {A.}~\bibnamefont {{Choplin}}},\ and\ \bibinfo {author} {\bibfnamefont {L.}~\bibnamefont {{Siess}}},\ }\bibfield  {title} {\bibinfo {title} {{Statistical framework for nuclear parameter uncertainties in nucleosynthesis modeling of r- and i-process}},\ }\href {https://doi.org/10.1140/epja/s10050-025-01510-y} {\bibfield  {journal} {\bibinfo  {journal} {\epja}\ }\textbf {\bibinfo {volume} {61}},\ \bibinfo {eid} {48} (\bibinfo {year} {2025})}\BibitemShut {NoStop}%
\bibitem [{\citenamefont {{Martinet}}\ and\ \citenamefont {{Goriely}}(2025)}]{2025A&A...694A.180M}%
  \BibitemOpen
  \bibfield  {author} {\bibinfo {author} {\bibfnamefont {S.}~\bibnamefont {{Martinet}}}\ and\ \bibinfo {author} {\bibfnamefont {S.}~\bibnamefont {{Goriely}}},\ }\bibfield  {title} {\bibinfo {title} {{The impact of mass uncertainties on r-process nucleosynthesis in neutron star mergers}},\ }\href {https://doi.org/10.1051/0004-6361/202451991} {\bibfield  {journal} {\bibinfo  {journal} {\aap}\ }\textbf {\bibinfo {volume} {694}},\ \bibinfo {eid} {A180} (\bibinfo {year} {2025})}\BibitemShut {NoStop}%
\bibitem [{\citenamefont {{Surman}}\ \emph {et~al.}(2014)\citenamefont {{Surman}}, \citenamefont {{Mumpower}}, \citenamefont {{Sinclair}}, \citenamefont {{Jones}}, \citenamefont {{Hix}},\ and\ \citenamefont {{McLaughlin}}}]{Surman:2014}%
  \BibitemOpen
  \bibfield  {author} {\bibinfo {author} {\bibfnamefont {R.}~\bibnamefont {{Surman}}}, \bibinfo {author} {\bibfnamefont {M.}~\bibnamefont {{Mumpower}}}, \bibinfo {author} {\bibfnamefont {R.}~\bibnamefont {{Sinclair}}}, \bibinfo {author} {\bibfnamefont {K.~L.}\ \bibnamefont {{Jones}}}, \bibinfo {author} {\bibfnamefont {W.~R.}\ \bibnamefont {{Hix}}},\ and\ \bibinfo {author} {\bibfnamefont {G.~C.}\ \bibnamefont {{McLaughlin}}},\ }\bibfield  {title} {\bibinfo {title} {{Sensitivity studies for the weak r process: neutron capture rates}},\ }\href {https://doi.org/10.1063/1.4867191} {\bibfield  {journal} {\bibinfo  {journal} {AIP Advances}\ }\textbf {\bibinfo {volume} {4}},\ \bibinfo {eid} {041008} (\bibinfo {year} {2014})}\BibitemShut {NoStop}%
\bibitem [{\citenamefont {{Vescovi}}\ \emph {et~al.}(2022)\citenamefont {{Vescovi}}, \citenamefont {{Reifarth}}, \citenamefont {{Cristallo}},\ and\ \citenamefont {{Couture}}}]{Vescovi+2022}%
  \BibitemOpen
  \bibfield  {author} {\bibinfo {author} {\bibfnamefont {D.}~\bibnamefont {{Vescovi}}}, \bibinfo {author} {\bibfnamefont {R.}~\bibnamefont {{Reifarth}}}, \bibinfo {author} {\bibfnamefont {S.}~\bibnamefont {{Cristallo}}},\ and\ \bibinfo {author} {\bibfnamefont {A.}~\bibnamefont {{Couture}}},\ }\bibfield  {title} {\bibinfo {title} {{Neutron-capture measurement candidates for the r-process in neutron star mergers}},\ }\href {https://doi.org/10.3389/fspas.2022.994980} {\bibfield  {journal} {\bibinfo  {journal} {Front. Astron. Space Sci.}\ }\textbf {\bibinfo {volume} {9}},\ \bibinfo {eid} {994980} (\bibinfo {year} {2022})}\BibitemShut {NoStop}%
\bibitem [{\citenamefont {{Nishimura}}\ \emph {et~al.}(2019)\citenamefont {{Nishimura}}, \citenamefont {{Rauscher}}, \citenamefont {{Hirschi}}, \citenamefont {{Cescutti}}, \citenamefont {{Murphy}},\ and\ \citenamefont {{Fr{\"o}hlich}}}]{Nishimura:2019}%
  \BibitemOpen
  \bibfield  {author} {\bibinfo {author} {\bibfnamefont {N.}~\bibnamefont {{Nishimura}}}, \bibinfo {author} {\bibfnamefont {T.}~\bibnamefont {{Rauscher}}}, \bibinfo {author} {\bibfnamefont {R.}~\bibnamefont {{Hirschi}}}, \bibinfo {author} {\bibfnamefont {G.}~\bibnamefont {{Cescutti}}}, \bibinfo {author} {\bibfnamefont {A.~S.~J.}\ \bibnamefont {{Murphy}}},\ and\ \bibinfo {author} {\bibfnamefont {C.}~\bibnamefont {{Fr{\"o}hlich}}},\ }\bibfield  {title} {\bibinfo {title} {{Uncertainties in {\ensuremath{\nu}}p-process nucleosynthesis from Monte Carlo variation of reaction rates}},\ }\href {https://doi.org/10.1093/mnras/stz2104} {\bibfield  {journal} {\bibinfo  {journal} {\mnras}\ }\textbf {\bibinfo {volume} {489}},\ \bibinfo {pages} {1379} (\bibinfo {year} {2019})}\BibitemShut {NoStop}%
\bibitem [{\citenamefont {{Bliss}}\ \emph {et~al.}(2020)\citenamefont {{Bliss}}, \citenamefont {{Arcones}}, \citenamefont {{Montes}},\ and\ \citenamefont {{Pereira}}}]{2020PhRvC.101e5807B}%
  \BibitemOpen
  \bibfield  {author} {\bibinfo {author} {\bibfnamefont {J.}~\bibnamefont {{Bliss}}}, \bibinfo {author} {\bibfnamefont {A.}~\bibnamefont {{Arcones}}}, \bibinfo {author} {\bibfnamefont {F.}~\bibnamefont {{Montes}}},\ and\ \bibinfo {author} {\bibfnamefont {J.}~\bibnamefont {{Pereira}}},\ }\bibfield  {title} {\bibinfo {title} {{Nuclear physics uncertainties in neutrino-driven, neutron-rich supernova ejecta}},\ }\href {https://doi.org/10.1103/PhysRevC.101.055807} {\bibfield  {journal} {\bibinfo  {journal} {\prc}\ }\textbf {\bibinfo {volume} {101}},\ \bibinfo {eid} {055807} (\bibinfo {year} {2020})}\BibitemShut {NoStop}%
\bibitem [{\citenamefont {{Psaltis}}\ \emph {et~al.}(2022)\citenamefont {{Psaltis}}, \citenamefont {{Arcones}}, \citenamefont {{Montes}}, \citenamefont {{Mohr}}, \citenamefont {{Hansen}}, \citenamefont {{Jacobi}},\ and\ \citenamefont {{Schatz}}}]{2022ApJ...935...27P}%
  \BibitemOpen
  \bibfield  {author} {\bibinfo {author} {\bibfnamefont {A.}~\bibnamefont {{Psaltis}}}, \bibinfo {author} {\bibfnamefont {A.}~\bibnamefont {{Arcones}}}, \bibinfo {author} {\bibfnamefont {F.}~\bibnamefont {{Montes}}}, \bibinfo {author} {\bibfnamefont {P.}~\bibnamefont {{Mohr}}}, \bibinfo {author} {\bibfnamefont {C.~J.}\ \bibnamefont {{Hansen}}}, \bibinfo {author} {\bibfnamefont {M.}~\bibnamefont {{Jacobi}}},\ and\ \bibinfo {author} {\bibfnamefont {H.}~\bibnamefont {{Schatz}}},\ }\bibfield  {title} {\bibinfo {title} {{Constraining Nucleosynthesis in Neutrino-driven Winds: Observations, Simulations, and Nuclear Physics}},\ }\href {https://doi.org/10.3847/1538-4357/ac7da7} {\bibfield  {journal} {\bibinfo  {journal} {\apj}\ }\textbf {\bibinfo {volume} {935}},\ \bibinfo {eid} {27} (\bibinfo {year} {2022})}\BibitemShut {NoStop}%
\bibitem [{\citenamefont {{Sprouse}}\ \emph {et~al.}(2024)\citenamefont {{Sprouse}}, \citenamefont {{Lund}}, \citenamefont {{Miller}}, \citenamefont {{McLaughlin}},\ and\ \citenamefont {{Mumpower}}}]{Sprouse+2024}%
  \BibitemOpen
  \bibfield  {author} {\bibinfo {author} {\bibfnamefont {T.~M.}\ \bibnamefont {{Sprouse}}}, \bibinfo {author} {\bibfnamefont {K.~A.}\ \bibnamefont {{Lund}}}, \bibinfo {author} {\bibfnamefont {J.~M.}\ \bibnamefont {{Miller}}}, \bibinfo {author} {\bibfnamefont {G.~C.}\ \bibnamefont {{McLaughlin}}},\ and\ \bibinfo {author} {\bibfnamefont {M.~R.}\ \bibnamefont {{Mumpower}}},\ }\bibfield  {title} {\bibinfo {title} {{Emergent Nucleosynthesis from a 1.2 s Long Simulation of a Black Hole Accretion Disk}},\ }\href {https://doi.org/10.3847/1538-4357/ad1819} {\bibfield  {journal} {\bibinfo  {journal} {\apj}\ }\textbf {\bibinfo {volume} {962}},\ \bibinfo {eid} {79} (\bibinfo {year} {2024})}\BibitemShut {NoStop}%
\bibitem [{\citenamefont {{Bernuzzi}}\ \emph {et~al.}(2025)\citenamefont {{Bernuzzi}}, \citenamefont {{Magistrelli}}, \citenamefont {{Jacobi}}, \citenamefont {{Logoteta}}, \citenamefont {{Perego}},\ and\ \citenamefont {{Radice}}}]{Bernuzzi+2025}%
  \BibitemOpen
  \bibfield  {author} {\bibinfo {author} {\bibfnamefont {S.}~\bibnamefont {{Bernuzzi}}}, \bibinfo {author} {\bibfnamefont {F.}~\bibnamefont {{Magistrelli}}}, \bibinfo {author} {\bibfnamefont {M.}~\bibnamefont {{Jacobi}}}, \bibinfo {author} {\bibfnamefont {D.}~\bibnamefont {{Logoteta}}}, \bibinfo {author} {\bibfnamefont {A.}~\bibnamefont {{Perego}}},\ and\ \bibinfo {author} {\bibfnamefont {D.}~\bibnamefont {{Radice}}},\ }\bibfield  {title} {\bibinfo {title} {{Long-lived neutron star remnants from asymmetric binary neutron star mergers: element formation, kilonova signals and gravitational waves}},\ }\href {https://doi.org/10.1093/mnras/staf1147} {\bibfield  {journal} {\bibinfo  {journal} {\mnras}\ }\textbf {\bibinfo {volume} {542}},\ \bibinfo {pages} {256} (\bibinfo {year} {2025})}\BibitemShut {NoStop}%
\bibitem [{\citenamefont {Sprouse}\ \emph {et~al.}(2021)\citenamefont {Sprouse}, \citenamefont {Mumpower},\ and\ \citenamefont {Surman}}]{PRISM}%
  \BibitemOpen
  \bibfield  {author} {\bibinfo {author} {\bibfnamefont {T.~M.}\ \bibnamefont {Sprouse}}, \bibinfo {author} {\bibfnamefont {M.~R.}\ \bibnamefont {Mumpower}},\ and\ \bibinfo {author} {\bibfnamefont {R.}~\bibnamefont {Surman}},\ }\bibfield  {title} {\bibinfo {title} {Following nuclei through nucleosynthesis: A novel tracing technique},\ }\href {https://doi.org/10.1103/PhysRevC.104.015803} {\bibfield  {journal} {\bibinfo  {journal} {Phys. Rev. C}\ }\textbf {\bibinfo {volume} {104}},\ \bibinfo {pages} {015803} (\bibinfo {year} {2021})}\BibitemShut {NoStop}%
\bibitem [{\citenamefont {{Lund}}\ \emph {et~al.}(2024)\citenamefont {{Lund}}, \citenamefont {{McLaughlin}}, \citenamefont {{Miller}},\ and\ \citenamefont {{Mumpower}}}]{2024ApJ...964..111L}%
  \BibitemOpen
  \bibfield  {author} {\bibinfo {author} {\bibfnamefont {K.~A.}\ \bibnamefont {{Lund}}}, \bibinfo {author} {\bibfnamefont {G.~C.}\ \bibnamefont {{McLaughlin}}}, \bibinfo {author} {\bibfnamefont {J.~M.}\ \bibnamefont {{Miller}}},\ and\ \bibinfo {author} {\bibfnamefont {M.~R.}\ \bibnamefont {{Mumpower}}},\ }\bibfield  {title} {\bibinfo {title} {{Magnetic Field Strength Effects on Nucleosynthesis from Neutron Star Merger Outflows}},\ }\href {https://doi.org/10.3847/1538-4357/ad25ef} {\bibfield  {journal} {\bibinfo  {journal} {\apj}\ }\textbf {\bibinfo {volume} {964}},\ \bibinfo {eid} {111} (\bibinfo {year} {2024})}\BibitemShut {NoStop}%
\bibitem [{\citenamefont {{Metzger}}\ and\ \citenamefont {{Fern{\'a}ndez}}(2014)}]{2014MNRAS.441.3444M}%
  \BibitemOpen
  \bibfield  {author} {\bibinfo {author} {\bibfnamefont {B.~D.}\ \bibnamefont {{Metzger}}}\ and\ \bibinfo {author} {\bibfnamefont {R.}~\bibnamefont {{Fern{\'a}ndez}}},\ }\bibfield  {title} {\bibinfo {title} {{Red or blue? A potential kilonova imprint of the delay until black hole formation following a neutron star merger}},\ }\href {https://doi.org/10.1093/mnras/stu802} {\bibfield  {journal} {\bibinfo  {journal} {\mnras}\ }\textbf {\bibinfo {volume} {441}},\ \bibinfo {pages} {3444} (\bibinfo {year} {2014})}\BibitemShut {NoStop}%
\bibitem [{\citenamefont {{Lippuner}}\ \emph {et~al.}(2017)\citenamefont {{Lippuner}}, \citenamefont {{Fern{\'a}ndez}}, \citenamefont {{Roberts}}, \citenamefont {{Foucart}}, \citenamefont {{Kasen}}, \citenamefont {{Metzger}},\ and\ \citenamefont {{Ott}}}]{2017MNRAS.472..904L}%
  \BibitemOpen
  \bibfield  {author} {\bibinfo {author} {\bibfnamefont {J.}~\bibnamefont {{Lippuner}}}, \bibinfo {author} {\bibfnamefont {R.}~\bibnamefont {{Fern{\'a}ndez}}}, \bibinfo {author} {\bibfnamefont {L.~F.}\ \bibnamefont {{Roberts}}}, \bibinfo {author} {\bibfnamefont {F.}~\bibnamefont {{Foucart}}}, \bibinfo {author} {\bibfnamefont {D.}~\bibnamefont {{Kasen}}}, \bibinfo {author} {\bibfnamefont {B.~D.}\ \bibnamefont {{Metzger}}},\ and\ \bibinfo {author} {\bibfnamefont {C.~D.}\ \bibnamefont {{Ott}}},\ }\bibfield  {title} {\bibinfo {title} {{Signatures of hypermassive neutron star lifetimes on r-process nucleosynthesis in the disc ejecta from neutron star mergers}},\ }\href {https://doi.org/10.1093/mnras/stx1987} {\bibfield  {journal} {\bibinfo  {journal} {\mnras}\ }\textbf {\bibinfo {volume} {472}},\ \bibinfo {pages} {904} (\bibinfo {year} {2017})}\BibitemShut {NoStop}%
\bibitem [{\citenamefont {{Roederer}}\ \emph {et~al.}(2022)\citenamefont {{Roederer}}, \citenamefont {{Lawler}}, \citenamefont {{Den Hartog}}, \citenamefont {{Placco}}, \citenamefont {{Surman}}, \citenamefont {{Beers}}, \citenamefont {{Ezzeddine}}, \citenamefont {{Frebel}}, \citenamefont {{Hansen}}, \citenamefont {{Hattori}} \emph {et~al.}}]{2022ApJS..260...27R}%
  \BibitemOpen
  \bibfield  {author} {\bibinfo {author} {\bibfnamefont {I.~U.}\ \bibnamefont {{Roederer}}}, \bibinfo {author} {\bibfnamefont {J.~E.}\ \bibnamefont {{Lawler}}}, \bibinfo {author} {\bibfnamefont {E.~A.}\ \bibnamefont {{Den Hartog}}}, \bibinfo {author} {\bibfnamefont {V.~M.}\ \bibnamefont {{Placco}}}, \bibinfo {author} {\bibfnamefont {R.}~\bibnamefont {{Surman}}}, \bibinfo {author} {\bibfnamefont {T.~C.}\ \bibnamefont {{Beers}}}, \bibinfo {author} {\bibfnamefont {R.}~\bibnamefont {{Ezzeddine}}}, \bibinfo {author} {\bibfnamefont {A.}~\bibnamefont {{Frebel}}}, \bibinfo {author} {\bibfnamefont {T.~T.}\ \bibnamefont {{Hansen}}}, \bibinfo {author} {\bibfnamefont {K.}~\bibnamefont {{Hattori}}}, \emph {et~al.},\ }\bibfield  {title} {\bibinfo {title} {{The R-process Alliance: A Nearly Complete R-process Abundance Template Derived from Ultraviolet Spectroscopy of the R-process-enhanced Metal-poor Star HD 222925}},\ }\href {https://doi.org/10.3847/1538-4365/ac5cbc} {\bibfield  {journal} {\bibinfo  {journal} {\apjs}\
  }\textbf {\bibinfo {volume} {260}},\ \bibinfo {eid} {27} (\bibinfo {year} {2022})}\BibitemShut {NoStop}%
\bibitem [{\citenamefont {{Surman}}\ and\ \citenamefont {{McLaughlin}}(2004)}]{2004ApJ...603..611S}%
  \BibitemOpen
  \bibfield  {author} {\bibinfo {author} {\bibfnamefont {R.}~\bibnamefont {{Surman}}}\ and\ \bibinfo {author} {\bibfnamefont {G.~C.}\ \bibnamefont {{McLaughlin}}},\ }\bibfield  {title} {\bibinfo {title} {{Neutrinos and Nucleosynthesis in Gamma-Ray Burst Accretion Disks}},\ }\href {https://doi.org/10.1086/381672} {\bibfield  {journal} {\bibinfo  {journal} {\apj}\ }\textbf {\bibinfo {volume} {603}},\ \bibinfo {pages} {611} (\bibinfo {year} {2004})}\BibitemShut {NoStop}%
\bibitem [{\citenamefont {{Cyburt}}\ \emph {et~al.}(2010)\citenamefont {{Cyburt}}, \citenamefont {{Amthor}}, \citenamefont {{Ferguson}}, \citenamefont {{Meisel}}, \citenamefont {{Smith}}, \citenamefont {{Warren}}, \citenamefont {{Heger}}, \citenamefont {{Hoffman}}, \citenamefont {{Rauscher}}, \citenamefont {{Sakharuk}}, \citenamefont {{Schatz}}, \citenamefont {{Thielemann}} \emph {et~al.}}]{cyburt2010jina}%
  \BibitemOpen
  \bibfield  {author} {\bibinfo {author} {\bibfnamefont {R.~H.}\ \bibnamefont {{Cyburt}}}, \bibinfo {author} {\bibfnamefont {A.~M.}\ \bibnamefont {{Amthor}}}, \bibinfo {author} {\bibfnamefont {R.}~\bibnamefont {{Ferguson}}}, \bibinfo {author} {\bibfnamefont {Z.}~\bibnamefont {{Meisel}}}, \bibinfo {author} {\bibfnamefont {K.}~\bibnamefont {{Smith}}}, \bibinfo {author} {\bibfnamefont {S.}~\bibnamefont {{Warren}}}, \bibinfo {author} {\bibfnamefont {A.}~\bibnamefont {{Heger}}}, \bibinfo {author} {\bibfnamefont {R.~D.}\ \bibnamefont {{Hoffman}}}, \bibinfo {author} {\bibfnamefont {T.}~\bibnamefont {{Rauscher}}}, \bibinfo {author} {\bibfnamefont {A.}~\bibnamefont {{Sakharuk}}}, \bibinfo {author} {\bibfnamefont {H.}~\bibnamefont {{Schatz}}}, \bibinfo {author} {\bibfnamefont {F.~K.}\ \bibnamefont {{Thielemann}}}, \emph {et~al.},\ }\bibfield  {title} {\bibinfo {title} {{The JINA REACLIB Database: Its Recent Updates and Impact on Type-I X-ray Bursts}},\ }\href {https://doi.org/10.1088/0067-0049/189/1/240} {\bibfield
  {journal} {\bibinfo  {journal} {\apjs}\ }\textbf {\bibinfo {volume} {189}},\ \bibinfo {pages} {240} (\bibinfo {year} {2010})}\BibitemShut {NoStop}%
\bibitem [{\citenamefont {{Kondev}}\ \emph {et~al.}(2021)\citenamefont {{Kondev}}, \citenamefont {{Wang}}, \citenamefont {{Huang}}, \citenamefont {{Naimi}},\ and\ \citenamefont {{Audi}}}]{2021ChPhC..45c0001K}%
  \BibitemOpen
  \bibfield  {author} {\bibinfo {author} {\bibfnamefont {F.~G.}\ \bibnamefont {{Kondev}}}, \bibinfo {author} {\bibfnamefont {M.}~\bibnamefont {{Wang}}}, \bibinfo {author} {\bibfnamefont {W.~J.}\ \bibnamefont {{Huang}}}, \bibinfo {author} {\bibfnamefont {S.}~\bibnamefont {{Naimi}}},\ and\ \bibinfo {author} {\bibfnamefont {G.}~\bibnamefont {{Audi}}},\ }\bibfield  {title} {\bibinfo {title} {{The NUBASE2020 evaluation of nuclear physics properties}},\ }\href {https://doi.org/10.1088/1674-1137/abddae} {\bibfield  {journal} {\bibinfo  {journal} {Chinese Physics C}\ }\textbf {\bibinfo {volume} {45}},\ \bibinfo {eid} {030001} (\bibinfo {year} {2021})}\BibitemShut {NoStop}%
\bibitem [{\citenamefont {{Wang}}\ \emph {et~al.}(2021)\citenamefont {{Wang}}, \citenamefont {{Huang}}, \citenamefont {{Kondev}}, \citenamefont {{Audi}},\ and\ \citenamefont {{Naimi}}}]{2021ChPhC..45c0003W}%
  \BibitemOpen
  \bibfield  {author} {\bibinfo {author} {\bibfnamefont {M.}~\bibnamefont {{Wang}}}, \bibinfo {author} {\bibfnamefont {W.~J.}\ \bibnamefont {{Huang}}}, \bibinfo {author} {\bibfnamefont {F.~G.}\ \bibnamefont {{Kondev}}}, \bibinfo {author} {\bibfnamefont {G.}~\bibnamefont {{Audi}}},\ and\ \bibinfo {author} {\bibfnamefont {S.}~\bibnamefont {{Naimi}}},\ }\bibfield  {title} {\bibinfo {title} {{The AME 2020 atomic mass evaluation (II). Tables, graphs and references}},\ }\href {https://doi.org/10.1088/1674-1137/abddaf} {\bibfield  {journal} {\bibinfo  {journal} {Chinese Physics C}\ }\textbf {\bibinfo {volume} {45}},\ \bibinfo {eid} {030003} (\bibinfo {year} {2021})}\BibitemShut {NoStop}%
\bibitem [{\citenamefont {{M{\"o}ller}}\ \emph {et~al.}(2012)\citenamefont {{M{\"o}ller}}, \citenamefont {{Myers}}, \citenamefont {{Sagawa}},\ and\ \citenamefont {{Yoshida}}}]{FRDM}%
  \BibitemOpen
  \bibfield  {author} {\bibinfo {author} {\bibfnamefont {P.}~\bibnamefont {{M{\"o}ller}}}, \bibinfo {author} {\bibfnamefont {W.~D.}\ \bibnamefont {{Myers}}}, \bibinfo {author} {\bibfnamefont {H.}~\bibnamefont {{Sagawa}}},\ and\ \bibinfo {author} {\bibfnamefont {S.}~\bibnamefont {{Yoshida}}},\ }\bibfield  {title} {\bibinfo {title} {{New Finite-Range Droplet Mass Model and Equation-of-State Parameters}},\ }\href {https://doi.org/10.1103/PhysRevLett.108.052501} {\bibfield  {journal} {\bibinfo  {journal} {\prl}\ }\textbf {\bibinfo {volume} {108}},\ \bibinfo {eid} {052501} (\bibinfo {year} {2012})}\BibitemShut {NoStop}%
\bibitem [{\citenamefont {{M{\"o}ller}}\ \emph {et~al.}(2019)\citenamefont {{M{\"o}ller}}, \citenamefont {{Mumpower}}, \citenamefont {{Kawano}},\ and\ \citenamefont {{Myers}}}]{2019ADNDT.125....1M}%
  \BibitemOpen
  \bibfield  {author} {\bibinfo {author} {\bibfnamefont {P.}~\bibnamefont {{M{\"o}ller}}}, \bibinfo {author} {\bibfnamefont {M.~R.}\ \bibnamefont {{Mumpower}}}, \bibinfo {author} {\bibfnamefont {T.}~\bibnamefont {{Kawano}}},\ and\ \bibinfo {author} {\bibfnamefont {W.~D.}\ \bibnamefont {{Myers}}},\ }\bibfield  {title} {\bibinfo {title} {{Nuclear properties for astrophysical and radioactive-ion-beam applications (II)}},\ }\href {https://doi.org/10.1016/j.adt.2018.03.003} {\bibfield  {journal} {\bibinfo  {journal} {Atomic Data and Nuclear Data Tables}\ }\textbf {\bibinfo {volume} {125}},\ \bibinfo {pages} {1} (\bibinfo {year} {2019})}\BibitemShut {NoStop}%
\bibitem [{\citenamefont {Ormand}(2021)}]{YAHFC}%
  \BibitemOpen
  \bibfield  {author} {\bibinfo {author} {\bibfnamefont {W.~E.}\ \bibnamefont {Ormand}},\ }\href {https://doi.org/10.2172/1808762} {\emph {\bibinfo {title} {Monte Carlo Hauser-Feshbach computer code system to model nuclear reactions: YAHFC}}},\ \bibinfo {type} {Tech. Rep.}\ (\bibinfo  {institution} {Lawrence Livermore National Laboratory (LLNL), Livermore, California, United States},\ \bibinfo {year} {2021})\BibitemShut {NoStop}%
\bibitem [{\citenamefont {Alhassid}(2015)}]{Alhassid:2015aa}%
  \BibitemOpen
  \bibfield  {author} {\bibinfo {author} {\bibfnamefont {Y.}~\bibnamefont {Alhassid}},\ }\bibfield  {title} {\bibinfo {title} {The shell model monte carlo approach to level densities: Recent developments and perspectives},\ }\href {https://doi.org/10.1140/epja/i2015-15171-3} {\bibfield  {journal} {\bibinfo  {journal} {\epja}\ }\textbf {\bibinfo {volume} {51}},\ \bibinfo {pages} {171} (\bibinfo {year} {2015})}\BibitemShut {NoStop}%
\bibitem [{\citenamefont {Hilaire}\ \emph {et~al.}(2023)\citenamefont {Hilaire}, \citenamefont {Goriely}, \citenamefont {P{\'e}ru},\ and\ \citenamefont {Gosselin}}]{Hilaire:23}%
  \BibitemOpen
  \bibfield  {author} {\bibinfo {author} {\bibfnamefont {S.}~\bibnamefont {Hilaire}}, \bibinfo {author} {\bibfnamefont {S.}~\bibnamefont {Goriely}}, \bibinfo {author} {\bibfnamefont {S.}~\bibnamefont {P{\'e}ru}},\ and\ \bibinfo {author} {\bibfnamefont {G.}~\bibnamefont {Gosselin}},\ }\bibfield  {title} {\bibinfo {title} {A new approach to nuclear level densities: The {QRPA} plus boson expansion},\ }\href {https://doi.org/10.1016/j.physletb.2023.137989} {\bibfield  {journal} {\bibinfo  {journal} {\plb}\ }\textbf {\bibinfo {volume} {843}},\ \bibinfo {pages} {137989} (\bibinfo {year} {2023})}\BibitemShut {NoStop}%
\bibitem [{\citenamefont {Goriely}\ \emph {et~al.}(2025)\citenamefont {Goriely}, \citenamefont {P{\'e}ru},\ and\ \citenamefont {Hilaire}}]{Goriely:25}%
  \BibitemOpen
  \bibfield  {author} {\bibinfo {author} {\bibfnamefont {S.}~\bibnamefont {Goriely}}, \bibinfo {author} {\bibfnamefont {S.}~\bibnamefont {P{\'e}ru}},\ and\ \bibinfo {author} {\bibfnamefont {S.}~\bibnamefont {Hilaire}},\ }\bibfield  {title} {\bibinfo {title} {{QRPA prediction of the nuclear level densities and de-excitation photon strength functions}},\ }\href {https://doi.org/10.1016/j.physletb.2025.139677} {\bibfield  {journal} {\bibinfo  {journal} {\plb}\ }\textbf {\bibinfo {volume} {868}},\ \bibinfo {pages} {139677} (\bibinfo {year} {2025})}\BibitemShut {NoStop}%
\bibitem [{\citenamefont {Sun}\ \emph {et~al.}(2025)\citenamefont {Sun}, \citenamefont {Wang}, \citenamefont {Dutta},\ and\ \citenamefont {Wang}}]{Sun:25}%
  \BibitemOpen
  \bibfield  {author} {\bibinfo {author} {\bibfnamefont {Y.}~\bibnamefont {Sun}}, \bibinfo {author} {\bibfnamefont {J.}~\bibnamefont {Wang}}, \bibinfo {author} {\bibfnamefont {S.}~\bibnamefont {Dutta}},\ and\ \bibinfo {author} {\bibfnamefont {L.-J.}\ \bibnamefont {Wang}},\ }\bibfield  {title} {\bibinfo {title} {A novel projected shell model method for nuclear level density},\ }\href {https://doi.org/10.1051/epjconf/202532902002} {\bibfield  {journal} {\bibinfo  {journal} {EPJ Web Conf.}\ }\textbf {\bibinfo {volume} {329}},\ \bibinfo {pages} {02002} (\bibinfo {year} {2025})}\BibitemShut {NoStop}%
\bibitem [{\citenamefont {{Gorton}}\ \emph {et~al.}(2026)\citenamefont {{Gorton}}, \citenamefont {{Kravvaris}}, \citenamefont {{Escher}},\ and\ \citenamefont {{Johnson}}}]{2026PhRvC.113d4327G}%
  \BibitemOpen
  \bibfield  {author} {\bibinfo {author} {\bibfnamefont {O.~C.}\ \bibnamefont {{Gorton}}}, \bibinfo {author} {\bibfnamefont {K.}~\bibnamefont {{Kravvaris}}}, \bibinfo {author} {\bibfnamefont {J.~E.}\ \bibnamefont {{Escher}}},\ and\ \bibinfo {author} {\bibfnamefont {C.~W.}\ \bibnamefont {{Johnson}}},\ }\bibfield  {title} {\bibinfo {title} {{Radiative strength functions from the energy-localized Brink-Axel hypothesis}},\ }\href {https://doi.org/10.1103/kg6r-t5d1} {\bibfield  {journal} {\bibinfo  {journal} {\prc}\ }\textbf {\bibinfo {volume} {113}},\ \bibinfo {eid} {044327} (\bibinfo {year} {2026})}\BibitemShut {NoStop}%
\bibitem [{\citenamefont {Bauge}\ \emph {et~al.}(2001)\citenamefont {Bauge}, \citenamefont {Delaroche},\ and\ \citenamefont {Girod}}]{Bauge:01}%
  \BibitemOpen
  \bibfield  {author} {\bibinfo {author} {\bibfnamefont {E.}~\bibnamefont {Bauge}}, \bibinfo {author} {\bibfnamefont {J.~P.}\ \bibnamefont {Delaroche}},\ and\ \bibinfo {author} {\bibfnamefont {M.}~\bibnamefont {Girod}},\ }\bibfield  {title} {\bibinfo {title} {Lane-consistent, semimicroscopic nucleon-nucleus optical model},\ }\href {https://doi.org/10.1103/PhysRevC.63.024607} {\bibfield  {journal} {\bibinfo  {journal} {Phys. Rev. C}\ }\textbf {\bibinfo {volume} {63}},\ \bibinfo {pages} {024607} (\bibinfo {year} {2001})}\BibitemShut {NoStop}%
\bibitem [{\citenamefont {Whitehead}\ \emph {et~al.}(2021)\citenamefont {Whitehead}, \citenamefont {Lim},\ and\ \citenamefont {Holt}}]{Whitehead:21}%
  \BibitemOpen
  \bibfield  {author} {\bibinfo {author} {\bibfnamefont {T.~R.}\ \bibnamefont {Whitehead}}, \bibinfo {author} {\bibfnamefont {Y.}~\bibnamefont {Lim}},\ and\ \bibinfo {author} {\bibfnamefont {J.~W.}\ \bibnamefont {Holt}},\ }\bibfield  {title} {\bibinfo {title} {Global microscopic description of nucleon-nucleus scattering with quantified uncertainties},\ }\href {https://doi.org/10.1103/PhysRevLett.127.182502} {\bibfield  {journal} {\bibinfo  {journal} {Phys. Rev. Lett.}\ }\textbf {\bibinfo {volume} {127}},\ \bibinfo {pages} {182502} (\bibinfo {year} {2021})}\BibitemShut {NoStop}%
\bibitem [{\citenamefont {Gilbert}\ and\ \citenamefont {Cameron}(1965)}]{gilbert1965composite}%
  \BibitemOpen
  \bibfield  {author} {\bibinfo {author} {\bibfnamefont {A.}~\bibnamefont {Gilbert}}\ and\ \bibinfo {author} {\bibfnamefont {A.~G.~W.}\ \bibnamefont {Cameron}},\ }\bibfield  {title} {\bibinfo {title} {A composite nuclear-level density formula with shell corrections},\ }\href {https://doi.org/10.1139/p65-139} {\bibfield  {journal} {\bibinfo  {journal} {Canadian Journal of Physics}\ }\textbf {\bibinfo {volume} {43}},\ \bibinfo {pages} {1446–1496} (\bibinfo {year} {1965})}\BibitemShut {NoStop}%
\bibitem [{\citenamefont {{Koning}}\ and\ \citenamefont {{Delaroche}}(2003)}]{2003NuPhA.713..231K}%
  \BibitemOpen
  \bibfield  {author} {\bibinfo {author} {\bibfnamefont {A.~J.}\ \bibnamefont {{Koning}}}\ and\ \bibinfo {author} {\bibfnamefont {J.~P.}\ \bibnamefont {{Delaroche}}},\ }\bibfield  {title} {\bibinfo {title} {{Local and global nucleon optical models from 1 keV to 200 MeV}},\ }\href {https://doi.org/10.1016/S0375-9474(02)01321-0} {\bibfield  {journal} {\bibinfo  {journal} {\nphysa}\ }\textbf {\bibinfo {volume} {713}},\ \bibinfo {pages} {231} (\bibinfo {year} {2003})}\BibitemShut {NoStop}%
\bibitem [{\citenamefont {Berryman}\ \emph {et~al.}(2026)\citenamefont {Berryman}, \citenamefont {Pruitt}, \citenamefont {Escher}, \citenamefont {Gorton}, \citenamefont {Holmbeck}, \citenamefont {Kravvaris}, \citenamefont {Sieverding}, \citenamefont {Cabrera~Garcia}, \citenamefont {Surman}, \citenamefont {Kedia},\ and\ \citenamefont {McLaughlin}}]{osti_3022089}%
  \BibitemOpen
  \bibfield  {author} {\bibinfo {author} {\bibfnamefont {J.}~\bibnamefont {Berryman}}, \bibinfo {author} {\bibfnamefont {C.}~\bibnamefont {Pruitt}}, \bibinfo {author} {\bibfnamefont {J.}~\bibnamefont {Escher}}, \bibinfo {author} {\bibfnamefont {O.}~\bibnamefont {Gorton}}, \bibinfo {author} {\bibfnamefont {E.}~\bibnamefont {Holmbeck}}, \bibinfo {author} {\bibfnamefont {K.}~\bibnamefont {Kravvaris}}, \bibinfo {author} {\bibfnamefont {A.}~\bibnamefont {Sieverding}}, \bibinfo {author} {\bibfnamefont {J.}~\bibnamefont {Cabrera~Garcia}}, \bibinfo {author} {\bibfnamefont {R.}~\bibnamefont {Surman}}, \bibinfo {author} {\bibfnamefont {A.}~\bibnamefont {Kedia}},\ and\ \bibinfo {author} {\bibfnamefont {G.}~\bibnamefont {McLaughlin}},\ }\href {https://doi.org/10.2172/3022089} {\emph {\bibinfo {title} {Computational Workflows for Uncertainty-Quantified Nuclear Reactions: From Nuclear Theory Inputs to Astrophysical Reaction Rates}}},\ \bibinfo {type} {Tech. Rep.}\ (\bibinfo  {institution} {Lawrence Livermore National
  Laboratory (LLNL), Livermore, California, United States},\ \bibinfo {year} {2026})\BibitemShut {NoStop}%
\bibitem [{\citenamefont {{Rauscher}}\ and\ \citenamefont {{Thielemann}}(2000)}]{non-smoker}%
  \BibitemOpen
  \bibfield  {author} {\bibinfo {author} {\bibfnamefont {T.}~\bibnamefont {{Rauscher}}}\ and\ \bibinfo {author} {\bibfnamefont {F.-K.}\ \bibnamefont {{Thielemann}}},\ }\bibfield  {title} {\bibinfo {title} {{Astrophysical Reaction Rates From Statistical Model Calculations}},\ }\href {https://doi.org/10.1006/adnd.2000.0834} {\bibfield  {journal} {\bibinfo  {journal} {Atomic Data and Nuclear Data Tables}\ }\textbf {\bibinfo {volume} {75}},\ \bibinfo {pages} {1} (\bibinfo {year} {2000})}\BibitemShut {NoStop}%
\bibitem [{\citenamefont {{Koning}}\ \emph {et~al.}(2023)\citenamefont {{Koning}}, \citenamefont {{Hilaire}},\ and\ \citenamefont {{Goriely}}}]{Koning+2023}%
  \BibitemOpen
  \bibfield  {author} {\bibinfo {author} {\bibfnamefont {A.}~\bibnamefont {{Koning}}}, \bibinfo {author} {\bibfnamefont {S.}~\bibnamefont {{Hilaire}}},\ and\ \bibinfo {author} {\bibfnamefont {S.}~\bibnamefont {{Goriely}}},\ }\bibfield  {title} {\bibinfo {title} {{TALYS: modeling of nuclear reactions}},\ }\href {https://doi.org/10.1140/epja/s10050-023-01034-3} {\bibfield  {journal} {\bibinfo  {journal} {\epja}\ }\textbf {\bibinfo {volume} {59}},\ \bibinfo {eid} {131} (\bibinfo {year} {2023})}\BibitemShut {NoStop}%
\bibitem [{\citenamefont {{Nikas}}\ \emph {et~al.}(2020)\citenamefont {{Nikas}}, \citenamefont {{Perdikakis}}, \citenamefont {{Beard}}, \citenamefont {{Surman}}, \citenamefont {{Mumpower}},\ and\ \citenamefont {{Tsintari}}}]{2020arXiv201001698N}%
  \BibitemOpen
  \bibfield  {author} {\bibinfo {author} {\bibfnamefont {S.}~\bibnamefont {{Nikas}}}, \bibinfo {author} {\bibfnamefont {G.}~\bibnamefont {{Perdikakis}}}, \bibinfo {author} {\bibfnamefont {M.}~\bibnamefont {{Beard}}}, \bibinfo {author} {\bibfnamefont {R.}~\bibnamefont {{Surman}}}, \bibinfo {author} {\bibfnamefont {M.~R.}\ \bibnamefont {{Mumpower}}},\ and\ \bibinfo {author} {\bibfnamefont {P.}~\bibnamefont {{Tsintari}}},\ }\bibfield  {title} {\bibinfo {title} {{Propagation of Hauser-Feshbach uncertainty estimates to r-process nucleosynthesis: Benchmark of statistical property models for neutron rich nuclei far from stability}},\ }\href {https://doi.org/10.48550/arXiv.2010.01698} {\bibfield  {journal} {\bibinfo  {journal} {arXiv e-prints}\ ,\ \bibinfo {eid} {arXiv:2010.01698}} (\bibinfo {year} {2020})}\BibitemShut {NoStop}%
\bibitem [{\citenamefont {{Rauscher}}\ \emph {et~al.}(1997)\citenamefont {{Rauscher}}, \citenamefont {{Thielemann}},\ and\ \citenamefont {{Kratz}}}]{1997PhRvC..56.1613R}%
  \BibitemOpen
  \bibfield  {author} {\bibinfo {author} {\bibfnamefont {T.}~\bibnamefont {{Rauscher}}}, \bibinfo {author} {\bibfnamefont {F.-K.}\ \bibnamefont {{Thielemann}}},\ and\ \bibinfo {author} {\bibfnamefont {K.-L.}\ \bibnamefont {{Kratz}}},\ }\bibfield  {title} {\bibinfo {title} {{Nuclear level density and the determination of thermonuclear rates for astrophysics}},\ }\href {https://doi.org/10.1103/PhysRevC.56.1613} {\bibfield  {journal} {\bibinfo  {journal} {\prc}\ }\textbf {\bibinfo {volume} {56}},\ \bibinfo {pages} {1613} (\bibinfo {year} {1997})}\BibitemShut {NoStop}%
\end{thebibliography}%

\end{document}